\documentclass[11pt]{article}
\usepackage[hyperfootnotes=false, colorlinks=true, allcolors=black]{hyperref}
   \usepackage{epsfig}
      \usepackage{amsmath}
      \usepackage{amssymb}
 
      \usepackage{caption}
   
\usepackage{subcaption}
\usepackage{cite}
\usepackage{mathtools}

\usepackage[margin=1 in]{geometry}
\usepackage{subcaption}

\usepackage{epstopdf}

\usepackage[normalem]{ulem}

\numberwithin{equation}{section}

\usepackage{amsmath}

\allowdisplaybreaks

\usepackage{setspace}
  \usepackage{graphicx}
  \usepackage{color}
\usepackage{comment}

 \newcommand{\be}{\begin{equation}}
\newcommand{\bea}{\begin{eqnarray}}
\newcommand{\eea}{\end{eqnarray}}
\newcommand{\beq}{\begin{equation}}
 \newcommand{\ee}{\end{equation}}

\def\nn{\nonumber}

\def\eff{\text{eff}}
\def\trumpet{\text{trumpet}}

\global\long\def\mC{\mathcal{C}}%
 
\global\long\def\mD{\mathcal{D}}%

\global\long\def\mF{\mathcal{F}}%

\global\long\def\mM{\mathcal{M}}%

\global\long\def\mW{\mathcal{W}}%

\global\long\def\mY{\mathcal{Y}}%
 
\global\long\def\mZ{\mathcal{Z}}%

\global\long\def\e{\epsilon}%
 
\global\long\def\ra{\rightarrow}%

\global\long\def\avg#1{\left\langle #1\right\rangle }%

\global\long\def\f#1#2{\frac{#1}{#2}}%
 
\global\long\def\del{\partial}%
 
\global\long\def\t{\theta}%
 
\global\long\def\a{\alpha}%
 
\global\long\def\b{\beta}%
 
\global\long\def\g{\gamma}%
 
\global\long\def\G{\Gamma}%
 
\global\long\def\s{\sigma}%
 
\global\long\def\r{\rho}%
 
\global\long\def\d{\delta}%
 
\global\long\def\Tr{\text{Tr}}%

\global\long\def\bra#1{\left|#1\right\rangle }%
 
\global\long\def\N{\mathbb{N}}%

\global\long\def\Z{\mathbb{Z}}%
 
\global\long\def\R{\mathbb{R}}%
 
\global\long\def\C{\mathbb{C}}%
 
\global\long\def\p{\varphi}%

\global\long\def\w{\omega}%
 
\global\long\def\D{\Delta}%

\global\long\def\l{\ell}%
 
\global\long\def\const{\text{const}}%

\global\long\def\app{\approx}%
\global\long\def\arcsinh{\text{arcsinh}}%
\global\long\def\arccosh{\text{arccosh}}%
\global\long\def\diag{\text{diag}}%

\begin{document}
  \renewcommand{\theequation}{\thesection.\arabic{equation}}

\begin{titlepage}

  \bigskip\bigskip\bigskip\bigskip

  \bigskip

\begin{center}
\Large \bf {An effective matrix model for dynamical end of the world branes in Jackiw-Teitelboim gravity}
\end{center}

    \bigskip

  \begin{center}

 \bf {Ping Gao$^a$, Daniel L. Jafferis$^b$ and David K. Kolchmeyer$^b$}
  \bigskip \rm
\bigskip

${}^a$ {\it  Center for Theoretical Physics, Massachusetts Institute of Technology,
Cambridge, MA 02139, USA}

${}^b$ {\it  Center for the Fundamental Laws of Nature, Harvard University, Cambridge, MA, USA}
\smallskip

\vspace{1cm}
  \end{center}

  \bigskip\bigskip

 \bigskip\bigskip
  \begin{abstract}

We study Jackiw-Teitelboim gravity with dynamical end of the world branes in asymptotically nearly AdS$_2$ spacetimes. We quantize this theory in Lorentz signature, and compute the Euclidean path integral summing over topologies including dynamical branes. The latter will be seen to exactly match with a modification of the SSS matrix model. The resolution of UV divergences in the gravitational instantons involving the branes will lead us to understand the matrix model interpretation of the Wilsonian effective theory perspective on the gravitational theory. We complete this modified SSS matrix model nonperturbatively by extending the integration contour of eigenvalues into the complex plane. Furthermore, we give a new interpretation of other phases in such matrix models. We derive an effective $W(\Phi)$ dilaton gravity, which exhibits similar physics semiclassically. In the limit of a large number of flavors of branes, the effective extremal entropy $S_{0,\eff}$ has the form of counting the states of these branes.

 \medskip
  \noindent
  \end{abstract}

  \end{titlepage}

  \tableofcontents

\section{Introduction}

Euclidean wormholes have been at the center of a spate of recent progress in understanding quantum gravity \cite{Almheiri:2019qdq, Penington:2019kki, Goto:2020wnk, Saad:2018bqo, Marolf:2020rpm, Engelhardt:2020qpv, Giddings:2020yes}. These non-perturbative contributions were shown to reproduce late time behavior of the spectral form associated to level repulsion in chaotic systems (referred to as the ramp) \cite{Saad:2018bqo}, and the Page curve for entanglement entropy of an AdS black hole evaporating into an external bath \cite{Almheiri:2019qdq, Penington:2019kki}. In pure Jackiw-Teitelboim (JT) gravity \cite{Jackiw:1984je, Teitelboim:1983ux} the non-perturbative Euclidean path integral can be computed exactly \cite{Saad:2019lba}, and precisely matched with a particular double-scaled matrix model. 

The latter fact highlights the curious aspect that the quantum mechanics dual is an ensemble average, rather than a particular Hamiltonian theory. This seems at odds with AdS/CFT in higher dimensions. In particular, although there are no further constraints of consistency on  positive hermitian Hamiltonians in quantum mechanics, conformal field theories in higher dimensions are highly constrained by microlocality and come only in sporadic families. Therefore there cannot exist continuous ensembles of conformal field theories with a number of  parameters that scales with the dimensionality of the Hilbert space, as happens in the matrix model for JT gravity,  otherwise any potential matrix model dual of higher dimensional AdS quantum gravity would allow nonlocal heavy operators that does not obey crossing symmetry because their matrix elements could be tuned arbitrarily by the overwhelming number of parameters, and that contradict with AdS/CFT because bulk dual of heavy operators are black holes localized in a finite region.

For these reasons it is of great interest to know if JT gravity itself can be the long distance effective theory of some UV modified theory that behaves more conventionally, with an ordered dual possessing a discrete spectrum, as might be expected to arise in a string compactification to 1+1 dimensions. We will not answer that question here, but will explore consequences of a minimal ingredient that such a UV ``completion'' of JT gravity would require: dynamical branes that can end spacetime. 

In 1+1 dimensions, the fundamental objects that could form the microstates of black holes are codimension 1 branes, in other words, end of the world (EOW) branes. Dynamical objects of that type are also required to solve the factorization problem \cite{Harlow:2018tqv} starting from canonical quantization of Lorentzian JT theory. At an even most basic level, there are simply no Lorentzian configurations in pure JT gravity with a single nearly AdS$_2$ boundary that could appear in the Hilbert space of a putative dual quantum mechanics. 

EOW branes could have a variety of microscopic realizations in a compactification to JT gravity, including intrinsic end of the world branes like the Horava-Witten brane in M-theory and Kaluza-Klein solitons where part of an internal manifold smoothly shrinks. In the long distance effective theory these can be treated as boundary conditions in JT gravity. We will study the simplest possibility, which is characterized simply by the brane tension (equivalently the rest mass of this 0+1 dimensional object). 

Branes of this type were used in \cite{Penington:2019kki} to model black hole microstates of pure JT gravity, however in that context they were not treated as dynamical objects added to the theory. Because of this, the loops of EOW branes were not considered in \cite{Penington:2019kki} even though the number of flavors $K$ was as large as $e^{S_0}$. The Page curve of \cite{Penington:2019kki} indicates a phase transition when $K$ is on the order of $e^{S_0}$.  Here we will be interested in the effects of summing over EOW branes in the path integral, which will be important when $K$ is of order $e^{S_0}$.

We will see that positive tension EOW branes are always cloaked behind horizons. Nevertheless, the EOW branes modify the 
spectral density because they act as gravitational instantons that correct the long distance effective action in the form of a more general $W(\Phi)$ gravity. 

Similarly to \cite{Harlow:2018tqv}, the classical phase space may be parametrized in a geodesic slice gauge in terms of the (renormalized) length $L$ of the spatial slice and its conjugate momentum, leading to a Morse potential quantum mechanics. The spectrum of the Hamiltonian is a continuum, so it is unsurprising that the theory with EOW branes remains an ensemble.


The continuous spectrum of the Morse potential quantum mechanics, which is associated to the $L \rightarrow \infty$ limit, leads to a logarithmic divergence in the contributions of gravitational instantons. The large $L$ limit can equivalently be understood as a $b \rightarrow 0$  limit of high temperature in the reference frame of the EOW brane, where $b$ is the length of the EOW brane loop. As such the calculation is sensitive to UV physics deep in the throat; the large red shift allows it to contribute at a fixed energy as measured at the AdS boundary. Therefore an EFT understanding is crucial for the proper treatment of this divergence. 

In the gravity calculation, we obtain finite results by using a regulated EOW brane such that the $b\rightarrow 0$ pole is cancelled and the theory becomes pure JT in the far UV. The low energy spectral density $\rho(E)$ becomes universal as the regulator is removed, up to a single parameter. One characterization of that UV sensitive parameter is the zero point energy, $E_0$, at which the spectrum begins. In this way, it behaves like a relevant parameter in the Wilsonian sense. 

The effect of the EOW branes in the dual quantum mechanics is to introduce $K$ vectors in the SSS matrix model \cite{Saad:2019lba} (as in \cite{Penington:2019kki}). These are the states in the Hilbert space produced by a given EOW brane. The branes here are dynamical objects in the gravity theory, and one must integrate over the vector degrees of freedom in the matrix model. 

For the purpose of computing the spectral density and its correlation functions, the vectors can be integrated out, leading to a modified matrix ensemble for the Hamiltonian, whose matrix potential differs from that of pure JT gravity by $\delta V(E)$. The deformation of the matrix model potential by various types of EOW branes has been considered in a recent work \cite{Goel:2020yxl}.

The same $\delta V$ can equally be computed from the trumpet partition function with fixed AdS boundary energy, which is the inverse Laplace transformation of $Z_\trumpet(b,\b)$. The full path integral is a sum over the number of closed loops of EOW brane boundaries, which is reproduced by the exponentiation of the additional potential, $e^{-K\Tr \delta V(H)}$. 

There is a trivial UV divergence in $\delta V$, which can be absorbed into the overall normalization of the ensemble measure. It is straightforward to compute the change in the tree-level spectral density $\delta \rho$ induced by the change in the potential $\delta V$. However, the large $E$ behavior of $\delta V(E)$ implies that the integral of $\delta \rho(E)$ over its support (which is noncompact) diverges. More severely, without the appropriate fine tuning of the UV sensitive parameter, $E_0$ would go to $-\infty$. This means that an infinite number of eigenvalues that would have been pushed to infinity in the double-scaling limit of the SSS matrix model remain at finite energies.


Regulating the theory at short distance leads to a modified $\delta V$ that decays at large energy, allowing a $\delta \rho$ that is normalizable. This is a special feature that is associated to this model's rapid approach to the pure JT spectrum at high energies. We exactly match the result with the gravitational calculation. More importantly, from the IR point of view, $E_0$ is a free parameter in both the matrix model and gravity sides, which is sensitive to the details of the bulk UV physics.

The contribution of near cusps, which arise in the limit of an infinitely long trumpet, corresponds to an effective $W(\Phi)\propto e^{-2\pi \Phi}$ in the general dilaton gravity considered in \cite{Witten:2020ert, Witten:2020wvy}. The matrix model potential in the double scaling limit is in fact unchanged from SSS, and the only difference is via the IR parameter $E_0$ which determines how the double scaling limit is taken. 

For large $K \gg e^{S_0}$ in the matrix model with heavy EOW branes, the effective extremal entropy $S_{0,\eff}$ defined as $\r(E)\app e^{S_{0,\eff}}\sqrt{E-E_0}$ for $E-E_0\ll 1$ scales as $\log K$. This is a form of induced gravity, in which $K$ flavors of heavy EOW branes count the microscopic states of black hole.

As pointed out in \cite{Saad:2019lba}, the matrix model dual to pure JT gravity is non-perturbatively unstable because the effective potential becomes arbitrarily negative away from the support of the spectrum. 
This non-perturbative instability causes negativity in the spectrum of JT gravity with too many flavors of deficit angles \cite{Maxfield:2020ale} and in JT gravity with EOW branes as well. To have a well-defined matrix model, one needs to extend the contour of each eigenvalue through the largest saddle point of the effective potential on the real axis and into the complex plane \cite{Saad:2019lba}. This completion promotes $H$ to a complex matrix because an order $~e^{-e^{S_0}}$ fraction of eigenvalues will be complex even in the pure JT model.

\begin{figure}
\begin{centering}
\includegraphics[width=7cm]{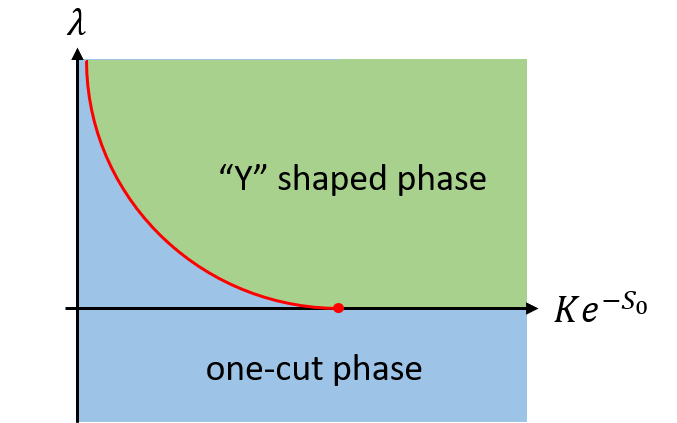}
\par\end{centering}
\caption{The phase diagram of matrix model of JT gravity with $K$ flavors of EOW branes. Green region is ``Y" shaped phase, blue region is one-cut phase and red curve is critical line. \label{fig:phase-diagram}}
\end{figure}

We will adopt this completion and study how the spectrum changes as we increase the number of flavors $K$ of EOW branes. As $K$ increases, the qualitative behavior depends on how the UV divergence of the EOW branes is renormalized, which comes with a renormalization parameter $\lambda$. There are three cases to consider, corresponding to the zero point energy $E_0$ being positive ($\lambda<0$), zero ($\lambda=0$) or negative ($\lambda>0$) when $K$ is slightly positive. See Fig. \ref{fig:phase-diagram} as the phase diagram. For the first case, there is no critical $K$ and no phase transition. For second case, a continuous phase transition occurs and $E_0$ moves to positive after $K$ going beyond a critical value. For the last case, instead of a negative spectrum, our non-perturbative completion implies that the spectrum undergoes a continuous phase transition to a ``Y" shaped on the complex energy plane when $K$ is over a critical value (see Fig. \ref{fig:The-support-of}). 

We interpret these $e^{S_0}$ order of complex eigenvalues as unstable black hole states that could decay to lower energy objects other than EOW branes. This matches the Lorentzian analysis in the effective $W(\Phi)$ dilaton gravity \cite{Witten:2020ert}, for which the spectrum is unbounded from below. This is also related to the existence of a minimal temperature below which no stable black hole exists. Because of this, we regard JT gravity with such EOW branes as incomplete, requiring other stable objects with lower energy (such as Dirichlet-Neumann branes \cite{Goel:2020yxl}) to which these unstable black holes decay. Using the effective $W(\Phi)$ gravity description, we explicitly show how this occurs as a Hawking-Page phase transition.

The paper is organized as follows. In Section \ref{sec:2}, we find the phase space of JT gravity with one dynamical EOW brane and quantize it canonically. We compute the partition function of this system and identify the measure of EOW branes in the Euclidean path integral. Using this measure to sum over arbitrary numbers of EOW branes, we compute the tree-level spectral density. For the $E_0=0$ case, we find the effective $W(\Phi)$ dilaton gravity by integrating out the EOW branes. In Section \ref{sec:effmatrixmodel}, we derive the change in the matrix model potential $\delta V$ induced by the EOW branes  and solve for the one-cut solution of the spectral density that matches with our gravitational computation. In Section \ref{sec:yshape}, we study the ``Y" shaped spectrum in the matrix model and the phase transition from the  one-cut solution when the number of flavors, $K$, of EOW branes exceeds a critical value. In Section \ref{sec:5}, we study the effective $W(\Phi)$ dilaton gravity for a gas of cusps that is related to $K$ heavy EOW branes. By requiring smoothness of the Euclidean metric, we find complex saddles when $K$ is beyond the critical value. These complex saddles exhibit similar ``Y" shaped spectra. We interpret the complex energies as unstable black holes and study the Hawking-Page phase transition after including lower energy Dirichlet-Neumann branes. We conclude in Section \ref{sec:disc} with a few discussions.

\section{Ends of the world in 2d gravity} \label{sec:2}

\subsection{Classical solution and phase space with a boundary brane}

The JT gravity \cite{Teitelboim:1983ux, Jackiw:1984je} action with boundaries is
\begin{equation}
S=S_{0}+\kappa\left[\f 12\int\Phi\sqrt{|g|}(R+2)+\int_{AdS}du\Phi\sqrt{-g_{uu}}(K-1)+\int_{brane}dv\sqrt{-g_{vv}}(\Phi K-\mu)\right]
\end{equation}
where $S_{0}=\f{\phi_{0}}{4G}$ is the extremal entropy and $\kappa=(8\pi G)^{-1}$.
In this paper, we will set $\kappa=1$ for notational simplicity.
$u$ and $v$ are affine parameters along the AdS boundary and EOW brane
respectively. As analyzed in Appendix \ref{sec:Variation-of-JT},
the equations of motion are
\begin{equation}
R+2=0,\qquad\nabla_{a}\nabla_{b}\Phi-g_{ab}\nabla^{2}\Phi+g_{ab}\Phi=0
\end{equation}
and the boundary conditions are
\begin{equation}
\begin{cases}
n^{c}\del_{c}\Phi=C\Phi+D & \text{or}\quad h^{ab}\d g_{ab}=0\\
\nabla_{a}n^{a}=C & \text{or}\quad\d\Phi=0
\end{cases},\;\begin{cases}
C=0,D=\mu & \text{brane}\\
C=1,D=0 & \text{AdS}
\end{cases}\label{eq:bc}
\end{equation}
For an AdS boundary, we fix the asymptotic metric $h^{ab}\d g_{ab}=0$
and the value of dilaton $\d\Phi=0$; for an EOW brane, we allow
the metric and dilaton to fluctuate but impose
\begin{equation}
\nabla_{a}n^{a}=0,\qquad n^{a}\del_{a}\Phi=\mu\label{eq:b1}
\end{equation}

\begin{figure}
\begin{centering}
\includegraphics[width=3.5cm]{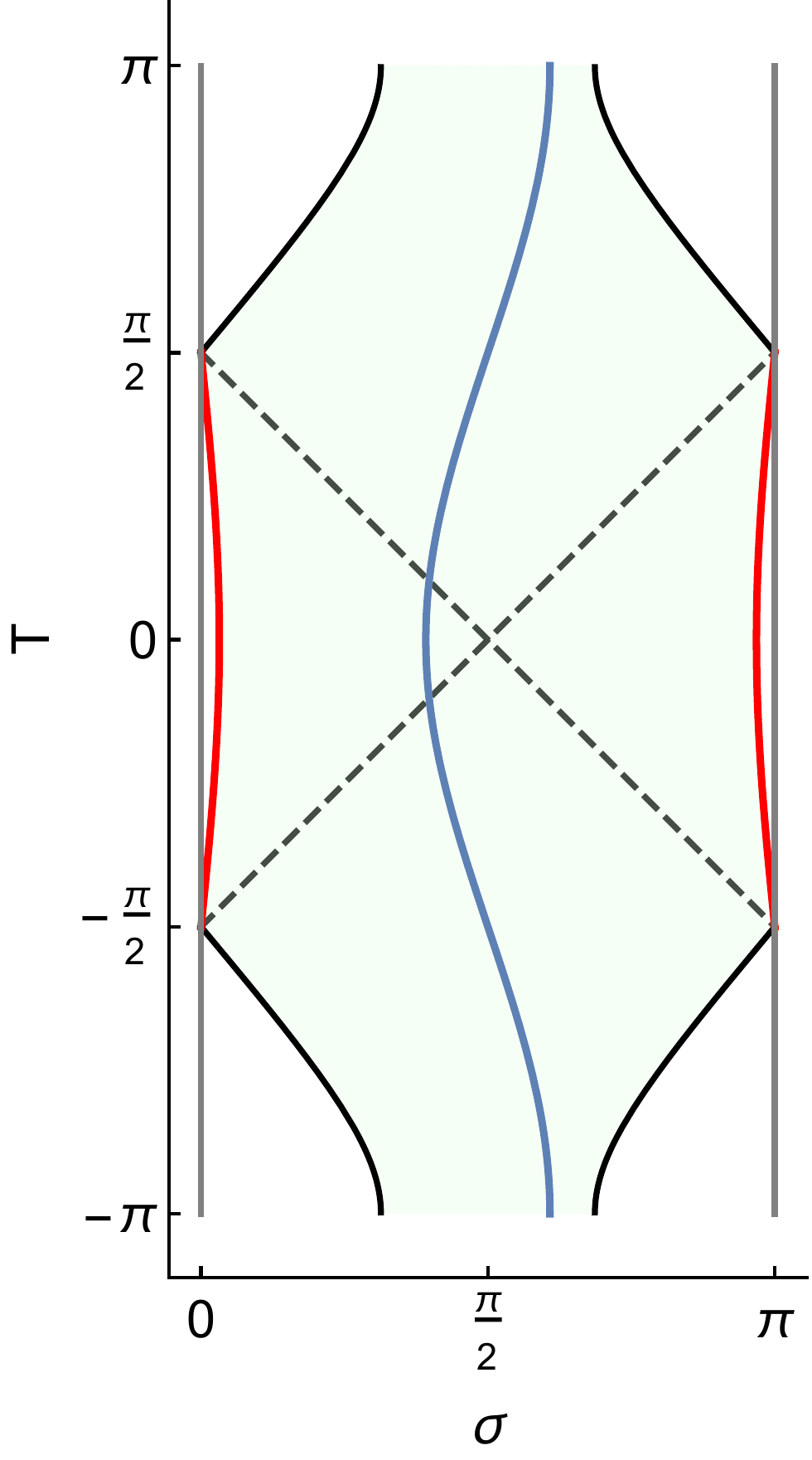}
\par\end{centering}
\caption{The solution of JT gravity. The light green region is physical in
which $\Phi+\phi_{0}\protect\geq0$. Red curves are two AdS boundaries
with $T\in[-\pi/2,\pi/2]$. The dashed lines are horizon. Blue curve
is the geodesic of EOW brane. \label{fig:The-solution-of}}
\end{figure}

The general solution to the equations of motion can be written as
\begin{equation}
ds^{2}=\f{-dT^{2}+d\s^{2}}{\sin^{2}\s},\quad\Phi=\f{\a\cos(T-\b)-\g\cos\s}{\sin\s}\label{eq:4}
\end{equation}
where $\s\in[0,\pi]$, $T\in\R$ and $\a,\b,\g$ are real numbers.
Physically, $\Phi+\phi_{0}$ must be nonnegative as it would represent the area of codimension 2 surfaces
in higher dimensions if JT gravity were obtained via dimensional reduction. As the solution is
periodic in $T$, the physical region $\Phi+\phi_{0}\geq0$ appears
periodically along $T$ axis. There is a $SL(2)\simeq SO(2,1)$ isometry
for AdS$_{2}$ metric. To see this, we can write AdS$_{2}$ as a hyperplane
in higher dimension
\begin{equation}
-Y_{-1}^{2}-Y_{0}^{2}+Y_{1}^{2}=-1,\quad ds^{2}=-dY_{-1}^{2}-dY_{0}^{2}+dY_{1}^{2}
\end{equation}
with coordinate transformation
\begin{equation}
Y_{-1}=\sin T\csc\s,\quad Y_{0}=\cos T\csc\s,\quad Y_{1}=-\cot\s
\end{equation}
Then the dilaton solution can be written in a simpler form
\begin{equation}
\Phi=V^{\mu}Y_{\mu},\quad V^{\mu}=(\a\sin\b,\a\cos\b,-\g)
\end{equation}
Under $SO(2,1)$ transformation, $\a,\b,\g$ will change according
to the action on $V^{\mu}$. However, there is one invariant describing the
solution, 
\begin{equation}
V^{\mu}V_{\mu}=\g^{2}-\a^{2}
\end{equation}
whose sign will separate the solutions into two types. Indeed, the
existence of a saddle for $\Phi$ depends on $\g^{2}-\a^{2}$. Differentiating with respect to $T$ and $\s$, we get
\begin{align}
\del_{T}\Phi & =0\implies T=\b+n\pi\\
\del_{\s}\Phi & =0\implies\cos\s=(-)^{n}\f{\g}{\a}
\end{align}
If $|\g|\leq|\a|$, the saddle exists and if $|\g|>|\a|$, it does
not. In this paper, we will only consider cases with a $\Phi$ saddle,
which is the analog of the near extremal horizon ``area'' in JT gravity.
The $SO(2,1)$ is a gauge symmetry and we fix it by choosing the dilaton solution to be
\begin{equation}
\Phi=\Phi_{h}\f{\cos T}{\sin\s},\qquad\Phi_{h}\geq0\label{eq:dilaton}
\end{equation}
The physical region with AdS boundary is chosen to be $T\in[-\pi/2,\pi/2]$
and $\Phi_{h}$ is the saddle value of $\Phi$ (see Fig. \ref{fig:The-solution-of}).

The AdS boundary condition is \cite{Maldacena:2016upp}
\begin{equation}
g_{uu}=-\f 1{\e^{2}},\;\Phi=A=\f{\phi_{b}}{\e},\;\e\ra0 \label{eq:adsbdc}
\end{equation}
For fixed $\text{\ensuremath{\Phi=A}}$ value, its solution is given
by
\begin{equation}
\cos T=\f A{\Phi_{h}}\sin\s\label{eq:ads-sl}
\end{equation}
In large $A$ limit, we would like to put AdS boundary near $\s=\pi$.
In small $\e$ expansion with fixed metric, we have
\begin{align}
\s & =\pi-\e T'+O(\e^{2})\\
\tan\f T2 & =\tanh\f{\Phi_{h}}{2\phi_{b}}(u-u_{0})\label{eq:15}
\end{align}

The EOW brane boundary condition is given in (\ref{eq:b1}). For any
brane world line $\s-f(T)=0$, the normal vector is
\begin{equation}
n^{T}=-\f{\sin\s f'(T)}{\sqrt{1-f'(T)^{2}}},\;n^{\s}=-\f{\sin\s}{\sqrt{1-f'(T)^{2}}}
\end{equation}
The normal derivative to $\Phi$ on the brane is 
\begin{equation}
n^{a}\del_{a}\Phi=\Phi_{h}\f{\cos T\cot f+\sin Tf'}{\sqrt{1-f'^{2}}}\label{eq:14}
\end{equation}
On the other hand, $\nabla^{a}n_{a}=0$ implies that it is a geodesic
whose general solution is
\begin{equation}
\s=f(T)=\arccos(r\cos(T-\theta)),\qquad\t\in[0,\pi),\qquad r\in\R\label{eq:16}
\end{equation}
Note that we require the world line to be timelike, namely $|f'(T)|<1$.
This restricts $r\in[-1,1]$. Using coordinate $Y_{\mu}$, we can
rewrite geodesic solution (\ref{eq:16}) as
\begin{equation}
U^{\mu}Y_{\mu}=0,\quad U^{\mu}\propto(r\sin\t,r\cos\t,1)
\end{equation}
where $U^{\mu}$ can be rescaled with any nonzero real number without
change the solution. As $SO(2,1)$ is a gauge symmetry, we
are free to choose $V^{\mu}$ as above to partly fix it.  In (\ref{eq:dilaton}), we rotate $V^{\mu}$
to $(0,\Phi_{h},0)$ and clearly there is an unfixed $SO(1,1)$ between the
$V^{-1}$ and $V^{1}$ components. This $SO(1,1)$ can be further
fixed by rotating the $U^{-1}$ and $U^{1}$ components to set $U^{-1}=0$
(note that $|r\sin\t|<1$ makes this always possible). This is equivalent
to set $\sin\t=0$ ($\t=0$). Hence, after fixing all gauge symmetries,
the solution for the geodesic is
\begin{equation}
\cos\s=r\cos T,\qquad r\in[-1,1]\label{eq:geod}
\end{equation}
Plugging this solution into $n^{a}\del_{a}\Phi=\mu$, we get
\begin{equation}
n^{a}\del_{a}\Phi=\f{r\Phi_{h}}{\sqrt{1-r^{2}}}=\mu\implies r=\f{\mu}{\sqrt{\Phi_{h}^{2}+\mu^{2}}}\label{eq:22}
\end{equation}

As shown above, the EOW brane boundary condition completely determines
its worldline from the parameters $\mu$ and $\Phi_{h}$. There
is another way to characterize the brane geodesic from the AdS boundary
point of view, which will be important for our  phase space description.
It is the length of a spacelike geodesic shooting from AdS boundary
at $u=0$ and ending normally  on the brane. It is simplest to calculate
this in the $Y_{\mu}$ coordinate. Note that different geodesics
can be transformed to each other via $SO(2,1)$ isometry. In particular,
the brane geodesic (\ref{eq:geod}) can be written as
\begin{equation}
U^{\mu}M_{\mu}^{\nu}Y_{\nu}=0,\quad U^{\mu}=(0,0,1),\quad M=\begin{pmatrix}1\\
 & \cosh\xi & \sinh\xi\\
 & \sinh\xi & \cosh\xi
\end{pmatrix},\quad\tanh\xi=r
\end{equation}
On the other hand, the geodesic $U^{\mu}Y_{\mu}=0$ is very simple,
namely $\s=\pi/2$. We can calculate the spacelike geodesic connecting
it with boundary and then do a $SO(2,1)$ transformation to find the
case related to (\ref{eq:geod}). Note that orthogonality and geodesic
length is preserved under isometry. 

It is clear that for $\s=\pi/2$, the orthogonal spacelike geodesic
is $T=T_{0}$ for any $T_{0}$. This spatial slice intersects the AdS
boundary at $(T_{0},\s_{0})$. The geodesic length from the brane to $(T_{0},\s_{0})$
is
\begin{equation}
L_{bare}=\int_{\pi/2}^{\s_{0}}\f{d\s}{\sin\s}=\f 12\log\f{1-\cos\s_{0}}{1+\cos\s_{0}}\label{eq:24}
\end{equation}
To get the spatial slice orthogonal to (\ref{eq:geod}), we simply
do the transformation
\begin{equation}
Y_{\mu}\ra M_{\mu}^{\nu}Y_{\nu}=(\sin T_{0}\csc\s_{0},\cos T_{0}\csc\s_{0}\cosh\xi-\cot\s_{0}\sinh\xi,\cos T_{0}\csc\s_{0}\sinh\xi-\cot\s_{0}\cosh\xi)
\end{equation}
Under this transformation, $\cos\s_{0}$ transforms as
\begin{equation}
\cos\s_{0}=-\f{Y_{1}}{\sqrt{Y_{-1}^{2}+Y_{0}^{2}}}\ra-\f{\cosh\xi\cos\s_{0}-\cos T_{0}\sinh\xi}{\sqrt{\sin^{2}T_{0}+(\cos T_{0}\cosh\xi-\cos\s_{0}\sinh\xi)^{2}}}
\end{equation}
The boundary location $(\s_{0},T_{0})$ is given by (\ref{eq:ads-sl})
with $A=\phi_{b}/\e$. Taking this into (\ref{eq:24}) gives divergent
result and needs to be renormalized as \cite{Harlow:2018tqv}
\begin{equation}
L=L_{bare}-\log A=\log\f{2(r+\sec T_{0})}{\Phi_{h}\sqrt{1-r^{2}}}
\end{equation}
Using the expression for $r$ in (\ref{eq:22}) and $T_{0}$ in (\ref{eq:15}),
we have
\begin{equation}
L=\log\f{2\mu+2\sqrt{\mu^{2}+\Phi_{h}^{2}}\cosh w}{\Phi_{h}^{2}},\quad w\equiv\f{\Phi_{h}}{\phi_{b}}u_{0}\label{eq:28}
\end{equation}
As the lower bound of $\Phi_{h}$ is zero, we must require $\mu\geq0$
for $L$ to be well-defined. 

The AdS boundary stress tensor is
\begin{equation}
T_{\del M}^{\mu\nu}=\f 1{\e^{3}}\f 2{\sqrt{|\g|}}\f{\d S}{\d\g_{\mu\nu}}
\end{equation}
where $\g_{\mu\nu}$ is the induced metric. In 2d case, there is only
one component of stress tensor, namely Hamiltonian. Using (\ref{eq:41}),
we get
\begin{equation}
H=T_{\del M}^{uu}=(n^{a}\del_{a}\Phi-\Phi)h^{uu}/\e^{3} \label{eq:30H}
\end{equation}
Using (\ref{eq:ads-sl}) (AdS boundary has $A=\phi_{b}/\e$) and (\ref{eq:14})
(here we need to flip the sign for $n_{a}$ pointing outward), we
can evaluate it on AdS boundary as
\begin{equation}
H=\Phi_{h}^{2}/(2\phi_{b})
\end{equation}
This Hamiltonian is nonnegative as the ADM energy of a black hole
should be. On the other hand, the brane Hamiltonian is vanishing because
we choose the brane boundary condition such that $T_{brane}^{vv}\propto n^{a}\del_{a}\Phi-\mu=0$. This is consistent with gravity being dynamical on the EOW brane. 

Similarly to the two-sided 2D JT gravity \cite{Harlow:2018tqv}, the phase space
is two dimensional, which is characterized by $u_{0}$ and $H$, where
$u_{0}$ is the boundary time constant corresponding to $T=0$ slice.
As $H$ is the Hamiltonian on AdS boundary, its canonical conjugate
is time translation. This implies the symplectic form in phase space
is
\begin{equation}
\w=\d u_{0}\wedge\d H
\end{equation}
The dynamics in phase space is given by
\begin{equation}
\dot{x}^{a}=(\w^{-1})^{ba}\del_{b}H\implies\dot{u}_{0}=1,\;\dot{H}=0
\end{equation}
The symplectic form can be written as
\begin{equation}
\w=\f{\Phi_{h}}{\phi_{b}}\d u_{0}\wedge\d\Phi_{h}=\d w\wedge\d\Phi_{h}
\end{equation}
where we see that $w$ is the conjugate coordinate for $\Phi_{h}$.
Note that using parameters $w$ and $\Phi_{h}$ is not quite a good
 description of the phase space because $\Phi_{h}$ is restricted
to be a nonnegative number. In order to find nice phase space coordinates
with range $\R^{2}$, we need to do a canonical transformation.

The canonical transformation to the $L$ variable is easy to find. Solving for $w$ in (\ref{eq:28}),
we find
\begin{equation}
w=\pm\arccosh\f{e^{L}\Phi_{h}^{2}-2\mu}{2\sqrt{\Phi_{h}^{2}+\mu^{2}}}
\end{equation}
where the sign of $w$ depends on that of $u_{0}$. This implies that
\begin{equation}
\d w\wedge\d\Phi_{h}=\pm\f{e^{L}\Phi_{h}}{\sqrt{e^{2L}\Phi_{h}^{2}-4\mu e^{L}-4}}\d L\wedge\d\Phi_{h}=\d L\wedge\d P,\quad P\equiv\pm\sqrt{\Phi_{h}^{2}-4\mu e^{-L}-4e^{-2L}}
\end{equation}
Solving for $\Phi_{h}^{2}$ leads to the Hamiltonian in terms of $L$ and
$P$
\begin{equation}\label{eq:2.37}
H=\f 2{\phi_{b}}\left[\f{P^{2}}4+\mu e^{-L}+e^{-2L}\right]
\end{equation}
This is the Hamiltonian with Morse potential. It is obvious that the Hamiltonian
is nonnegative for all $L$ only when $\mu>0$. Indeed, for $\mu<0$,
we can do a similar phase space analysis and end up with the same
Hamiltonian. In that case, Hamiltonian is negative for $L$ larger
than a critical value but still lower bounded. In particular, there
is a stable minimal energy point located at $P=0,\quad L=-\log|\mu|$
with ground energy $E=-\mu^{2}/\phi_{b}$. Such Hamiltonian allows
bound states, whose geometric meaning is a naked EOW brane, rather than a black hole. In this
paper, we will mainly focus on $\mu>0$ case.

\subsection{Quantization with a boundary brane}

\label{sec:EOWbranequantization}

As the Hamiltonian is simply a particle in the Morse potential, its quantization
is straightforward by replacing $P\ra-i\del_{L}$. It follows that the
energy eigen-functions $f_{E}(L)$ obey
\begin{equation}
\left[-\del_{L}^{2}+4\mu e^{-L}+4e^{-2L}\right]f_{E}(L)=2\phi_{b}Ef_{E}(L)
\end{equation}
Using a new variable $z=4e^{-L}$, we can rewrite the equation as
\begin{equation}
\left[-z(\del_{z}+z\del_{z}^{2})+\mu z+\f 14z^{2}\right]f_{E}(z)=2\phi_{b}Ef_{E}(z)
\end{equation}
For $\mu>0$, the spectrum of solutions is continuous. 

The general solution is given by Whittaker function
\begin{equation}
f_{k}(z)=N_{k}z^{-1/2}W_{-\mu,ik}(z),\quad E=\f{k^{2}}{2\phi_{b}},\quad k\geq0
\end{equation}
Note that here we restrict $k\geq0$ because of identity $W_{a,b}(z)=W_{a,-b}(z)$.
The normalization is given by a flat measure integration over $L\in\R$
\begin{equation}
\d(k-k')=N_{k}N_{k'}\lim_{\e\ra0}\int_{0}^{\infty}\f{dz}{z^{1+\e}}\f 1zW_{-\mu,ik}(z)W_{-\mu,ik'}(z)
\end{equation}
After some algebra, we can work out the normalized eigen-function
\begin{equation}
\Psi_{k}(z)=\sqrt{\p(k)}z^{-1/2}W_{-\mu,ik}(z),\qquad\p(k)\equiv\f{|\G(\f 12+\mu-ik)|^{2}k\sinh2\pi k}{\pi^{2}}\label{eq:63}
\end{equation}

The propagator for the Morse potential quantum mechanics is
\begin{align}
G_{\mu,\b}(L_{2},L_{1})&=\int dke^{-\b k^{2}/(2\phi_{b})}\avg{L_{2}|k}\avg{k|L_{1}}\nn\\
&=\int dke^{-\b k^{2}/(2\phi_{b})}\p(k)(z_{1}z_{2})^{-1/2}W_{-\mu,ik}(z_{1})W_{-\mu,ik}(z_{2})\label{eq:11}
\end{align}
If we set $L_{1}=L_{2}$ and integrate it over $\R$, we will get
the partition function of JT gravity with a single EOW brane, which is divergent. This is because the Morse potential quantum mechanics has a continuous spectrum. As we will now show, we may alternatively interpret this divergence as a UV divergence associated to an EOW brane loop that shrinks to zero size. To see this, use identity (6.647-1 in \cite{gradshteyn2014table}) to write (\ref{eq:11})
as
\begin{equation}
G_{\b}(L_{2},L_{1})=\int dke^{-\b k^{2}/(2\phi_{b})}\p(k)\int_{0}^{\infty}dx\f{2x^{\mu-1/2}e^{-(z_{1}+z_{2})/2}e^{-\f{z_{1}+z_{2}}{2\sqrt{z_{1}z_{2}}}x}}{|\G(\f 12+\mu-ik)|^{2}(2\sqrt{z_{1}z_{2}}+x)^{\mu+1/2}}K_{2ik}(\sqrt{x(2\sqrt{z_{1}z_{2}}+x)})
\end{equation}
which holds only when $ \mu > - \frac{1}{2}$. Set $z_{1}=z_{2}$ and perform
$\int_{-\infty}^{+\infty}dL=\int_{0}^{\infty}\f{dz}z$ integration.
The two $(x,z)$ variable integration can be computed using alternative
variables
\begin{equation}
v=\sqrt{x(2z+x)},\qquad w=\sqrt{\f x{2z+x}}\implies dxdz=\f vwdvdw
\end{equation}
It follows that the partition function with a single EOW brane is
\begin{align}
Z_{EOW}(\b) & =\int dke^{-\b k^{2}/(2\phi_{b})}\f{4k\sinh2\pi k}{\pi^{2}}\int_{0}^{1}\f{dww^{2\mu}}{1-w^{2}}\int_{0}^{\infty}\f{dv}ve^{-(w+w^{-1})v/2}K_{2ik}(v)\nonumber \\
 & =\f 2{\pi}\int dke^{-\b k^{2}/(2\phi_{b})}\int_{0}^{1}\f{dww^{2\mu}}{1-w^{2}}\cos\left(4k\arcsinh\sqrt{\f 12(\f 12(w+1/w)-1)}\right)
\end{align}
where we used identity (6.621-3 in \cite{gradshteyn2014table}). Redefining $w=e^{-b/2}$
for positive $b$, we have
\begin{align}
Z_{EOW}(\b)&=\f 1{2\pi}\int dke^{-\b k^{2}/(2\phi_{b})}\int_{0}^{\infty}\f{db}{\sinh(b/2)}\cos(kb)e^{-\mu b}\nn\\
&=\int_{0}^{\infty}dbZ_{\trumpet}(\b,b)\f{e^{-\mu b}}{2\sinh(b/2)},\label{eq:50}
\end{align}
where $Z_{\trumpet}(\b,b)$ is the Euclidean path integral of the trumpet
bounded by one AdS boundary of regularized length $\b$ and one geodesic
boundary of length $b$ in \cite{Saad:2019lba},
\begin{equation}
Z_{\trumpet}(\b,b)=\f{\phi_{b}^{1/2}}{(2\pi)^{1/2}\b^{1/2}}e^{-\phi_{b}b^{2}/(2\b)}=\f 1{2\pi}\int dk\cos\f{kb}2e^{-\b k^{2}/(8\phi_{b})}\label{eq:48}.
\end{equation}
The integral in \eqref{eq:50} represents a Euclidean spacetime that ends on a geodesic EOW brane in the bulk, and the length of the EOW brane $b$ is integrated with a measure $\mM(b)$ given by
\begin{equation}
\mM(b)=\f{db \, e^{-\mu b}}{2\sinh(b/2)}=\sum_{n=0}^{\infty}dbe^{-(n+1/2+\mu)b}\label{eq:49}.
\end{equation}
This integral clearly diverges due to its behavior as $b \rightarrow 0$. 

To obtain a well-defined path integral, we need to regulate the small $b$ behavior of the measure $\mM(b)$ such that \eqref{eq:50} converges. In particular, we require that the regulated $\mM(b)$ is bounded as $b \rightarrow 0$. 
The UV divergence in \eqref{eq:50} can be cancelled by a cusp-like counterterm, which physically corresponds to a closed geodesic whose length is zero in the limit that the regulator is removed.
This is a modification of the UV physics that regulates the contributions of EOW branes to the path integral of the effective IR theory. 
The specific form of the regulator and the cusp-like counterterm is far from unique. One option is to regulate the UV divergence of \eqref{eq:50} by cutting off the $b$ integral at $b = \epsilon$, adding a counterterm proportional to a trumpet smeared over $b\in[0,\e]$, and then taking $\epsilon \rightarrow 0$. The regulated path integral with one AdS boundary and one EOW brane becomes
\begin{equation}
\lim_{\epsilon \rightarrow 0} \left[\int_{\epsilon}^{\infty}dbZ_{\trumpet}(\b,b)\f{e^{-\mu b}}{2\sinh(b/2)} + \f {\lambda + \log \epsilon} \e \int_0^\e db Z_{\trumpet}(\beta,b)  \right],
\label{eq:regulatedZEOW}
\end{equation}
where $\lambda$ represents the finite part of the cusp-like counterterm. We can thus make the following replacement in the definition of $\mM(b)$,
\begin{equation}
\label{eq:regmeasure}
    \frac{e^{- \mu b}}{2 \sinh(b/2)} \rightarrow \frac{e^{- \mu b}}{2 \sinh(b/2)} \theta(b - \epsilon) + \f {\lambda + \log \epsilon} \e (\theta(\e-b)-\theta(-b))
\end{equation}
which is bounded for all $b\geq 0$ as required. As usual, the renormalization procedure forces us to view $\lambda$ as an additional parameter of the theory. The value of $\lambda$ depends on UV physics that becomes relevant in the small $b$ limit. Modifying this UV physics corresponds to modifying the large-eigenvalue behavior of the potential of the dual matrix model, as we
will discuss in Section \ref{sec:determinE0}.

We will find it convenient to express our results in terms of the inverse Laplace transform of $\mM(b)$, which we denote by $m(\a)$. We have that
\begin{equation}
\mM(b)=\int_{\mD} d\a \, m(\a) \, e^{-\a b},\label{eq:50-1}
\end{equation}
where $\mD$, the support of $m(\a)$, could take complex values. The boundedness of $\mM(b)$ at $b=0$ implies that 
\be \label{eq:boundm}
\left|\int_\mD d\a \,  m(\a)\right|<\infty.
\ee
It is clear that the unregulated
EOW brane measure \eqref{eq:49} could be written as
\begin{equation}
\label{eq:malpha}
m(\a)=\sum_{n=0}^{\infty}\d(\a-(n+1/2+\mu)).
\end{equation}
The $m(\a)$ corresponding to the regulated measure \eqref{eq:regmeasure} could be computed similarly, though its explicit form is not important for our calculations. We will replace the regulated measure with the unregulated measure whenever doing so leads to a finite result. The property \eqref{eq:boundm} allows us to exchange the order of a limit and integral in \eqref{eq:347}.

\subsection{Dynamical branes in path integral}
\label{sec:gravitycomputation}

\begin{figure}
\begin{centering}
\includegraphics[width=12cm]{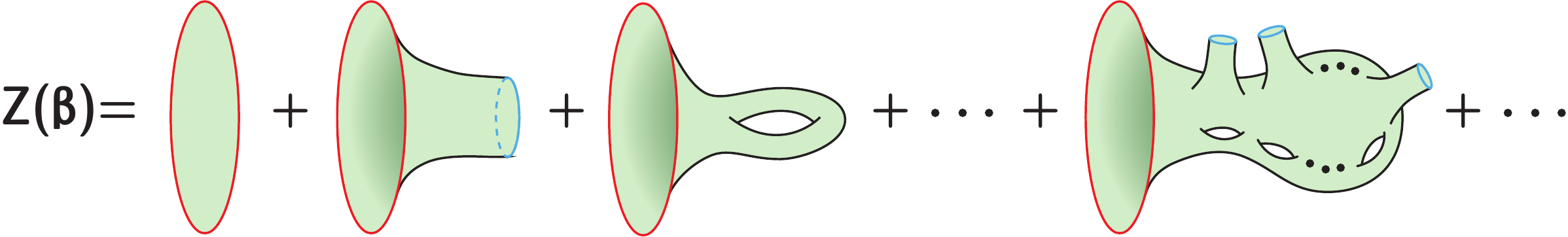}
\par\end{centering}
\caption{The partition function $Z(\protect\b)$ in which all genera and EOW
brane loops are summed. Red curve is AdS boundary and blue curves
are EOW branes. \label{fig:The-zero-genus}}
\end{figure}

In this section, we compute the Euclidean path integral with one AdS boundary and arbitrary numbers of EOW branes at genus zero. In general, given a fixed number of AdS boundaries, we are required to sum over all geometries with arbitrary numbers of handles and EOW branes (see Fig. \ref{fig:The-zero-genus}). As shown in \cite{Saad:2019lba},
path integrals can be computed by gluing trumpets to hyperbolic Riemann surfaces with closed geodesic boundaries. An EOW brane corresponds to a geodesic boundary whose length $b$ is integrated over with measure $\mM(b)$, given in \eqref{eq:49}. The $b \rightarrow 0$ divergence can be regulated as shown in \eqref{eq:regulatedZEOW}. In the Euclidean action, there is an Einstein-Hilbert
term $-S_{0}\chi$, where $\chi$ is Euler characteristic $\chi=2-2g-n$
of Riemann surface, where $g$ is genus and $n$ is number of boundaries.
This topological term endows every Riemann surface with a weight $e^{S_{0}\chi}$.
For fixed genus, adding one more EOW brane means increasing $n$ by
one and leads to a factor of $e^{-S_{0}}$. On the other hand, if
we have $K$ flavors of EOW brane, each loop will contribute a factor
of $K$ (modulo a possible permutation symmetry that we will specify
later). Therefore, for an $O(1)$ number of flavors, any effects of the EOW
brane loops are suppressed by $e^{-S_{0}}$. To enhance the effects of the EOW branes, we will assume
$K\sim O(e^{S_{0}})$. Furthermore, we will assume that $S_0$ is large so that we may restrict our attention to genus zero surfaces. For simplicity, we assume all flavors
of EOW branes have the same value of $\mu$. Our method for resumming EOW branes in the Euclidean path integral will closely follow the method used in \cite{Maxfield:2020ale}.

\begin{figure}
\begin{centering}
\includegraphics[width=6.5cm]{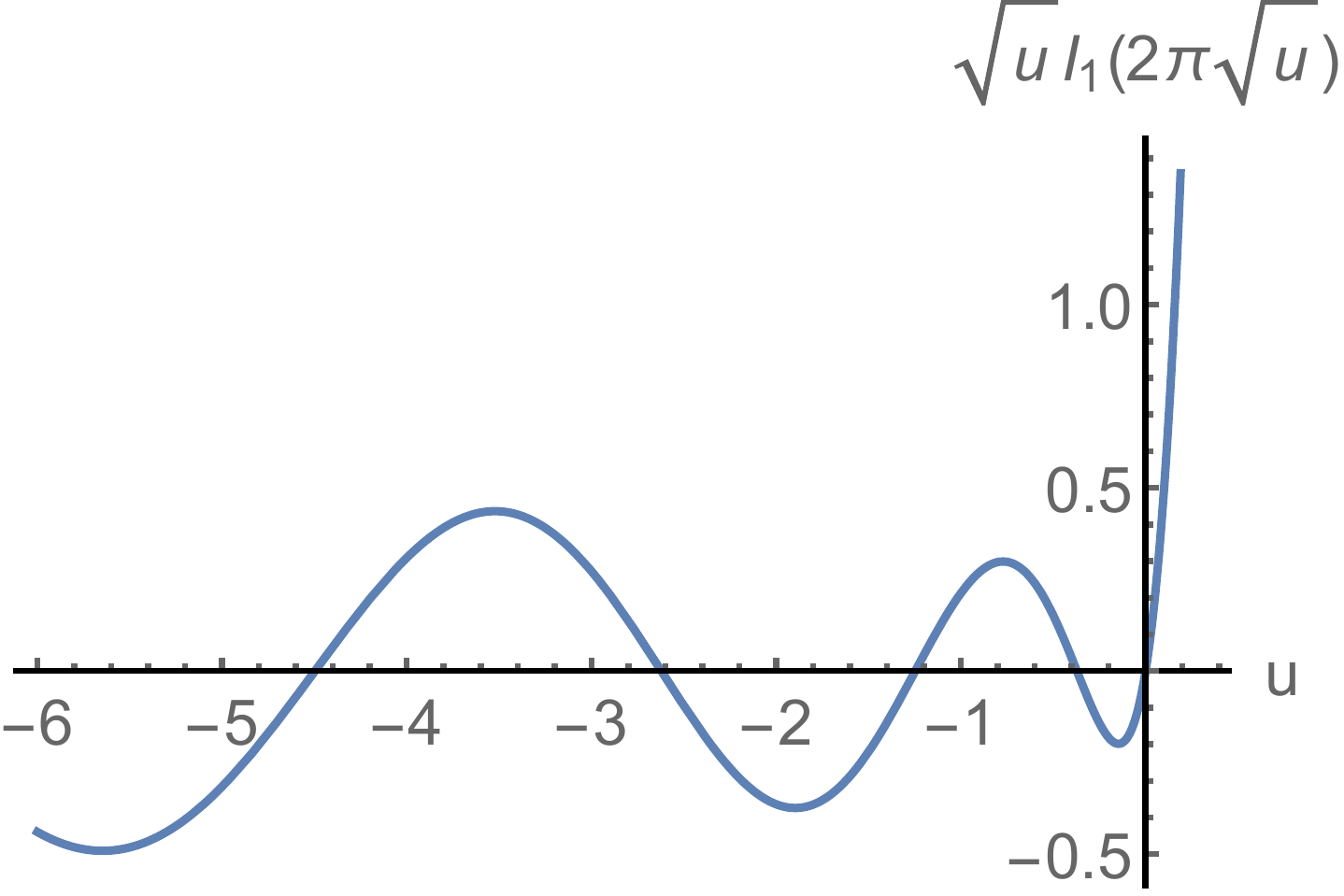}
\par\end{centering}
\caption{The plot of $\sqrt{u}I_{1}(2\pi\sqrt{u})$ as a function of $u$.
\label{fig:The-plot-of}}
\end{figure}

The leading term is a disk where there is no EOW brane. As computed
in \cite{Saad:2019lba}, the disk partition function is
\begin{equation}
Z_{\text{disk}}(\b)=\f{e^{S_{0}}\phi_{b}^{3/2}}{(2\pi)^{1/2}\b^{3/2}}e^{2\pi^{2}\phi_{b}/\b}
\end{equation}
The zero genus partition function is a series 
\begin{equation}
Z_{0}(\b)=Z_{\text{disk}}(\b)+KZ_{EOW}(\b)+\sum_{n=2}^{\infty}e^{-S_{0}(n-1)}Z_{n}(\b)\label{eq:51}
\end{equation}
where $Z_{n}(\b)$ is the Euclidean path integral over the surface
with $n$ EOW brane loops and one AdS boundary 
\begin{equation}
Z_{n}(\b)=\int bdb \, Z_{\trumpet}(\b,b)\int\prod_{i=1}^{n}db_{i}V_{0,n+1}(b,b_{1},\cdots,b_{n})C_{n}\prod_{i=1}^{n}\mM(b_{i}).\label{eq:52}
\end{equation}
In this formula, $V_{0,n+1}(b,b_{1},\cdots,b_{n})$ is the WP volume
for genus zero and $n+1$ holes computed in \cite{Mertens:2020hbs},
\begin{equation}
V_{0,n}(b_{1},\cdots,b_{n})=\lim_{x\ra0}-\f 12\del_{x}^{n-3}\left[u'(x)\prod_{i=1}^{n}J_{0}(b_{i}\sqrt{u(x)})\right]\label{eq:53},
\end{equation}
where $J_{0}(x)$ is Bessel function of first kind and $u(x)$ is
defined implicitly via
\begin{equation}
\sqrt{u(x)}I_{1}(2\pi\sqrt{u(x)})=-2\pi x\label{eq:4-1},
\end{equation}
where $I_{1}(x)$ is modified Bessel function of first kind. For a
given $x$, this equation has infinitely many solutions. We
take the largest $u$ because we need $u(0)=0$ in (\ref{eq:53}). This defines $u(x)$ smoothly in a neighborhood of $x = 0$, and for all
negative $x$. However, for negative $u$, $\sqrt{u} I_1(2 \pi \sqrt{u})$ oscillates (see Fig. \ref{fig:The-plot-of}) with
increasing amplitude as $|u|^{1/4}$, which leads to a piecewise continuous
function $u(x)$. As we will show later, for certain values of the parameters, this discontinuity will imply that the sum over EOW branes is divergent. 
This multi-valueness of $u(x)$ also leads to non-perturbative instability as analyzed in \cite{Johnson:2020lns}.
In (\ref{eq:52}), $C_{n}$ is a symmetry factor that accounts for identical EOW branes. We assume EOW branes to
be indistinguishable for the same flavor but distinguishable for different
flavors. This leads to
\begin{equation}
C_{n}=\sum_{\mathclap{\substack{\sum_{i = 1}^K \ell_{i}=n \\ \ell_i \ge 0}}}\f 1{\l_{1}!\cdots\l_{K}!}=\f 1{n!}K^{n}.
\end{equation}
Let us denote the last infinite sum in (\ref{eq:51}) as $\hat{Z}(\b)$.
Using (\ref{eq:53}), it can be written as
\begin{equation}
\hat{Z}(\b)=e^{S_{0}}\sqrt{\f{\phi_{b}}{2\pi\b}}\sum_{n=2}^{\infty}\f 1{n!}\del_{x}^{n-2}\left\{ \left[e^{-\f{\b u(x)}{2\phi_{b}}}\right]'\left(Ke^{-S_{0}}f(u(x))\right)^{n}\right\} _{x=0}\label{eq:8}
\end{equation}
where we have used
\begin{align}
\int_0^\infty bdb \, Z_{\trumpet}(\b,b)J_{0}(b\sqrt{u(x)}) & =\sqrt{\f{\b}{2\pi\phi_{b}}}e^{-\f{\b u(x)}{2\phi_{b}}}\label{eq:59}
\end{align}
and defined
\begin{align}
f(u) \equiv \int db \, \mM(b)J_{0}(b\sqrt{u})=\int_{\mD}d\a \,  m(\a)\int db \, e^{-\a b}J_{0}(b\sqrt{u}) & =\int_{\mD}d\a\f{m(\a)}{\sqrt{\a^{2}+u}}\label{eq:60},
\end{align}
where the $\int db \, e^{-\a b}J_{0}(b\sqrt{u})$ integral converges only when $\Re(\a\pm i\sqrt{u})\geq0$
for all $\a\in\mD$. Note that the boundedness of the regulated measure $\mM(b)$ for small $b$ also guarantees a finite $f(u)$ whereas using unregulated measure \eqref{eq:49} leads to divergent $f(u)$. To sum this series to a closed form, we need
to use the Lagrange inversion theorem (see Appendix \ref{sec:Lagrange-inversion-theorem}).
It turns out that
\begin{equation}
\hat{Z}(\b)=e^{S_{0}}\sqrt{\f{\phi_{b}}{2\pi\b}}\int_{a}^{0}dx\left[e^{-\f{\b u(x+g(x))}{2\phi_{b}}}-e^{-\f{\b u(x)}{2\phi_{b}}}-\left[e^{-\f{\b u(x)}{2\phi_{b}}}\right]'Ke^{-S_{0}}f(u(x))\right]\label{eq:12}
\end{equation}
where $g(x)$ is defined implicitly as
\begin{equation}
g(x)=Ke^{-S_{0}}f(u(g(x)+x))\label{eq:13}
\end{equation}
and $a$ must be a parameter such that $u(a)\ra+\infty$. Clearly,
we choose $a=-\infty$.

It turns out that the second and third terms in (\ref{eq:12}) are
exactly first two terms in (\ref{eq:51}). For the second term, we
can change the variable $x$ to $u$ to get
\begin{equation}
e^{S_{0}}\sqrt{\f{\phi_{b}}{2\pi\b}}\int_{-\infty}^{0}dx e^{-\f{\b u(x)}{2\phi_{b}}}=e^{S_{0}}\sqrt{\f{\phi_{b}}{2\pi\b}}\int_{\infty}^{0}due^{-\f{\b u}{2\phi_{b}}}x'(u)\label{eq:14-1}
\end{equation}
From (\ref{eq:4-1}), we have
\begin{equation}
x'(u)=-\f 1{2\pi}\del_{u}(\sqrt{u}I_{1}(2\pi\sqrt{u}))=-\f 12I_{0}(2\pi\sqrt{u})
\end{equation}
and (\ref{eq:14-1}) becomes
\begin{align}
e^{S_{0}}\sqrt{\f{\phi_{b}}{2\pi\b}}\int_{-\infty}^{0}dx e^{-\f{\b u(x)}{2\phi_{b}}}&=e^{S_{0}}\sqrt{\f{\phi_{b}}{2\pi\b}}\times\f 12\int_{0}^{\infty}due^{-\f{\b u}{2\phi_{b}}}I_{0}(2\pi\sqrt{u})\nn\\
&=\f{e^{S_{0}}\phi_{b}^{3/2}}{(2\pi)^{1/2}\b^{3/2}}e^{2\pi^{2}\phi_{b}/\b}=Z_{\text{disk}}(\b)
\end{align}
For the third term, we can recover the Bessel functions using (\ref{eq:59})
and (\ref{eq:60}), and change integration variable to $u$ to get
\begin{equation}
\sqrt{\f{\phi_{b}}{2\pi\b}}\int_{-\infty}^{0}dx\left[e^{-\f{\b u(x)}{2\phi_{b}}}\right]'Kf(u(x))=\f K2\int_{0}^{\infty}du\int b_{1}db_{1}Z_{\trumpet}(\b,b_{1})J_{0}(b_{1}\sqrt{u})\int db\mM(b)J_{0}(b\sqrt{u})\label{eq:17}
\end{equation}
Changing variable $u\ra x^2$ and using identity
\begin{equation}
\int_{0}^{\infty}dxxJ_{0}(ax)J_{0}(bx)=\f 1b\d(a-b),\qquad(a,b\geq0)
\end{equation}
we can rewrite it as
\begin{equation}
\sqrt{\f{\phi_{b}}{2\pi\b}}\int_{-\infty}^{0}dx\left[e^{-\f{\b u(x)}{2\phi_{b}}}\right]'Kf(u(x))=K\int dbZ_{\trumpet}(\b,b)\mM(b)=KZ_{EOW}(\b)
\end{equation}

Redefining $\xi(-x)\equiv u(x+g(x))$ and using
(\ref{eq:4-1}) and (\ref{eq:13}), we can write the partition function
as
\begin{equation}
Z(\b)=e^{S_{0}}\sqrt{\f{\phi_{b}}{2\pi\b}}\int_{0}^{\infty}dx \, e^{-\f{\b\xi(x)}{2\phi_{b}}}\label{eq:20}
\end{equation}
where $\xi(x)$ satisfies 
\begin{equation}
\sqrt{\xi(x)}\f{I_{1}(2\pi\sqrt{\xi(x)})}{2\pi}+Ke^{-S_{0}}f(\xi(x))=x.
\label{eq:stringeq}
\end{equation}
Equation \eqref{eq:stringeq} is the ``string equation'' that shows how the inclusion of EOW branes affects the partition function. Starting from the regulated and renormalized measure in \eqref{eq:regmeasure} and taking the $\epsilon \rightarrow 0$ limit, we find that $f(u)$ becomes $f_\lambda(u)$, which is defined by
\begin{align} 
f_\lambda(u)=&\lambda-\int_0^\infty db \log\left( 4\tanh\f b 4\right) \del_b\left( e^{-\mu b} J_0(b\sqrt{u}) \right)\nn\\
=& \lambda-\lambda_0 +\sum_{n=0}^\infty \left( \f 1 {\sqrt{(1/2+\mu+n)^2+u}}-\f 1 {1/2+\mu+n} \right) \label{eq:regflam}
\end{align}
where
\be
\lambda_0\equiv-\mu\int_0^\infty db \log\left( 4\tanh\f b 4\right) e^{-\mu b}=H(\mu-1/2)
\ee
and $H(x)$ is the Harmonic number. As the constant $\lambda_0$ has no physical meaning, we will shift $\lambda\ra \lambda+\lambda_0$ in the rest of paper for notational simplicity. 

In the limit that the regulator is removed, any regulated measure leads to the same $f_\lambda(u)$ for some value of renormalization parameter $\lambda$. For example, we used the cusp-like counterterm such that $f_{\lambda}(0) = \lambda$ in \eqref{eq:regflam}.
The derivative of $f_\lambda(u)$ agrees with the derivative of \eqref{eq:60} with $m(\alpha)$ taken to be the unregulated measure \eqref{eq:malpha}. In the limit that the regulator is removed, the string equation becomes 
\begin{equation}
\sqrt{\xi(x)}\f{I_{1}(2\pi\sqrt{\xi(x)})}{2\pi}+Ke^{-S_{0}}f_\lambda(\xi(x)) =x.
\label{eq:regstringeq}
\end{equation}
It is clear from the string equation that our theory contains three independent parameters: $K$, $\mu$, and $\lambda$. 

\subsection{Spectral density} \label{sec:gravspectrum}

\begin{figure}
\begin{centering}
\begin{subfigure}{.32\textwidth}
\includegraphics[height=3.5cm]{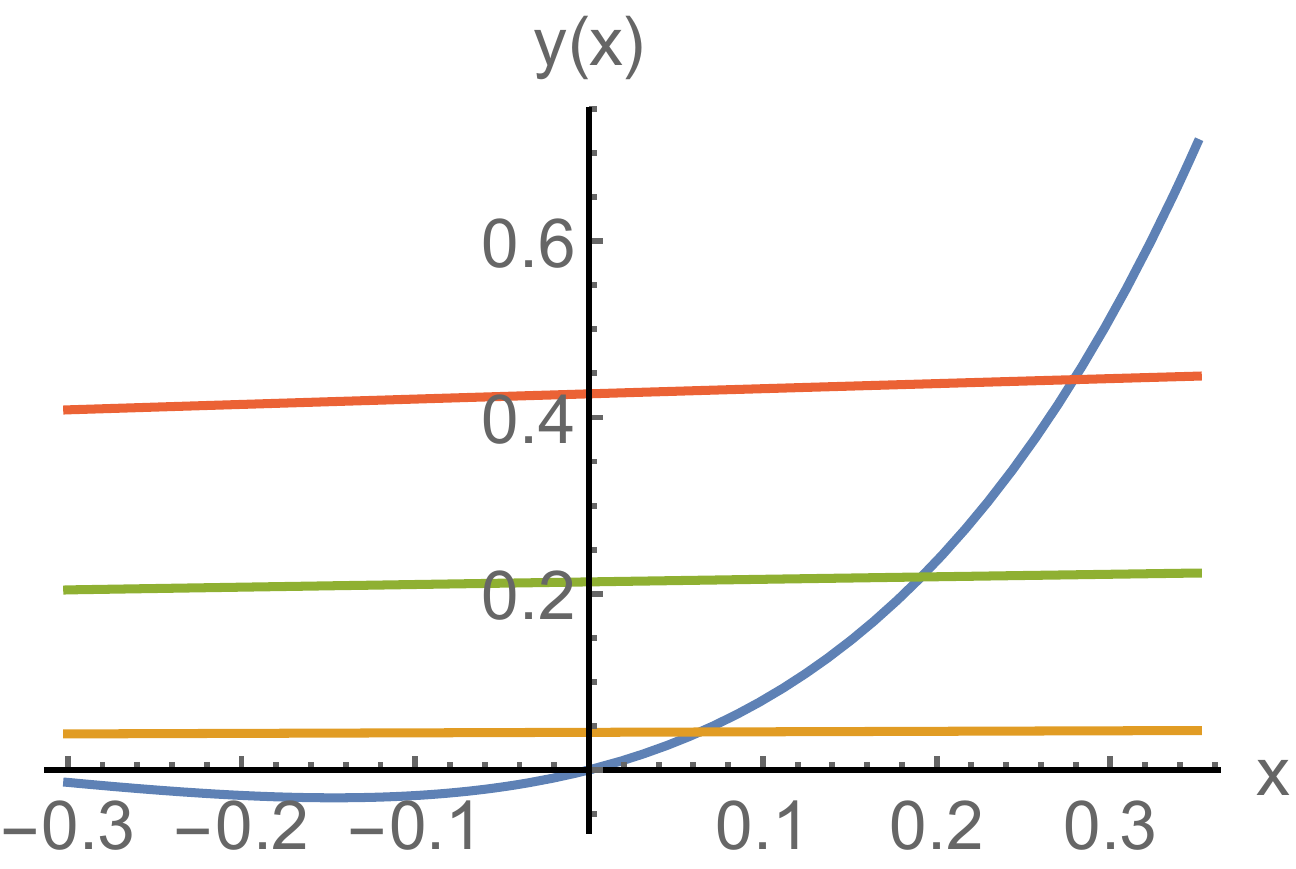}
\caption{$\lambda<0$}
\label{fig:4-a}
\end{subfigure}
\begin{subfigure}{.32\textwidth}
\includegraphics[height=3.5cm]{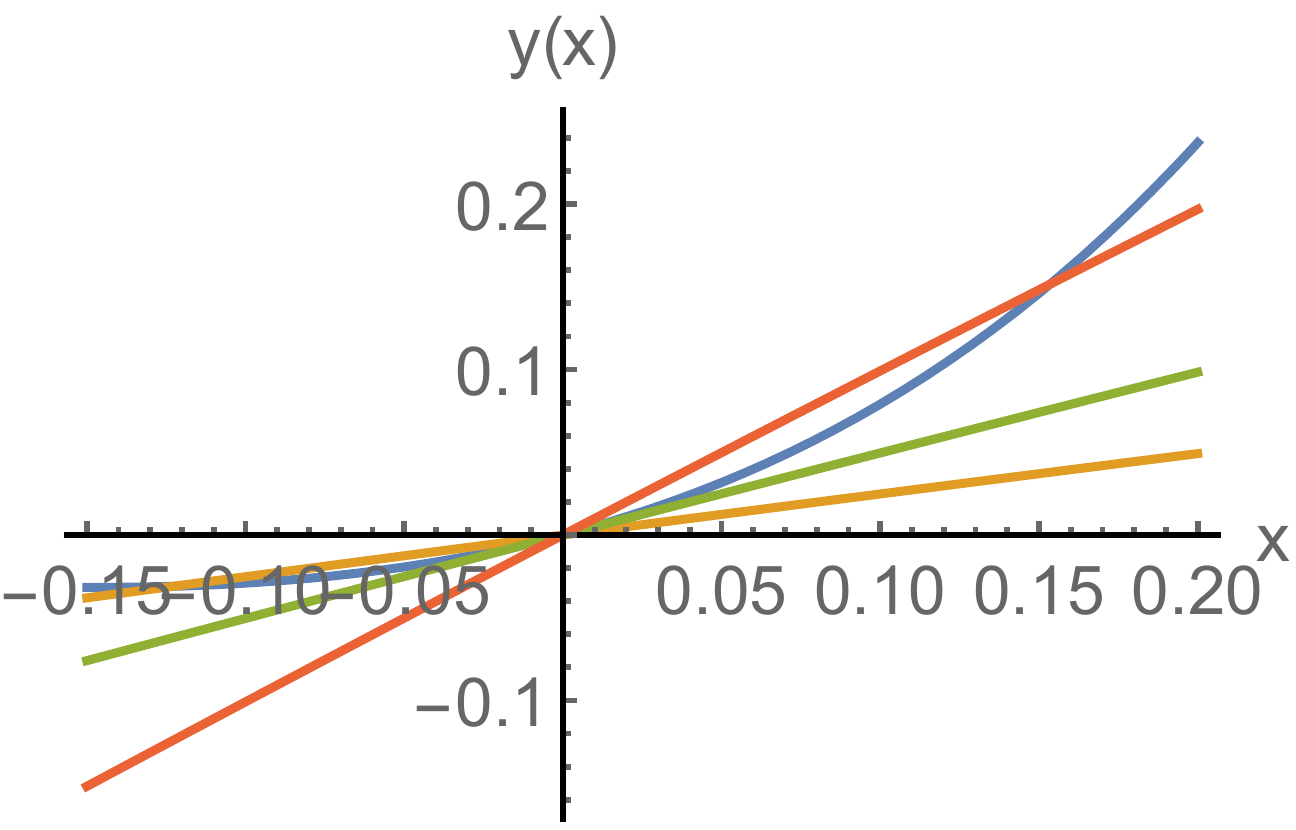}
\caption{$\lambda=0$}
\label{fig:4-b}
\end{subfigure}
\begin{subfigure}{.32\textwidth}
\includegraphics[height=3.5cm]{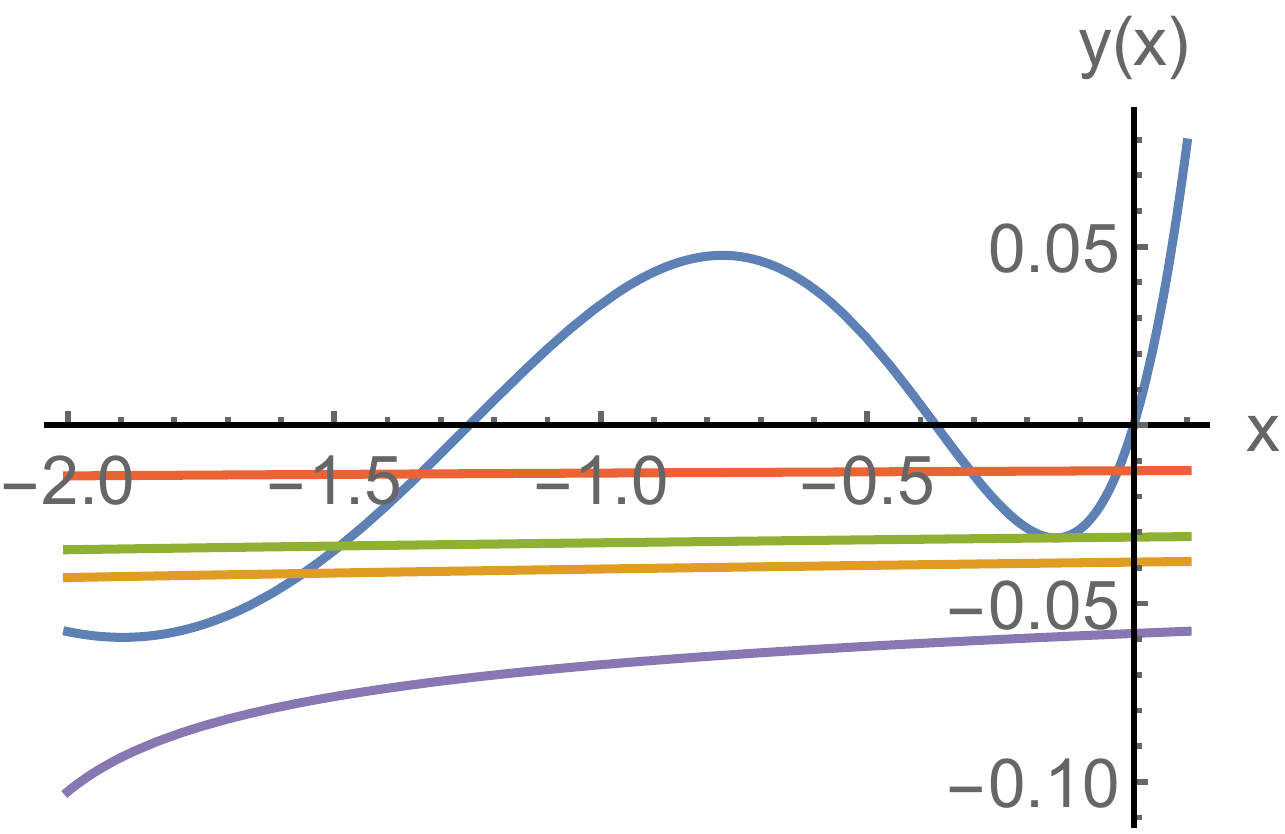}
\caption{$\lambda>0$}
\label{fig:4-c}
\end{subfigure}

\end{centering}
\caption{Plot of $\sqrt{x}I_1(2\pi\sqrt{x})/(2\pi)$ (blue) and $-Ke^{-S_0}f_\lambda(x)$ (other colors) for different $\lambda$. The largest intersection point gives zero point energy $E_0$. (a) yellow, green and red means increasing $K$ and $E_0$ moves rightward as $K$ increases; (b) yellow is $K<K_{cr}^=$, green is critical $K=K_{cr}^=$ and red is $K>K_{cr}^=$. $E_0=0$ when $K\leq K_{cr}^=$, and $E_0>0$ when $K>K_{cr}^=$; (c) yellow is $K>K_{cr}^>$, green is critical $K=K_{cr}^>$, red is $K<K_{cr}^>$ and purple is for $K$ too large such that no intersection exists (this purple curve also has smaller $\mu$ than the other three). $E_0$ moves leftward when $K$ increases and has a jump when $K>K_{cr}^>$. In the plot, we set $2\phi_b =1$ and $e^{S_0}=1$.} \label{fig:4}
\end{figure}

We perform an inverse Laplace transform to get the spectral density
\begin{align}
\r(E) & =\f{e^{S_{0}}\phi_{b}}{\pi}\int_{0}^{\infty}dx\f{\theta[2\phi_{b}E-\xi(x)]}{\sqrt{2\phi_{b}E-\xi(x)}}\nonumber \\
 & =\f{e^{S_{0}}\phi_{b}}{2\pi}\int_{2\phi_{b}E_{0}}^{2\phi_{b}E}\f{d\xi}{\sqrt{2\phi_{b}E-\xi}}\left(I_{0}(2\pi\sqrt{\xi})+2Ke^{-S_{0}}f'_\lambda(\xi)\right)\label{eq:23}
\end{align}
where the zero point energy $E_{0}$ is given by $x=0$ in \eqref{eq:regstringeq}
\begin{equation}
\sqrt{2\phi_{b}E_{0}}\f{I_{1}(2\pi\sqrt{2\phi_{b}E_{0}})}{2\pi}+Ke^{-S_{0}}f_\lambda(2\phi_{b}E_{0})=0\label{eq:23-1}
\end{equation}
As $f_\lambda(u)$ has singularities at $-(1/2+\mu+n)^2$ for $n\in\N$, we will assume $2\phi E_0 >-(1/2+\mu)^2$ throughout this paper. Using \eqref{eq:60} and \eqref{eq:regmeasure}, we can also write the second term in \eqref{eq:23} as
\begin{align}
\r(E) & \supset-\f{K\phi_{b}}{2\pi}\int_{\mD}d\a m(\a)\int_{2\phi_{b}E_{0}}^{2\phi_{b}E}\f{d\xi}{\sqrt{2\phi_{b}E-\xi}(\a^{2}+\xi)^{3/2}}\nonumber \\
 & =-\f{K\phi_{b}}{\pi}\int_{\mD}d\a m(\a)\f{\sqrt{2\phi_{b}(E-E_{0})}}{(\a^{2}+2\phi_{b}E)\sqrt{\a^{2}+2\phi_{b}E_{0}}}\label{eq:36-1}
\end{align}
In $\epsilon \ra 0$ limit, the derivative $f_\lambda^\prime(u)=f^\prime(u)$ is finite, which means that the dependence of $\rho(E)$ on the renormalization parameter $\lambda$ is through the zero point energy $E_0$ only. We are interested in how the spectrum changes when $K$ is increased from zero. For $K=0$, the spectrum reduces to the pure JT gravity result
\begin{equation}
\r_{JT}(E)=\r_{K=0}(E)=\f{e^{S_{0}}\phi_{b}}{2\pi^{2}}\sinh(2\pi\sqrt{2\phi_{b}E})\label{eq:rhoJT}
\end{equation}
For nonzero $K$, because $\sqrt{u}I_1(2\pi\sqrt{u})$ passes through the origin as shown in Fig. \ref{fig:The-plot-of}, we can organize the problem into three cases based on the value of $\lambda$: $f_\lambda(0)<0$, $f_\lambda(0)=0$ and $f_\lambda(0)>0$.

\begin{figure}
\begin{centering}
\begin{subfigure}{.32\textwidth}
\includegraphics[height=3.5cm]{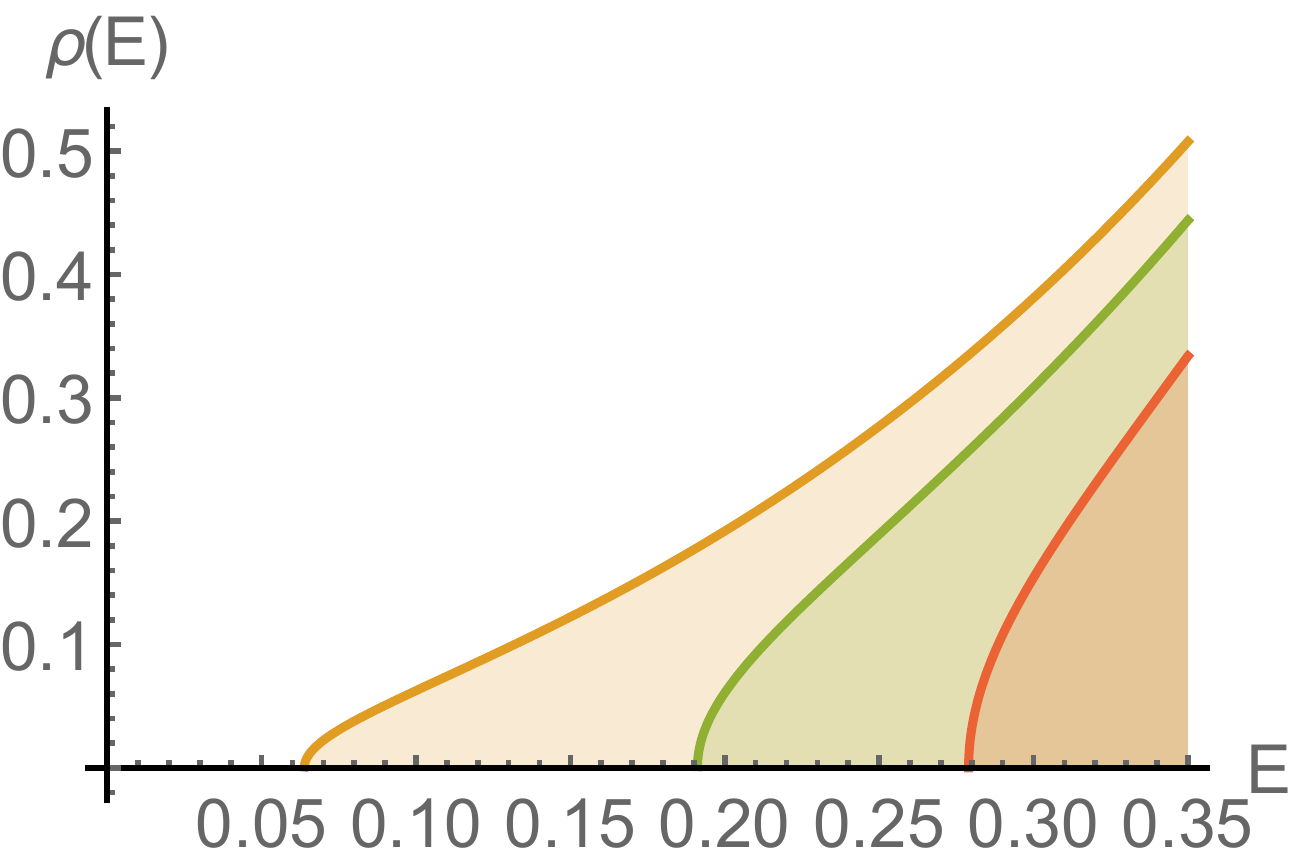}
\caption{$\lambda<0$}
\label{fig:5-a}
\end{subfigure}
\begin{subfigure}{.32\textwidth}
\includegraphics[height=3.5cm]{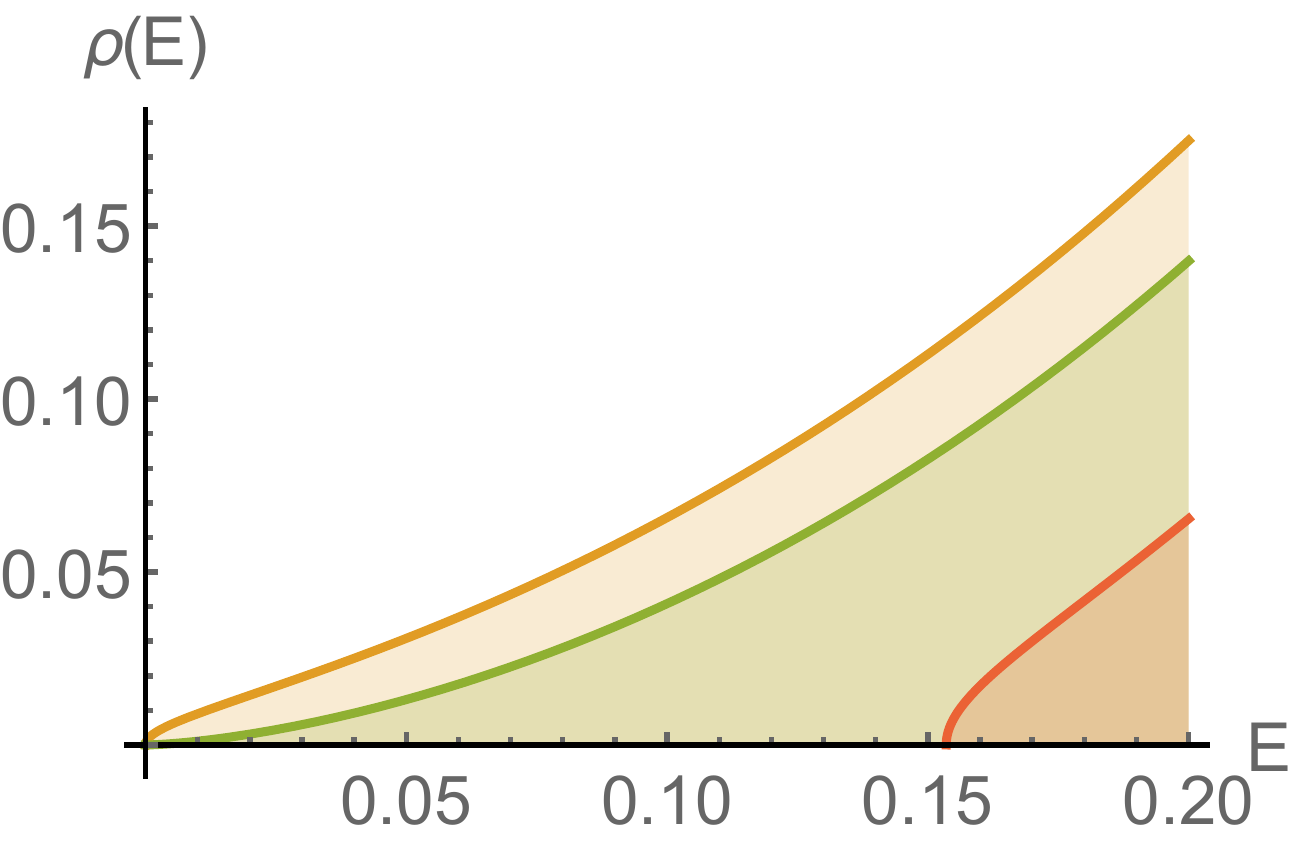}
\caption{$\lambda=0$}
\label{fig:5-b}
\end{subfigure}
\begin{subfigure}{.32\textwidth}
\includegraphics[height=3.5cm]{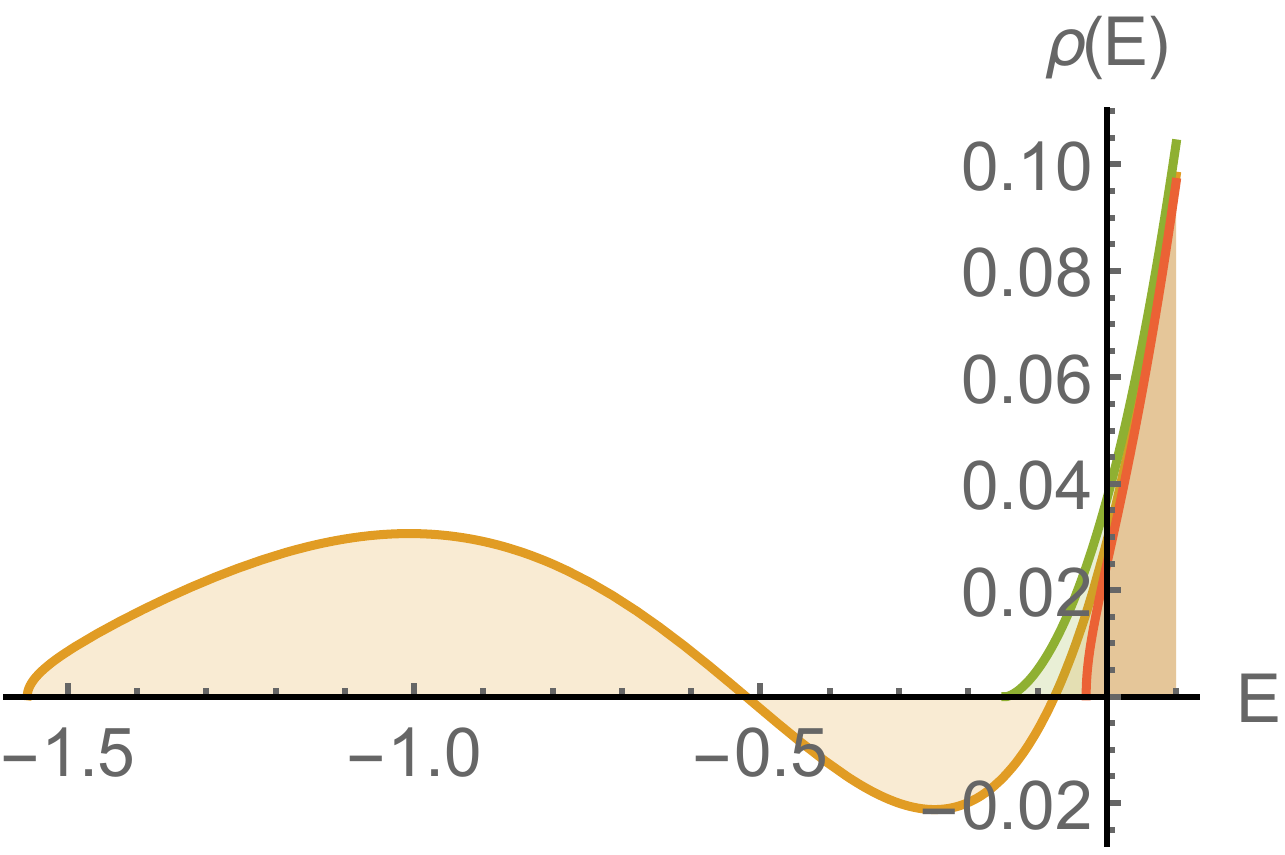}
\caption{$\lambda>0$}
\label{fig:5-c}
\end{subfigure}

\end{centering}
\caption{Plot of spectral density $\rho(E)$ for different $\lambda$. Each color represents taking the $K$ value corresponding to the same color curve in Fig. \ref{fig:4}. All parameters are identical to Fig. \ref{fig:4}.}
\end{figure}

Note that $f_\lambda(u)$ is a monotonically decreasing function. When $f_\lambda(0)<0$, $\lambda<0$, the largest solution to \eqref{eq:23-1} is positive, which implies $E_0>0$ (see Fig. \ref{fig:4-a}). As $K$ increases, $E_0$ increases as well and the spectrum is pushed to right (see Fig. \ref{fig:5-a}). 

When $f_\lambda(0)=0$, $\lambda=0$, the largest solution to \eqref{eq:23-1} has a non-smooth but continuous change. Indeed, it suffices to compare the slope of the two pieces in \eqref{eq:23-1}. The slope of $\sqrt{u}I_1(2\pi\sqrt{u})/(2\pi)$ is 1/2 at $u=0$ and that of $f_\lambda(u)$ is
\be
\del_u f_\lambda(u)|_{u=0}=\f 1 4 \psi^{(2)}(1/2 + \mu)<0,~~~~(\mu>-1/2)
\ee
where $\psi^{(n)}$ is $n$-th order polygamma function. When $K<-2e^{S_0}/\psi^{(2)}(1/2 + \mu)$, the slope of $-Ke^{-S_0}f(x)$ is smaller than that of $\sqrt{u}I_1(2\pi\sqrt{u})/(2\pi)$ and  the largest solution to \eqref{eq:23-1} is $E_0=0$ (see the yellow curve in Fig. \ref{fig:4-b}); when $K>-2e^{S_0}/\psi^{(2)}(1/2 + \mu)$, the largest solution to \eqref{eq:23-1} is at some $E_0>0$ (see the red curve in Fig. \ref{fig:4-b}). It turns out that $K=K_{cr}^= \equiv-2e^{S_0}/\psi^{(2)}(1/2 + \mu)$ is the critical point for a phase transition of spectrum, where two curves tangentially intersects at $x=0$ (see the green curve in Fig, \ref{fig:4-b}).

At this critical point, the spectral density near the edge scales like $\rho(E)\sim E^{3/2}$ rather than the generic case where $\rho(E)\sim (E-E_0)^{1/2}$ as shown in Fig. \ref{fig:5-b}. To see this, let us define
\begin{equation}
G(x)\equiv\sqrt{x}\f{I_{1}(2\pi\sqrt{x})}{2\pi}+Ke^{-S_{0}}f_\lambda(x)\label{eq:119}
\end{equation}
At the critical point, we have $G(0)=G'(0)=0$.
Close to the critical point, expanding $E$ around $E_{0}$, we have
\begin{equation}
G(2\phi_{b}E)\sim a_{1}(E-E_{0})+a_{2}(E-E_{0})^{2}+O((E-E_{0})^{3}),\qquad a_{1}\geq0,\quad a_{2}>0
\end{equation}
where $a_{1}=0$ at the critical point. As we discussed before, $E_0>0$ for $K>K_{cr}^=$, and $E_0=0$ for $K<K_{cr}^=$. Taking this into (\ref{eq:23}),
we find that the spectral edge scales as
\begin{equation}
\r(E)\sim\int_{2\phi_{b}E_{0}}^{2\phi_{b}E}\f{d\xi}{\sqrt{2\phi_{b}E-\xi}}\left[a_{1}+2a_{2}(\xi-E_{0})\right]\sim2(E-E_{0})^{1/2}\left[a_{1}+\f{4a_{2}}3(E-E_{0})\right]\label{eq:114-1}
\end{equation}
where the leading order scaling is $(E-E_0)^{1/2}$ unless $a_1=0$.

When $f_\lambda(0)>0$, $\lambda>0$, the largest solution to \eqref{eq:23-1} is negative, which implies $E_0<0$ (see Fig. \ref{fig:4-c}). As $K$ increases, $E_0$ decreases as well and the spectrum is pushed to left (see Fig. \ref{fig:5-c}). However, as $\sqrt{x}I_1(2\pi\sqrt{x})/(2\pi)$ is an oscillatory function when $x<0$, there will be another critical point $K=K_{cr}^{>}$ when $\sqrt{x}I_1(2\pi\sqrt{x})/(2\pi)$ tangentially intersects with $-Ke^{-S_0}f_\lambda(x)$ at some negative $x$. Similar to \eqref{eq:114-1}, we can show that at this critical point the near edge spectrum scales as $(E-E_0)^{3/2}$. Though $K_{cr}^>$ does not have an analytic closed form, we can easily confirm that $E_0$ would have a jump when $K>K_{cr}^>$ because of the oscillatory feature of  $\sqrt{x}I_1(2\pi\sqrt{x})/(2\pi)$ (see Fig. \ref{fig:4-c}).

However, such a jump of $E_0$ leads to an unphysical spectrum because there is a range of $E$ where $\r(E)<0$ (see the yellow curve in Fig. \ref{fig:5-c}). Such a negative spectrum has been observed also in JT gravity with deficit angles \cite{Maxfield:2020ale, Witten:2020wvy} and it indicates a breakdown of the gravitational computation. To be precise, the gravitational sum over loops of EOW branes, as a perturbative series in $K$, will not be convergent for $K > K_{cr}^>$ because $u(x)$, defined in \eqref{eq:4-1}, has a branch cut beginning at a positive value of $x$.\footnote{The Lagrange inversion theorem, which is used to perform the sum, requires $u(x)$ to be analytic in a sufficiently large domain.} Indeed, when $K$ increases past $K_{cr}^>$, the largest two real solutions of \eqref{eq:23-1} collide and then split into two complex conjugate values, which are separated by the branch cut. Therefore, the rule to take $2\phi_b E_0$ as largest real solution to \eqref{eq:23-1} is not analytic.  

From the dual matrix model point of view, this is interpreted as a nonperturbative instability \cite{Saad:2019lba, Johnson:2020lns} that requires a nonperturbative completion of the divergent sum to yield a reasonable spectral density. In Sections \ref{sec:effmatrixmodel} and \ref{sec:yshape}, we will consider a nonperturbatively well-defined matrix model 
whose tree-level spectral density agrees with the results in this section for $K < K_{cr}^{>}$ but remains well-defined for $K > K_{cr}^{>}$. We will find that beyond this critical point, the spectrum undergoes a special phase transition to include both real and complex energies. 

Before we move on to next section, we will point out a worse problem for the spectrum formula given by \eqref{eq:23} and \eqref{eq:23-1} when $K>K_{cr}^>$. Our regulated $f_\lambda(x)$ in \eqref{eq:regflam} has singularities at $x=-(1/2+\mu+n)^2$ for all nonnegative integer $n$.
If we take $K$ large enough, there could be no real solution to \eqref{eq:23-1} at all because $f_\lambda(x)$ blows up near its first singularity too fast (see the purple curve in Fig. \ref{fig:4-c}). This feature does not exist in JT gravity with deficit angles \cite{Maxfield:2020ale, Witten:2020wvy}, where real solution of $E_0$ always exists. This is another evidence that gravitational computation breaks down when we include too many EOW branes if $\lambda>0$. 

\subsection{Integrating out EOW branes}\label{sec:2.5intEOW}

In this subsection we discuss the effective action for the metric and dilaton that remains after integrating out $K$ species of EOW branes with mass $\mu$. Having calculated the change of the spectral density due to the EOW branes, we now want to find a dilaton gravity theory whose disk path integral reproduces the total spectral density. Because the effective action associated with cusps is already known \cite{Maxfield:2020ale,Witten:2020wvy}, we will focus on the theory where $E_0 = 0$, or where $\lambda = 0$ in \eqref{eq:regflam}. For small enough $K$, before any phase transition occurs, the spectral density follows from \eqref{eq:36-1},
\begin{equation}
  \rho_K(E) = - K \frac{\phi_b}{\pi} \int_0^\infty db \, \frac{\sin^2(\frac{b}{2} \sqrt{2 \phi_b E})}{\sqrt{2 \phi_b E}} \frac{e^{- \mu b}}{\sinh \frac{b}{2}}.
  \label{eq:EOWcorrection}
\end{equation}
In \cite{Witten:2020ert}, Witten provides a formula for the spectral density $\rho(E;U)$ associated with the dilaton potential $W(\Phi) = 2 \Phi + U(\Phi)$, for a certain class of potentials. The result is
\begin{equation}
    \rho(E;U) - \rho_{JT}(E) = e^{S_0}\frac{\phi_b}{4 \pi \sqrt{2 \phi_b E}} \left( e^{2 \pi \sqrt{2 \phi_b E}} U(\sqrt{2 \phi_b E}) + e^{-2 \pi \sqrt{2 \phi_b E}} U(-\sqrt{2 \phi_b E}) \right).
    \label{eq:wittensformula}
\end{equation}
This result was derived for a class of potentials given by
\begin{equation}
    U(\Phi) = 2 \sum_{i = 1}^r \epsilon_i \, e^{- \alpha_i \Phi}, \quad \pi < \alpha_i < 2 \pi, \quad U(0) = 0.
    \label{eq:specialclass}
\end{equation}
For certain potentials, \eqref{eq:wittensformula} predicts that the spectral density can be negative for some energies, similar to Fig. \ref{fig:5-c}. In this case, we expect that \eqref{eq:wittensformula} breaks down. The authors of \cite{Johnson:2020lns} studied an example where a naive application of \eqref{eq:wittensformula} predicts negativity in the spectral density, while the correct spectral density has $E_0 \neq 0$. For now, we assume that $K e^{-S_0}$ is sufficiently small such that $\rho_{JT}(E) + \rho_K(E)$ is positive everywhere.

If we take $U(\Phi)$ to be
\begin{equation}
    U(\Phi) = 2 K e^{- S_0} e^{- 2 \pi \Phi} \int_0^\infty db \, (\cos(b \Phi) - 1) \frac{e^{- \mu b}}{2 \sinh \frac{b}{2}},
    \label{eq:uphi}
\end{equation}
then \eqref{eq:wittensformula} coincides with \eqref{eq:EOWcorrection}. We claim that \eqref{eq:uphi} is the correction to the JT action from integrating out the EOW branes. However, \eqref{eq:uphi} is not manifestly of the form of \eqref{eq:specialclass}. In the spirit of effective field theory, we will expand \eqref{eq:uphi} in a power series in $\frac{1}{\mu}$. The first few orders are given by
\begin{equation}
    U(\Phi) = 2 K e^{- S_0} e^{- 2 \pi \Phi} \left[ - \frac{\Phi^2}{2 \mu^2} +\frac{\Phi^2 + 2 \Phi^4}{8 \mu^4} + O(\frac{1}{\mu^6}) \right].
    \label{eq:expansion}
\end{equation}
Each term in the above expansion takes the form of \eqref{eq:specialclass} for a particular limit of the parameters. Thus, \eqref{eq:wittensformula} can be used to compute the spectral density associated with \eqref{eq:uphi} order by order in the $\frac{1}{\mu}$ expansion. Alternatively, we can use the result that the string equation associated to a potential of the form \eqref{eq:specialclass} is \cite{Witten:2020ert,Maxfield:2020ale}
\begin{equation}
    \frac{\sqrt{\xi(x)}}{2\pi} I_1(2 \pi \sqrt{\xi(x)}) + \sum_i \epsilon_i I_0\left( \left(2 \pi - \alpha_i \right) \sqrt{\xi(x)} \right) = x.
\end{equation}
It follows that the string equation that corresponds to \eqref{eq:expansion} is
\begin{equation}
    \frac{\sqrt{\xi(x)}}{2\pi} I_1(2 \pi \sqrt{\xi(x)}) + 2 K e^{- S_0} \left[ - \frac{\xi}{8 \mu^2} + \frac{2\xi +  3 \xi^2}{64 \mu^4} + O(\frac{1}{\mu^6}) \right] = x.
\end{equation}
This agrees with \eqref{eq:regflam} for $\lambda = 0$ to the same order in $\frac{1}{\mu}$.

When $\mu$ is large and $K e^{- S_0}$ is order one, the EOW branes have a small effect on the spectral density, and one cannot reach a phase transition. As shown in Figure \ref{fig:4-b}, there is a phase transition when $K > K_{cr}^{=}$, which is $\mu$-dependent. To investigate the validity of \eqref{eq:uphi} across this phase transition, we note that for sufficiently large $K$, the largest zero of $W(\Phi)$, which we denote $\phi_*$, discontinuously jumps from zero to a positive value. As $K$ increases further, $\phi_*$ increases monotonically. The zero temperature entropy of a black hole, evaluated from the on-shell action, is given by $2 \pi \phi_*$. We can compare this semiclassical entropy with our exact calculations by examining the low-energy behavior of the spectral density, \eqref{eq:23}. For $2\phi_b(E-E_0)\ll 1$, \eqref{eq:23} approximately becomes
\begin{equation}
    \rho(E) \approx \frac{e^{S_0} \phi_b }{\pi} \sqrt{2 \phi_b(E - E_0)}\left(I_0(2 \pi \sqrt{2 \phi_b E_0}) + 2 K e^{-S_0} f^\prime_{0}(2 \phi_b E_0) \right).
\end{equation}
We define the effective zero-temperature entropy (or effective extremal entropy), $S_{0,\eff}$, by the coefficient of $\sqrt{E-E_0}$ in $\r(E)$
\begin{equation}
    e^{S_{0,\eff}}\equiv \f{\sqrt{2\phi_b}\phi_b e^{S_0}}{\pi} \left(I_0(2 \pi \sqrt{2 \phi_b E_0}) + 2 K e^{-S_0} f^\prime_{0}(2 \phi_b E_0) \right).
    \label{eq:seff}
\end{equation}
The effective zero-temperature entropy, which we simply read off from the normalization of the edge of the spectrum, represents the change in the zero-temperature entropy due to integrating out EOW branes. For order one values of $K e^{-S_0}$, $S_{0,\eff}$ and $2 \pi \phi_*$ disagree; indeed, for $K < K_{cr}^=$, $S_{0,\eff}$ is nonzero while $2 \pi \phi_*$ is zero.\footnote{This is not a surprise because \eqref{eq:uphi} does not have a simple weak coupling limit. To restore $G$, we replace $\phi \rightarrow \frac{\phi}{8 \pi G}$. Because $\frac{1}{G}$ does not appear in the action as an overall prefactor, $G \rightarrow 0$ is not a weak coupling limit. Furthermore, near the phase transition, \eqref{eq:seff} is not a good measure of the zero temperature entropy because the coefficient of the $E^{1/2}$ term in the spectral density is suppressed relative to the $E^{3/2}$ term.} However, for large values of $K$ both $S_{0,\eff}$ and $2 \pi \phi_*$ diverge as $\log K$. This provides a qualitative check of \eqref{eq:uphi} for $K > K_{cr}^=$.

The fact that $S_{0,\eff}$ increases as $\log K$ for large $K$ suggests that in this limit, the zero-temperature entropy is counting the number of species of EOW branes. Thus, we can view JT gravity with EOW branes as a model of induced gravity, since part of the black hole entropy in the effective dilaton gravity theory is induced by dynamical EOW branes that have been integrated out. 

Indeed, such induced gravity interpretation is a general feature for heavy EOW branes in large $K$. For $\lambda<0$ and large $\mu$ limit, we can use \eqref{eq:23-1} and \eqref{eq:seff} to see the same scaling $S_{0,\eff}\sim\log K$ when $Ke^{-S_0}\gg 1$ even though it does not undergo any phase transition. For $\lambda>0$ where the gravitational computation breaks down when $K>K_{cr}^>$, we will see the same scaling in Section \ref{sec:4.3yshape} and \ref{subsec:-dilaton-JT} for $K\gg K_{cr}^>$.

\section{Effective matrix model}\label{sec:effmatrixmodel}

The duality between JT gravity and a double-scaled matrix model \cite{Saad:2019lba} can be expressed as the following identity
\begin{equation}
Z_{n}(\b_{1},\cdots,\b_{n})=\avg{\Tr \, e^{-\b_{1}H}\cdots\Tr \,  e^{-\b_{n}H}}\equiv\lim_{\text{\ensuremath{\substack{\text{double}\\
\text{scaling}
}
}}}\f 1{\mZ}\int dH \, e^{-N\Tr V(H)} \Tr \, e^{-\b_{1}H}\cdots\Tr \, e^{-\b_{n}H}\label{eq:74}
\end{equation}
where $H$ is a $N\times N$ hermitian matrix. In this equation, $Z_{n}(\b_{1},\cdots,\b_{n})$
is the Euclidean path integral over all Riemann surfaces with $n$ AdS boundaries with inverse temperatures $\b_{1}$ to $\b_{n}$, and
$\avg{\Tr e^{-\b_{1}H}\cdots\Tr e^{-\b_{n}H}}$ is the expectation
value of $n$ operator insertions $\Tr e^{-\b_{i}H}$ in the matrix
model in the double scaling limit. $\mZ$ is the matrix model partition
function with no insertion of operators. The double scaling is the
special limit in which we zoom in the lower edge of spectrum in the large
$N$ limit while fixing the total number of eigenvalues $\sim e^{S_{0}}$
in any finite energy range. On both sides of \eqref{eq:74}, we have a topological
expansion, in which each order is related to a genus $g$ Riemann
surface weighted by $e^{(2-2g-n)S_{0}}$. The beautiful work in \cite{Saad:2019lba}
derives (\ref{eq:74}) by showing that both topological expansions
are equivalent order by order. In this section, we will work out how
dynamical EOW branes modify the matrix model potential $V(H)$ and the genus zero spectral density.

\subsection{Potential deformation by EOW branes}\label{sec:3.1}

Our strategy is to find the ``inverse trumpet'' $\tilde{Z}(\beta)$ such that\footnote{A recent work \cite{Goel:2020yxl} also studied matrix model dual to branes with
different boundary condition via the inverse trumpet.}
\begin{equation}
\int d\b Z_{\trumpet}(b,\b)\tilde{Z}(\b)=\mM(b)/b.
\end{equation}
Using $\tilde{Z}(\b)$, we can compute a path integral with $n$ loops
of EOW branes by integrating the path integral of pure JT gravity with at least $n$ AdS boundaries against $n$ inverse trumpets. Invoking the duality between pure JT gravity and a matrix integral, the Euclidean path integral over all surfaces with $m$ AdS boundaries and $n$ EOW brane loops is given by
\begin{equation}
    \mZ_{n}(\beta_1,\ldots,\beta_m) = \frac{K^n}{n!} \avg{\left[\prod_{j = 1}^m \text{Tr } e^{-\beta_j H}\right] \left[\int d\tilde{\beta}  \Tr e^{-\tilde{\beta} H}  \tilde{Z}(\tilde{\beta}) \right]^n}.
\end{equation}
If we define
\begin{equation}
    \d V(H)\equiv-\int d\b e^{-\b H}\tilde{Z}(\b),
\end{equation}
then the sum over all geometries with $m$ AdS boundaries and any number of EOW branes becomes
\begin{equation}
\mZ(\beta_1,\ldots,\beta_m)=\sum_{n=0}^{\infty}\mZ_{n}(\beta_1,\ldots,\beta_m)=\lim_{\text{\ensuremath{\substack{\text{double}\\
\text{scaling}
}
}}}\f 1{\mZ}\int dH      \prod_{j = 1}^m \text{Tr } e^{-\beta_j H} \, e^{-N\Tr V(H)-K\Tr\d V(H)}.
\label{eq:mmEOW}
\end{equation}
Equation \eqref{eq:mmEOW} includes unwanted terms that may be interpreted as disconnected vacuum bubble geometries with EOW branes; these may be cancelled by multiplying $\mZ$ by an overall constant.

%
%

Finding the inverse trumpet $\tilde{Z}(\beta)$ is straightforward. The inverse Laplace
transformation of trumpet is given by (\ref{eq:48}), which written
in energy variable is
\begin{equation}
Z_{\trumpet}(b,\b)=\int_{0}^{\infty}dE\r_{\trumpet}(b,E)e^{-\b E},\qquad\r_{\trumpet}(b,E)=\sqrt{\f{\phi_{b}}{2E}}\f{\cos(\sqrt{2\phi_{b}E}b)}{\pi}\label{eq:51-1}
\end{equation}
It leads to 
\begin{equation}
\int d\b Z_{\trumpet}(b,\b)\tilde{Z}(\b)=-\int_{0}^{\infty}dE\r_{\trumpet}(b,E)\d V(E)=\mM(b)/b
\end{equation}
Using the orthogonality of cosine function
\begin{equation}
\int_{0}^{\infty}dx\cos ax\cos bx=\f{\pi}2(\d(a+b)+\d(a-b))
\end{equation}
we have
\begin{equation}
\d V(E)=-2\int_{0}^{\infty}db\cos(\sqrt{2\phi_{b}E}b)\mM(b)/b\label{eq:84}
\end{equation}
This integral is not well-defined near $b\sim0$. We can regularize
it by, for example, replacing $1/b$ with $b^{\a}$ and taking $\a\ra-1+\e$ in the last
step. Using the identity
\begin{equation}
\int_{0}^{\infty}dbe^{-xb}b^{\a}=x^{-1-\a}\G(1+\a)
\end{equation}
we find for (\ref{eq:49})
\begin{equation}
\d V(E)=-[\zeta(1+\a,1/2+\mu+i\sqrt{2\phi_{b}E})+c.c]\G(1+\a)=\f{2\mu}{\e}-\log\left[\f{e^{2\mu\g_{E}}}{2\pi}|\G(1/2+\mu+i\sqrt{2\phi_{b}E})|^{2}\right]+o(\e)
\end{equation}
Note that the leading divergence and constant $\log\f{e^{2\mu\g_{E}}}{2\pi}$
can be absorbed into the normalization of the matrix model partition function, and thus does not affect any correlation functions. Therefore, we find that summing over dynamical EOW brane loops in the path integral amounts to shifting the matrix model potential by $K \delta V$, with
\begin{equation}
\boxed{\d V(E)=-\log\left[|\G(1/2+\mu+i\sqrt{2\phi_{b}E})|^{2}\right].\label{eq:dv}}
\end{equation}
From this matrix model analysis, it appears that we do not need to introduce any counterterm to the matrix model potential because the $b \rightarrow 0$ divergence of \eqref{eq:84} does not seem to affect any normalized correlators. However, just as the computations in Section \ref{sec:2} required regularization, the analysis of the double-scaled matrix model with the potential deformed by \eqref{eq:dv} requires additional input to be well-defined. In Section \ref{sec:determinE0}, we will explain why and interpret the meaning of the cusp counterterm introduced in section \ref{sec:gravitycomputation} from the perspective of the matrix model.

The derivative of $\delta V(E)$ is a sum of poles:
\begin{equation}
\label{eq:sumofpoles}
\delta V^\prime(E) = \sum_{n = 0}^\infty \frac{1}{E + \frac{(\frac{1}{2} +  \mu + n)^2}{2 \phi_b}}. 
\end{equation}
Throughout this paper, we will assume that the support of the spectrum of the matrix model does not contain any of the singularities in \eqref{eq:sumofpoles}.

The deformation of potential \eqref{eq:dv} could also be rewritten as an integral over complex vector fields. Using identity $\Tr\log A=\log\det A$, we have
\begin{align}
e^{K\Tr\log\left[|\G(1/2+\mu+i\sqrt{2\phi_{b}H})|^{2}\right]}&=\left[\det|\G(1/2+\mu+i\sqrt{2\phi_{b}H})|^{-2}\right]^{-K}\nn\\
&=\int DQ_i^\dagger DQ_i e^{-\sum_{i=1}^K Q_i^\dagger|\G(1/2+\mu+i\sqrt{2\phi_{b}H})|^{-2} Q_i}
\end{align}
where each $Q_i$ is a $N$-dimensional complex vector for $i=1,\cdots,K$. Alternatively to a deformation of the potential, we could understand each flavor of EOW loop as being dual to a coupling $Q_i^\dagger|\G(1/2+\mu+i\sqrt{2\phi_{b}H})|^{-2} Q_i$ of complex vectors in the matrix model. Each of the vectors $Q_i$ has the interpretation of a state with a single EOW brane in the gravity theory.


\subsection{Saddle equation for the matrix model with deformed potential} \label{sec:3.2}

In this section, we discuss how the shift in the potential $\delta V$ affects the tree-level spectral density of the matrix model. Instead of using the explicit form of $\delta V$ given in \eqref{eq:dv}, we will use the more general formula \eqref{eq:84} that applies for general geodesic length measures. This is because it will turn out to be necessary to use a regulated $\mM(b)$ in \eqref{eq:84} when determining the location of the lower endpoint of the spectrum.

The total potential of the matrix model
is
\begin{equation}
NV=NV_{JT}+K\d V\label{eq:88}
\end{equation}
where $V_{JT}$ is the potential of the matrix model dual to pure JT gravity
and $\d V$ is given in (\ref{eq:84}). 
Next, we define the effective potential
\begin{equation}
NV_{\text{eff}}(E) \equiv NV(E)-\int_{D_{\r}}d\lambda\r(\lambda)\log(\lambda-E)^{2},\label{eq:89}
\end{equation}
and the action \begin{equation} I=N\int_{D_{\r}}d\lambda\r(\lambda)V_{\text{eff}}(\lambda), \end{equation}
where $D_{\r}$ is the support of $\r(\lambda)$. As in \cite{Eynard:2015aea},
we call the second term in (\ref{eq:89}) the Coulomb gas replusive
potential between pairs of eigenvalues. In the large $N$ limit the
saddle equation for variation of each $\lambda\in D_{\r}$ gives $V_{\text{eff}}'(E)=0$
for $E\in D_{\r}$, namely
\begin{equation}
NV_{JT}'(E)+K\d V'(E)=2\int_{D_{\r}}d\lambda\f{\r(\lambda)}{E-\lambda},\qquad E\in D_{\r}\label{eq:7}
\end{equation}
where the integral is understood as its principal value. Here $\r(\lambda)$
is normalized as $\int_{D_{\r}}d\lambda\r(\lambda)=N$. This means
that the effective potential on the support of spectrum is a constant.
Moreover, this constant is the minimum value of the
effective potential along the path over which all eigenvalues are integrated. Physically, this means that all eigenvalues tend to stay in
the lowest energy configuration which is a balance between the external
force $-V^\prime$ and the internal Coulomb repulsion for each eigenvalue.

Let us define $\r(\lambda)=\r_{JT}(\lambda)+\d\r(\lambda)$ for $\lambda\in D_{\r}$
where $\r_{JT}(\lambda)$ is the spectral density in the original SSS
matrix model. Note that $\r(\lambda)$ and $\r_{JT}(\lambda)$ may
have different support and thus $\d\r(\lambda)$ could be a continuous but only piecewise
differentiable function. Define the resolvent as
\begin{equation}\label{eq:dfnR}
R(E)=\int_{D_{\r}}d\lambda\f{\r(\lambda)}{E-\lambda}
\end{equation}
One can show that $R(E)$ is a double cover map from $E\in\C\cup\{\infty\}$
to $\C$ with a branch cut along $D_{\r}$. It follows from the definition
and (\ref{eq:7}) that
\begin{equation}
\r(E)=-\f 1{2\pi i}(R(E+i\e)-R(E-i\e)),\qquad NV'(E)=R(E+i\e)+R(E-i\e),\qquad E\in D_{\r}
\end{equation}
As the double cover map has a sign ambiguity along the cut, for
a sensible matrix model solution, we need to assign the phase around
the cut such that the measure $\r(E)dE$ is nonnegative along $D_{\r}$.
For branch cut of $s$ pieces $\cup_{i=1}^{s}[a_{2i-1},a_{2i}]\subset\R$,
we can write 
\begin{equation}\label{eq:3.20}
\r(E)=\f 1{2\pi}M(E)\sqrt{-\s(E)},\qquad\s(E)=\prod_{i=1}^{2s}(E-a_{i}),\qquad E\in D_{\r}
\end{equation}
for some analytic function $M(E)$ ($M(x)$ is a polynomial of degree $d-s$ when potential $V(x)$ is a polynomial of degree $d+1$). Then it is easy to see that
\begin{equation}\label{eq:3.21}
R(x)=\f 12(NV'(x)-M(x)\sqrt{\s(x)})\implies NV_{\text{eff}}'(x)=i\pi(\r(x+i\e)-\r(x-i\e)),\qquad x\in\R
\end{equation}
Applying this to SSS matrix model, we can write (\ref{eq:7}) as
\begin{align}
K\d V'(E)+i\pi(\r_{JT}(E+i\e)-\r_{JT}(E-i\e))+2\int_{D_{*}}d\lambda\f{\r_{JT}(\lambda)}{E-\lambda} & =2\int_{D_{\r}}d\lambda\f{\d\r(\lambda)}{E-\lambda},\quad E\in D_{\r}\label{eq:13-1}
\end{align}
where $D_{*}\equiv D_{\r_{JT}}\backslash(D_{\r}\cap D_{\r_{JT}})$.
This is a nice formula in which the LHS is known except the range
of $D_{\r}$, and RHS has the same form as (\ref{eq:7}). We can define the LHS in terms of a new potential $N\tilde{V}'(E)$ and write (\ref{eq:13-1})
as
\begin{equation}
N\tilde{V}'(E)=2\int_{D_{\r}}d\lambda\f{\d\r(\lambda)}{E-\lambda}
\end{equation}
This is the saddle-point equation for a single-matrix model with potential $N \tilde{V}^\prime(E)$. The normalization condition for $\delta \rho$ is
\begin{equation}
\int_{D_{\r}}d\lambda\d\r(\lambda)=\int_{D_*} d \lambda \rho_{JT}(\lambda)\label{eq:normalization}
\end{equation}

Standard techniques of solving for the resolvent and spectral density can
be applied to the new potential $\tilde{V}$. For example, we can define the
variation of the resolvent
\begin{equation}
\d R(E)=2\int_{D_{\r}}d\lambda\f{\d\r(\lambda)}{E-\lambda}
\end{equation}
and it follows that
\begin{equation}
\d\r(E)=-\f 1{2\pi i}(\d R(E+i\e)-\d R(E-i\e)),\qquad N\tilde{V}'(E)=\d R(E+i\e)+\d R(E-i\e),\qquad E\in D_{\r}\label{eq:17-1}
\end{equation}
One can also use the Tricomi relation \cite{Eynard:2015aea} to write $\d R$ in terms
of contour integral of the potential
\begin{equation}
\d R(E)=\f 1{2\pi i}\oint_{E}d\lambda\f{\d R(\lambda)}{\lambda-E}\sqrt{\f{\s(E)}{\s(\lambda)}}=\f 1{2\pi i}\oint_{D_{\r}}d\lambda\f{\d R(\lambda)}{E-\lambda}\sqrt{\f{\s(E)}{\s(\lambda)}}=\f 1{4\pi i}\oint_{D_{\r}}d\lambda\f{N\tilde{V}'(\lambda)}{E-\lambda}\sqrt{\f{\s(E)}{\s(\lambda)}}\label{eq:18}
\end{equation}
where in the second step we used (\ref{eq:17-1}). In the double scaling
limit, the rightmost end $a_{2s}\ra+\infty$, and the
formula reduces to
\begin{equation}
\d R(E)=\f 1{4\pi i}\oint_{D_{\r}}d\lambda\f{N\tilde{V}'(\lambda)}{E-\lambda}\sqrt{\f{\s_{*}(E)}{\s_{*}(\lambda)}},\qquad\s_{*}(E)=(a_{2s-1}-E)\prod_{i=1}^{2s-2}(E-a_{i})\label{eq:101-dR}
\end{equation}

\subsection{One-cut solution of spectral density} \label{sec:3.3-one-cut}

To simplify the computation of the tree-level spectral density of JT gravity coupled to EOW branes, let us assume that the lower edge of the spectrum $E_0$ is negative. This simplifies (\ref{eq:13-1})
as the third term in LHS vanishes due to $D_{*}=\emptyset$. 

By (\ref{eq:101-dR}), $\d R$ is linear in $N\tilde{V}'$ and thus
is the sum of two terms that correspond to the first two terms of $N\tilde{V}'$
in (\ref{eq:13-1}). We call the first term the potential related
piece (with subscript ``$K$'') and the second term the universal piece
(with subscript ``$U$'') respectively. For the potential related piece,
we will consider a general measure for the EOW branes (\ref{eq:50-1})
giving an extra force
\begin{equation}
-\d V'(E)=-\sqrt{\f{2\phi_{b}}E}\int_{0}^{\infty}db\sin(\sqrt{2\phi_{b}E}b)\mM(b)=-2\phi_{b}\int_{\mD}d\a\f{m(\a)}{\a^{2}+2\phi_{b}E}\label{eq:102}
\end{equation}
For the unregulated measure \eqref{eq:malpha}, we recover \eqref{eq:sumofpoles}. Because it is a convergent sum, using any regulated measure leads to the same result in $\e\ra 0$ limit. In the rest of this paper unless specified, we will use unregulated measure \eqref{eq:malpha} in all integrals over $\alpha$ whenever it is convergent (and thus it makes no difference which regulated one is used, in the limit the regulator is removed).

The minus sign in \eqref{eq:102} shows that the force is leftward for $2\phi_b E>-(1/2+\mu)^2$ and one might have conclude $E_0<0$. However, although $\d V'(E)$ is independent of regularization and $\lambda$ in the $\e\ra 0$ limit, $E_0$ still depends on the regularization in a subtle way that we will explain in Section \ref{sec:determinE0}. 

Let us still assume here that 
$E_{0}<0$ for computational simplicity. In the end, the result \eqref{eq:63-1-1} and \eqref{eq:32} has an analytic form that can be easily continued to $E_{0}>0$ case. On the other hand, solving \eqref{eq:13-1} with $E_0>0$ requires dealing with the third term in \eqref{eq:13-1} as $D_* \neq \emptyset$, but this leads to the same result.

\subsubsection{Universal piece}

As $\r_{JT}(E)$ in (\ref{eq:rhoJT}) has a branch cut along the negative
real axis, the universal piece of $N\tilde{V}'(\lambda)$ is proportional
to $\sin(2\pi\sqrt{2\phi_{b}(-\lambda)})$ and only supported for
$\lambda<0$. This implies that the contour integral of (\ref{eq:101-dR})
can be reduced to surrounding $[E_{0},0]$
\begin{equation}
\d R_{U}(E)=\f{e^{S_{0}}\phi_{b}}{4\pi^{2}i}\oint_{[E_{0},0]}\f{d\lambda\sin(2\pi\sqrt{2\phi_{b}(-\lambda)})}{\lambda-E}\f{\sqrt{E_{0}-E}}{\sqrt{E_{0}-\lambda}}
\end{equation}
Expand the sine function in Taylor series, in which each term can be evaluated by moving the contour to infinity. It turns out that the integral
becomes a sum over residue at $E$ and $\infty$,
\begin{equation}
\d R_{U}(E)=-\f{e^{S_{0}}\phi_{b}}{2\pi}\left[\sin(2\pi\sqrt{2\phi_{b}(-E)})-\f 1{2\pi i}\sum_{n=0}^{\infty}\f{(2\pi\sqrt{2\phi_{b}})^{2n+1}}{(2n+1)!}\oint_{\infty}d\lambda\f{\lambda^{n+1/2}}{\lambda-E}\f{\sqrt{E_{0}-E}}{\sqrt{\lambda-E_{0}}}\right]
\end{equation}
The first term has a branch cut on the positive real axis.
Thus, along the negative real axis, $\d\r_{U}(E)=0$.
For $E>0$, the contribution to $\d\r_{U}(E)=-\r_{JT}(E)$ which cancels
out $\r_{JT}(E)$. Therefore the spectral density comes from
the second integral around infinity. We can do a coordinate transformation
$\lambda\ra1/z$, and the integral becomes
\begin{align}
R_{U}(E)\simeq & \f{e^{S_{0}}\phi_{b}}{4\pi^{2}i}\sum_{n=0}^{\infty}\f{(2\pi\sqrt{2\phi_{b}})^{2n+1}}{(2n+1)!}\oint_{0}\f{dzz^{-n-1}}{1-Ez}\f{\sqrt{E_{0}-E}}{\sqrt{1-E_{0}z}}\nonumber \\
= & \f{e^{S_{0}}\phi_{b}\sqrt{E_{0}-E}}{2\pi}\sum_{n=0}^{\infty}\f{(2\pi\sqrt{2\phi_{b}})^{2n+1}E^{n}}{(2n+1)!}\sum_{n_{0}=0}^{n}\f{(1/2)_{n_{0}}}{n_{0}!}\left(\f{E_{0}}E\right)^{n_{0}}
\end{align}
where in the second step we used Taylor expansion and picked out the
coefficient of $z^{n-1}$. Here ``$\simeq$'' means ignoring the analytic
part of $R_{U}(E)$ that does not contribute to spectral density.
Let us denote the sum over $n_{0}$ as $S_{n}(E_{0}/E)$. Expanding
$(E_{0}/E)^{n_{0}}=((E_{0}/E-1)+1)^{n_{0}}$ in powers of $(E_{0}/E-1)$,
we have
\begin{align}
S_{n}(E_{0}/E) & =\sum_{n_{0}=0}^{n}\sum_{k_{0}=0}^{n_{0}}\f{(1/2)_{n_{0}}}{(n_{0}-k_{0})!k_{0}!}(E_{0}/E-1)^{k_{0}}\nonumber \\
 & =\sum_{k_{0}=0}^{n}\f{(1/2)_{k_{0}}}{k_{0}!}(E_{0}/E-1)^{k_{0}}\sum_{s_{0}=0}^{n-k_{0}}\f{(1/2+k_{0})_{s_{0}}}{s_{0}!}\nonumber \\
 & =\int_{0}^{1}dt_{0}\sum_{k_{0}=0}^{n}\f{\G(n+3/2)}{\sqrt{\pi}(n-k_{0})!k_{0}!}(E_{0}/E-1)^{k_{0}}t_{0}^{k_{0}-1/2}\nonumber \\
 & =\f{\G(n+3/2)}{\sqrt{\pi}n!}\int_{0}^{1}dt_{0}\f{(1+t_{0}(E_{0}/E-1))^{n}}{t_{0}^{1/2}}
\end{align}
The sum over $n$ is straightforward
\begin{equation}
\sum_{n=0}^{\infty}\f{(2\pi\sqrt{2\phi_{b}})^{2n+1}E^{n}}{(2n+1)!}S_{n}(E_{0}/E)=\pi\sqrt{2\phi_{b}}\int_{0}^{1}dt_{0}\f{I_{0}\left(2\pi\sqrt{2\phi_{b}[t_{0}E_{0}+(1-t_{0})E]}\right)}{t_{0}^{1/2}}
\end{equation}
As $I_{0}(x)$ is an entire function, the integrand is analytic for
all $E$. Therefore, the discontinuity of $\d R_{U}(E)$ at $E_{\pm}$
is purely determined by the factor $\sqrt{E_{0}-E}$. It follows that
\begin{equation}
\r_{U}(E)=\f{e^{S_{0}}\phi_{b}\sqrt{2\phi_{b}(E-E_{0})}}{2\pi}\int_{0}^{1}dt_{0}\f{I_{0}\left(2\pi\sqrt{2\phi_{b}[E+t_{0}(E_{0}-E)]}\right)}{t_{0}^{1/2}}\label{eq:63-1-1}
\end{equation}
Defining $\xi=2\phi_{b}[E+t_{0}(E_{0}-E)]$, we can easily see that
(\ref{eq:63-1-1}) matches with the first term of gravity calculation
(\ref{eq:23}).

\subsubsection{Potential related piece}

Using (\ref{eq:102}) and (\ref{eq:101-dR}), we have
\begin{align}
\d R_{K}(E) & =-\f{K\phi_{b}}{2\pi i}\int_{\mD}d\a m(\a)\oint_{[E_{0},+\infty)}\f{d\lambda}{(\lambda-E)(\a^{2}+2\phi_{b}\lambda)}\f{\sqrt{E_{0}-E}}{\sqrt{E_{0}-\lambda}}\nonumber \\
 & =K\phi_{b}\int_{\mD}d\a m(\a)\left[\f 1{\a^{2}+2\phi_{b}E}-\f{\sqrt{E_{0}-E}}{(\a^{2}+2\phi_{b}E)\sqrt{E_{0}+\a^{2}/(2\phi_{b})}}\right]\label{eq:32}
\end{align}
where in second line we deformed the contour to poles at $\lambda=E$
and $\lambda=-\a^{2}/(2\phi_{b})$ and computed it using the residue theorem. This is justified by our assumption below \eqref{eq:sumofpoles} that $2 \phi_bE_0 > - (1/2+\mu)^2$. 
It is clear that only the second term of \eqref{eq:32} contributes to $\r_{K}$
\begin{equation}
\r_{K}(E)=-\f{K\phi_{b}}{\pi}\int_{\mD}d\a m(\a)\f{\sqrt{2\phi_{b}(E-E_{0})}}{(\a^{2}+2\phi_{b}E)\sqrt{\a^{2}+2\phi_{b}E_{0}}}\label{eq:36}
\end{equation}
which exactly matches with (\ref{eq:36-1}).

\subsubsection{Determining $E_{0}$} \label{sec:determinE0}

In the gravity computation of section \ref{sec:gravitycomputation}, the ambiguity in the finite part of the cusp-like counterterm leads to a free parameter $\lambda$. In the following, we will show that this ambiguity naturally corresponds to an ambiguity in the zero point energy $E_0$ of the tree-level spectrum of the matrix model, which is sensitive to additional UV deformations to potential \eqref{eq:dv}. Such UV deformations are not unique but play an equivalent role to the cusp-like counterterm in gravitational computation.


From the string equation \eqref{eq:regstringeq}, we see that $\lambda$ can be adjusted to set $E_0$ to a desired value. At first glance, this seems to be at odds with our understanding of matrix models. Ordinarily, the endpoints of a single-cut spectrum in a matrix model are not free parameters. They are fixed by the condition that the spectral density is normalized to the appropriate value. From \eqref{eq:normalization} and the simplifying assumption that $E_0 \leq 0$, the normalization condition is
\begin{equation}
    \int_{D_\rho} d\lambda \, \delta \rho(\lambda) = 0.
    \label{eq:normcondition}
\end{equation}
As detailed in the previous subsection, $\delta \rho$ is a sum of two terms, $\delta \rho_U = \rho_U - \rho_{JT}$ and $\delta \rho_K$. From \eqref{eq:63-1-1}, \eqref{eq:36}, and \eqref{eq:malpha}, we will have a different large $E$ behavior of $\delta \rho_U$ and $\delta \rho_K$, respectively, as $O(E^{-1/2})$ and $O(E^{-1/2})+O(E^{-1/2} \log E) $.\footnote{While \eqref{eq:36} cannot be evaluated analytically for $E_0 \neq 0$, one can check that it has the advertised large $E$ falloff for $E_0 = 0$ and that the large $E$ behavior of its $E_0$ derivative is subleading.}  
Thus, \eqref{eq:normcondition} cannot be satisfied for any choice of $E_0$ because $\delta \rho_U$ and $\delta \rho_K$ have different asymptotic behavior. The upshot is that the normalization condition can no longer determine the tree-level spectral density of the double-scaled matrix model, where the spectrum is assumed to have noncompact support on $[E_0,\infty)$. 
On the other hand, in a non-double-scaled matrix model, where the spectral density has compact support, the potential and normalization condition uniquely determine the spectrum. Thus, to determine $E_0$, we must also specify how the change in the potential $\delta V$ approaches \eqref{eq:dv} in the double scaling limit.

\begin{figure}
\begin{centering}
\includegraphics[height=4cm]{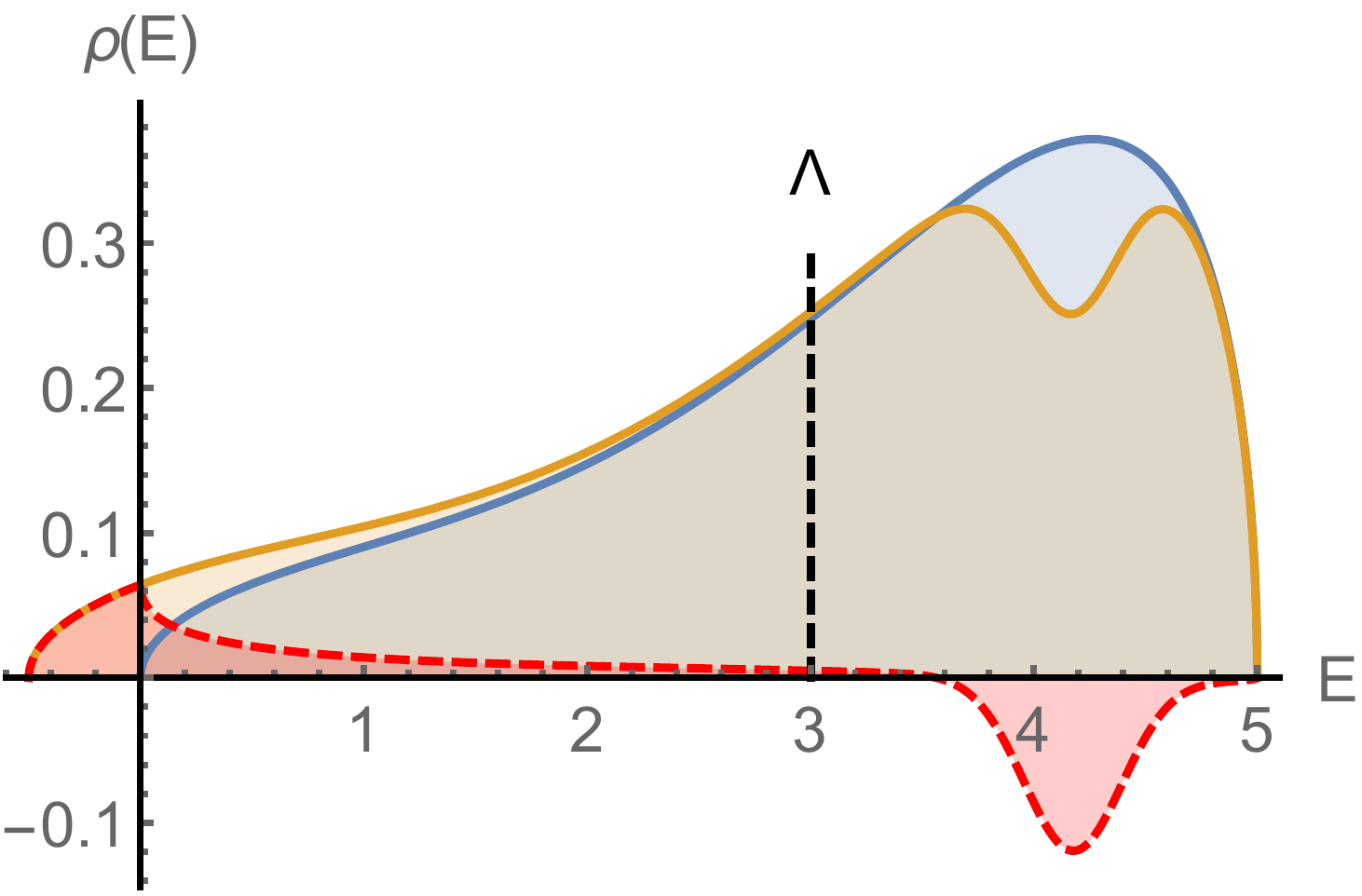}
\par\end{centering}
\caption{The blue curve is spectral density $\rho_\Lambda(E)$ that matches with $\r_{JT}(E)$ for $E<\Lambda$; the yellow curve is full spectral density $\tilde \r_\Lambda(E)$ with $E_0<0$; the red curve is the difference $\d \r_\Lambda(E)$. Due to normalization \eqref{eq:normcond2}, $\d\r_\Lambda(E)$ has a large negative ``bump" above $\Lambda$. \label{fig:uvbump}}
\end{figure}

We claim that for any $E_0$, it is possible to add a term to the potential of the matrix model dual to JT gravity that converges pointwise to \eqref{eq:dv} in the double scaling limit such that the resulting spectral density is supported on $[E_0,\infty)$ and given by $\rho_{JT} + \delta \rho,$ where $\delta \rho = \delta \rho_U + \delta \rho_K$. There is one exception: when the spectral density corresponding to a given value of $E_0$ has negativity, that value of $E_0$ is ruled out. 

To see this, consider a family of matrix models labelled by $\Lambda$ whose tree-level spectra $\rho_\Lambda$ have compact support and converge pointwise to $\rho_{JT}$ in the double scaling limit $\Lambda \rightarrow \infty$. For concreteness, we may assume that $\rho_\Lambda(E)$ agrees with $\rho_{JT}(E)$ for $E < \Lambda$. Next, define $\delta \rho_\Lambda(E)$ to agree with $\delta \rho$ for $E_0 < E < \Lambda$.\footnote{We will still assume that $E_0 < 0$ to be consistent with our earlier computations of $\delta \rho_U$ and $\delta \rho_K$, although this condition is not essential for our present argument.} For $E > \Lambda$, let $\delta \rho_\Lambda$ be a smooth function that interpolates between $\delta \rho(\Lambda)$ and $0$ in a way such that the normalization condition
\begin{equation}
\label{eq:normcond}
\int_{D_{\tilde\r_\Lambda}} dE \, \delta \rho_\Lambda(E) = 0    
\end{equation}
is obeyed, where $\tilde\r_\Lambda\equiv\r_\Lambda+\d\r_\Lambda$ is the full spectral density. See Figure \ref{fig:uvbump} for an illustration of $\delta \rho_\Lambda$. Similar to \eqref{eq:13-1}, the saddle point equation for $\delta \rho_\Lambda$ is
\begin{equation}
    2\int_{D_{\tilde\r_\Lambda}} d\lambda \, \frac{\delta \rho_\Lambda(\lambda)}{E - \lambda} - i \pi (\rho_{\Lambda}(E + i \epsilon) - \rho_{\Lambda}(E - i \epsilon))+ \delta F_\Lambda(E) = 0, \quad E \in D_{\tilde\r_\Lambda}
    \label{eq:3.45}
\end{equation}
where $\delta F_\Lambda$ is the extra force that must be applied to the eigenvalues such that the resulting spectrum is $\tilde\r_\Lambda$. Given $\delta \rho_\Lambda$, one simply uses the above equation to calculate $\delta F_\Lambda$. Let $a_{\Lambda+}$ be the location of the upper end of $D_{\tilde\r_\Lambda}$. Then \eqref{eq:3.45} becomes
\begin{equation}
    2\int_{E_0}^\Lambda d\lambda \, \frac{\delta \rho(\lambda)}{E - \lambda} +2\int_{\Lambda}^{a_{\Lambda +}} d\lambda \, \frac{\delta \rho_\Lambda(\lambda)}{E - \lambda} - i \pi (\rho_{\Lambda}(E + i \epsilon) - \rho_{\Lambda}(E - i \epsilon))+ \delta F_\Lambda(E) = 0, \quad E \in D_{\tilde\r_\Lambda},
    \label{eq:3.46}
\end{equation}
while \eqref{eq:normcond} becomes
\begin{equation}
\label{eq:normcond2}
    \int_{E_0}^\Lambda dE \, \delta \rho(E) + \int_\Lambda^{a_{\Lambda +}} dE \, \delta \rho_\Lambda(E) = 0.
\end{equation}
From the large $E$ behavior of $\delta \rho(E)$, we see that the first term above diverges as $\Lambda^{1/2} \log \Lambda$. This controls how large the ``bump'' in Figure \ref{fig:uvbump} can be. This justifies that normalization does not hold within the double-scaled regime. At fixed $E$, the second term of \eqref{eq:3.46} goes to zero in the large $\Lambda$ limit. We are left with
\begin{equation}
    2\int_{E_0}^\infty d\lambda \, \frac{\delta \rho(\lambda)}{E - \lambda}  - i \pi (\rho_{JT}(E + i \epsilon) - \rho_{JT}(E - i \epsilon)) = -\lim_{\Lambda \rightarrow \infty} \delta F_\Lambda(E), \quad E > E_0.
    \label{eq:3.48}
\end{equation}
We have thus demonstrated that it is possible to pick a $\delta F_\Lambda$ that converges pointwise to the derivative of \eqref{eq:dv} such that $E_0$ takes on any desired value (modulo the possibility that the resulting spectrum might have negativity). Thus, $E_0$ should be viewed as a free parameter, just as $\lambda$ is a free parameter in \eqref{eq:regstringeq}. The above argument mirrors in the dual matrix model our treatment of JT gravity with EOW branes as an effective field theory. The fact that $\delta F_\Lambda(E)$ can be tuned for $E > \Lambda$ to achieve different values of $E_0$ reflects the concept that the UV physics of the model can be tuned to achieve a desired IR theory. Of course, the exact shape of the bump in Figure \ref{fig:uvbump} is unimportant; different choices represent irrelevant UV modifications that belong to the same IR universality class.

As explained above, we have a family of double-scaled matrix models, each of which is labelled by a free IR parameter $E_0$. The spectral density of each matrix model agrees with the result of a gravitational calculation, \eqref{eq:23}. 
To match a particular matrix model to a choice of the cusp counterterm $\lambda$ in the gravity theory, we may use \eqref{eq:23-1} to match the spectral density on both sides. This IR perspective neglects the UV details that determine these parameters.\footnote{In the matrix model, $E_0$ is determined by the normalization of the spectrum before the double-scaling limit is taken. This is sensitive to UV deformations of the potential. It would be very interesting to understand if such non-double-scaled matrix model has any geometric dual. In particular, such dual, if exists, cannot be asymptotically AdS because the non-double-scaled matrix model has an upper bound of energy.} In section \ref{sec:yshape}, we will consider the regime $\lambda > 0$ and $K > K^>_{cr}$, where the gravity computation breaks down but the matrix model computation does not (once we specify a nonperturbative completion of the model, which is not unique). Thus, it is also worthwhile to consider a UV perspective, where we explicitly consider how the IR parameter $E_0$ is determined by UV data. The previous paragraph implies that this UV data is equivalent to the details of how the double-scaling limit is taken, and $E_0$ is determined by the normalization condition \eqref{eq:normcond}. To simplify computations, we have identified two alternative methods to determine $E_0$ from UV data that are easier to use in practice. First, instead of deforming the potential by \eqref{eq:dv}, one can use \eqref{eq:102} to define a $\delta V$ associated with a regulated measure (such as \eqref{eq:regmeasure}), and then impose the condition that $\delta \rho$ is a normalizable function. The UV data is contained in the details of the regulated measure. The second method is to take $\delta V$ to be \eqref{eq:dv}, but impose a condition on the asymptotic behavior of $\delta \rho$ that fixes $E_0$ (we will specify this below). The benefit of these alternative conditions is that they allow us to bypass the use of a non-double-scaled matrix model to define what we mean by the UV data of the matrix model. Also, they explicitly relate the UV data of the matrix model to the choice of cusp counterterm $\lambda$ in the gravity theory. However, these alternative methods can only be fully justified by checking that they reproduce \eqref{eq:23-1}, which determines $E_0$ in the gravity computation. We explain these two alternative methods below in more detail.




Let us first expand $\d \r(E)$ in large $E$. Note that $\rho_{JT}(E)$ corresponds
to $E_{0}=0$ and $K=0$ 
\begin{equation}
\r_{JT}(E)=\f{e^{S_{0}}\phi_{b}}{2\pi}\int_{0}^{2\phi_{b}E}\f{d\xi}{\sqrt{2\phi_{b}E-\xi}}I_{0}(2\pi\sqrt{\xi})
\end{equation}
We have $\d\r(E)$ in large $E$ limit
\begin{equation}\label{eq:3.40}
\d\r(E)=E^{-1/2}\cdot\left[\f{e^{S_{0}}\phi_{b}^{1/2}}{2\sqrt{2}\pi}\int_{2\phi_{b}E_{0}}^{0}d\xi I_{0}(2\pi\sqrt{\xi})-\f{K}{\pi}\int_{\mD}d\a \f{\sqrt{2}\phi_{b}^{3/2}Em(\a)}{(\a^{2}+2\phi_{b}E)\sqrt{\a^{2}+2\phi_{b}E_{0}}}\right]+o(E^{-1})
\end{equation}
If we use a regulated measure, we may take the large $E$ limit inside the integral of $\a$ and have
\be
\d\r(E)=E^{-1/2}\cdot\left[\f{e^{S_{0}}\phi_{b}^{1/2}}{2\sqrt{2}\pi}\int_{2\phi_{b}E_{0}}^{0}d\xi I_{0}(2\pi\sqrt{\xi})-\f{K\phi_b^{1/2}}{\sqrt{2}\pi}\int_{\mD}d\a \f{m(\a)}{\sqrt{\a^{2}+2\phi_{b}E_{0}}}\right]+o(E^{-1})
\label{eq:347}
\ee
where the $\a$ integral is just $f_\lambda(2\phi_b E_0)$ by \eqref{eq:60}, which is finite and involves $\lambda$ from the regulated measure. The $E^{-1/2}$ order integrates to $E^{1/2}$ divergence for large $E$ that violates \eqref{eq:normcondition} within the double scaling limit. Imposing that $\d\r(E)$ is normalizable in the double scaling limit requires that the order $E^{-1/2}$ term in \eqref{eq:347} vanishes. Using $I_{0}(a\sqrt{x})=\f 2a\del_{x}\sqrt{x}I_{1}(a\sqrt{x})$, the same zero point equation \eqref{eq:23-1} follows after taking $\e\ra 0$ in last step.

Note that the final result for $\d\r(E)$ after the double-scaling limit is taken is normalizable but does not obey $\int dE \d\r(E)=0$. In general, the normalization could be nonzero (but finite in the double-scaling limit). Given zero point equation \eqref{eq:23-1} before removing the regulator, one can show that
\be
\int_{2\phi_bE_{0}}^{\infty}\d\r(E)dE=\f{2\phi_b K}{\pi}\left(\lim_{\xi\ra\infty}\sqrt{\xi}f_\lambda(\xi)-\int_{2\phi_bE_{0}}^{0}d\xi\sqrt{-\xi}f'_\lambda(\xi)\right)
\ee
It is clear that normalization depends on $f_\lambda(\infty)$. From \eqref{eq:boundm} and \eqref{eq:60}, we know that $f_\lambda(\xi)\sim\xi^{-1/2}$ in large $\xi$. Normalization holds exactly only for specially designed $f_\lambda(x)$ such that
\be
\int_{2\phi_b E_{0}}^{0}d\xi\sqrt{-\xi}f'_\lambda(\xi)=\lim_{\xi\ra\infty}\sqrt{\xi}f_\lambda(\xi)
\ee

On the other hand, if we use unregulated measure \eqref{eq:malpha} in \eqref{eq:3.40}, taking the large $E$ limit in the integral of $\a$ is illegal because it is divergent. As we mentioned before, this term has an $O(E^{-1/2})+O(E^{-1/2}\log E)$ divergence. To match with the gravity result, we can specify the UV data of dual matrix model as that for $E\ra \infty$
\begin{align}
\d \r(E)\ra &E^{-1/2}\cdot\f{K\phi_b^{1/2}}{\sqrt{2}\pi}\left[\lambda-\int_{\mD}d\a \f{2\phi_bEm(\a)}{(\a^{2}+2\phi_{b}E)\a}\right]+o(E^{-1})\nn\\
= &E^{-1/2}\cdot\f{K\phi_b^{1/2}}{\sqrt{2}\pi} \left(\lambda+\lambda_0-\log e^{\g_E}\sqrt{2\phi_b E}\right)+o(E^{-1}) \label{eq:uvdefn}
\end{align}
where $\g_E$ is Euler constant. This requirement leads to
\begin{equation}\label{eq:3.42zero}
\f 1{2\pi}\int_{2\phi_{b}E_{0}}^{0}d\xi\del_{\xi}(\sqrt{\xi}I_{1}(2\pi\sqrt{\xi}))-Ke^{-S_{0}}\left[\lambda+\int_\mD d\a \left( \f 1 {\sqrt{\a^2+2\phi_b E_0}}-\f 1 {\a} \right)\right]=0
\end{equation}
which fixes $E_0$ exactly as in gravitational answer in (\ref{eq:23-1}).

\section{A ``Y'' shaped phase} \label{sec:yshape}

As shown in Fig. \ref{fig:5-c}, the one-cut
solution with $\lambda>0$ is unphysical for $K>K_{cr}^>$ as it has negative spectral density. On the gravity side, there is no obvious
way to obtain a sensible result beyond the critical point. Fortunately, the dual matrix model can be non-perturbatively well-defined, and the method to compute spectral density
for a given potential is known. It turns out to has a ``Y'' shaped
cut in the complex plane as we will see shortly.

\subsection{Matrix model on complex contours}\label{sec:4.1}

Before we discuss the phase transition, let us review some basic
features of the non-perturbative definition of matrix models. The partition function of a matrix model of $N\times N$ hermitian matrices is defined as
\begin{equation}
\mZ=\int dHe^{-N\Tr V(H)}=\int_{\R^{N}}\prod_{i}d\lambda_{i}\D(\lambda)^{2}e^{-N\sum_{i=1}^{N}V(\lambda_{i})}
\end{equation}
where $\D(\lambda)=\prod_{i<j}(\lambda_{i}-\lambda_{j})$ is Vandermonde
determinant and domain of integration of all eigenvalues $\lambda_{i}$ is $\R^{N}$. 

We can generalize this definition to $N$ contours in the complex plane
$\G=\otimes_{i=1}^{N}\G_{i}\in\C^{N}$ and define the ensemble of
normal matrices with eigenvalues on $\G$ \cite{Eynard:2015aea}
\begin{equation}
\mathbb{E}(\G)=\{V\Lambda V^{-1}|V\in U(N),\Lambda=\diag(\lambda_{1},\cdots,\lambda_{N}),\lambda_{i}\in\G_{i}\}
\end{equation}
The partition function of this matrix model is
\begin{equation}
\mZ_{\G}=\int_{\G}\prod_{i}d\lambda_{i}\D(\lambda)^{2}e^{-N\sum_{i=1}^{N}V(\lambda_{i})}\label{eq:117}
\end{equation}
It is clear that the hermitian matrix ensemble is just the special case
$\G_{i}=\R$ for all $i$. Absolute convergence for all eigenvalues is required to have a well-defined integral. With this condition, we
restrict  each integration contour $\G_{i}$ to
end at poles of $V$ with an appropriate angle such that $\Re V(x)\ra+\infty$
for $x\ra\del\G_{i}$, where $\del\G_{i}$ means the two ends of $\G_{i}$. Here we assume $V(x)$ tends to its singularities faster than a logarithm so that the Vandermonde determinant (scaling as $e^{2N\log x}$ for each eigenvalue $x\ra\infty$) is always subleading and does not affect the integral's convergence.
Though it seems that there are infinitely many ways to choose contours,
contours that differ by smooth deformations do not change the integral due to analyticity. It follows that the space of independent contours is isomorphic to the homology space $H_{1}(e^{-V(x)}dx)$. We can pick a basis $\g_i$ for this homology space, and for each eigenvalue $\lambda_i$, its integration contour can be chosen as an integer coefficient linear combination of this basis
\be 
\G_i=\sum_{ij} c_{ij}\g_j,~~c_{ij}\in\Z,~~ 1\leq j \leq \dim H_1(e^{-V(x)dx})
\ee
where $c_{ij}$ is the number of times $\lambda_i$ is integrated along $\g_i$ and the sign defines the direction of integration.

\begin{figure}
\begin{centering}
\includegraphics[height=4cm]{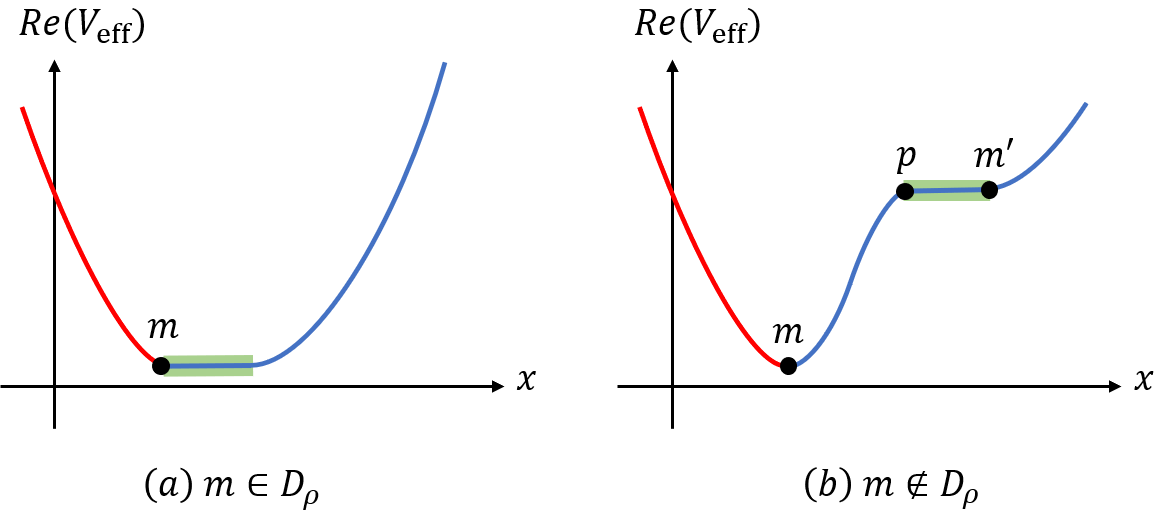}
\par\end{centering}
\caption{The real part of effective potential along a descent-ascent road parametrized by $x$. Red and blue curves are two different ascending roads $\g(m)$ for a saddle point $m$ of $V_\eff$. Green shaded region is (part of) the support of spectrum $D_\r$. In (a) $m$ is one edge of $D_\r$ and in (b) $m$ is not on $D_\r$. In (b) the blue ascending road hits at a point $p\in D_\r$, then follows along $D_\r$ to reach an edge $m'$ of $D_\r$ before extending to $\infty$.
}\label{fig:veff-illus}
\end{figure}

For example, if $V(x)=x^{4}$, its pole is at infinity and there are
four directions approaching infinity with $\Re V(x)\ra+\infty$ (i.e.
$\arg x\in(n\pi/2-\pi/8,n\pi/2+\pi/8)$ and $|x|\ra\infty$ for $n=0,1,2,3$),
which leads to a basis of three independent integration contours, say $\g_{n,n+1}$ for $n=0,1,2$, denoting contours connecting $e^{in\pi/2}\infty$ to $e^{i(n+1)\pi/2}\infty$. There are only three basis contours because $\g_{0,1}+\g_{1,2}+\g_{2,3}+\g_{3,0}=0$ by analyticity.
In general, for a potential with a finite total degree of its poles, the dimension
of $H_{1}(e^{-V(x)}dx)$ is finite, while for potentials with infinite 
degree of poles (infinitely many poles or an essential singularity), the
dimension of $H_{1}(e^{-V(x)}dx)$ is countably infinite. For the $N$
dimensional integration in $\G$, to properly count the number of basis contours, we also need to quotient by  permutations  of the eigenvalues. This leads to a basis of dimension $\dim(H_{1}(e^{-V(x)}dx)^{\otimes N}/V_{sym})$, where $V_{sym}$
is the volume of the permutation symmetry group of the eigenvalues.

\begin{figure}
\begin{centering}

\begin{subfigure}{.4\textwidth}
\includegraphics[height=4cm]{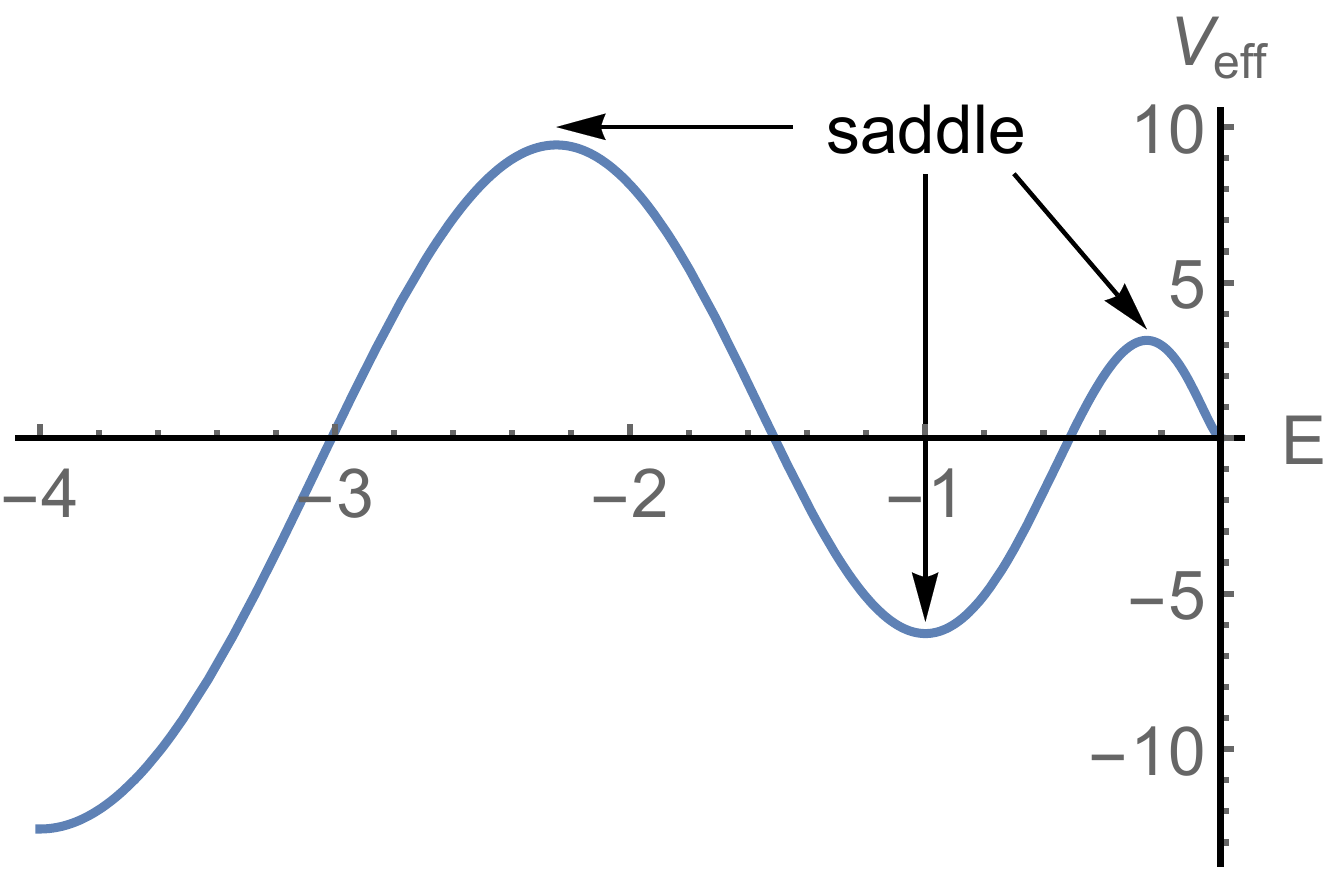}
\caption{}
\label{fig:5a}
\end{subfigure}
\begin{subfigure}{.59\textwidth}
\includegraphics[height=4cm]{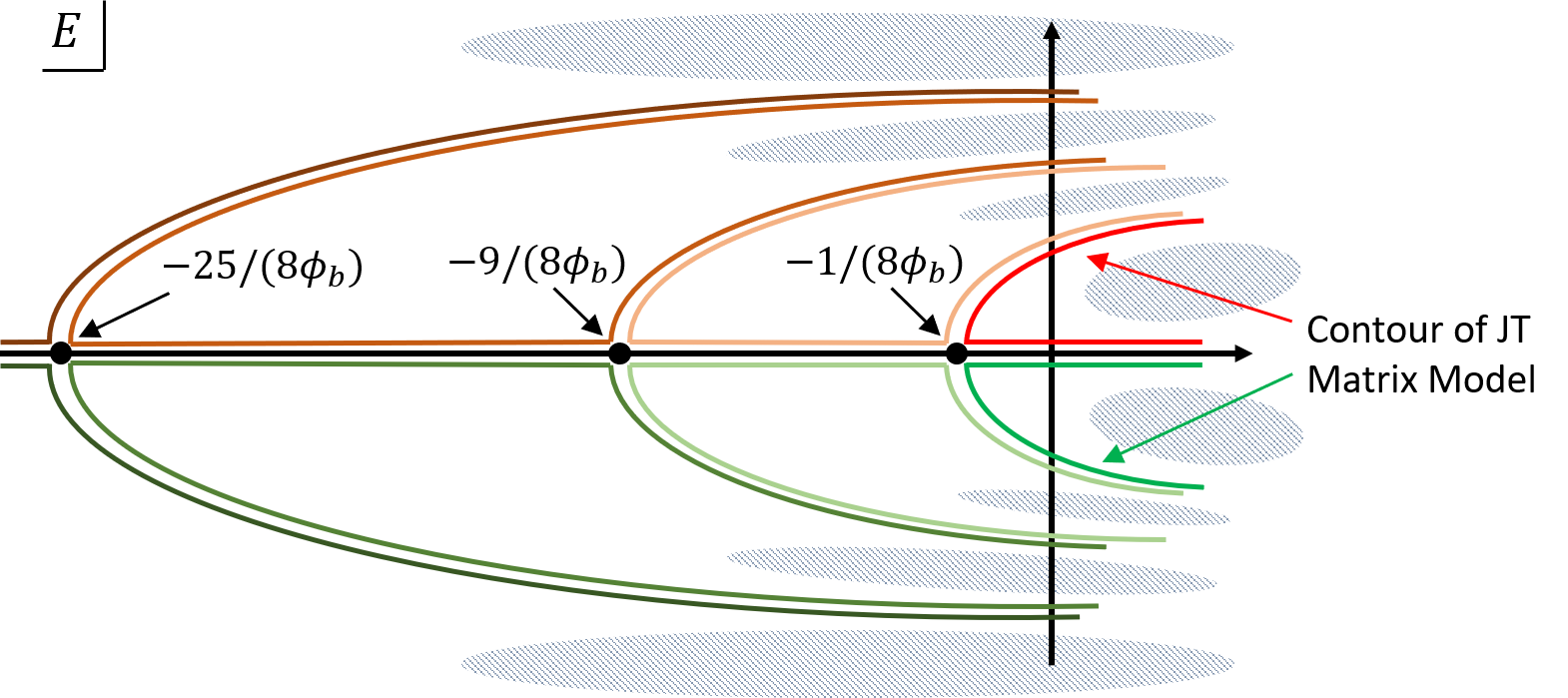}
\caption{}
\label{fig:5b}
\end{subfigure}

\par\end{centering}
\caption{(a) The effective potential of SSS potential for $E<0$ where the
first saddle is at $E_{saddle}=-1/(8\phi_{b})$. (b) First a few basis $\g_i$ (that are infinite many) on complex $E$ plane for the matrix
model dual to JT gravity. Here we use different colors to distinguish
independent basis. Along each $\g_i$ to infinity $\Re V_{\text{eff}}\ra +\infty$ (except being constant along positive real axis $D_{\r_{JT}}$) and along the shaded regions between contours $\Re V_{\text{eff}}\ra -\infty$. To have the spectrum $\protect\r_{JT}(E)$ starting
at origin, we must choose the defining contour $\G_i$ for every eigenvalue $\lambda_i$ on one of the rightmost two contours that are labeled by bright red and green respectively.
\label{fig:The-effective-potential-1}}
\end{figure}

Given a matrix model potential and assignment of contours $\G$ from this basis of $H_1(e^{-V(x)}dx)$, for each eigenvalue $\lambda_i\in\G_i$, one can use the saddle equation $V_\eff '(\lambda_i)=0$ (and thus setting $\Re V_\eff(\lambda_i)=0$) for $\lambda_i\in D_\r$ and $\Re V_\eff(\lambda_i)\geq 0$ for $\lambda_i\in \G_i\backslash D_\r$ to solve for the tree-level spectral density $\r$. Here we implicitly used the fact that saddle contour $\G_i$ must have nonzero overlap with $D_\r$. If the spectrum is multi-cut, we also need to use the information of how much fraction of eigenvalues lying on each basis $\g_i$ to determine the spectral density on each cut. 

Conversely, given a tree-level spectral density $\r$, we can derive an effective potential using analytic continuation of $\r$ (see \eqref{eq:3.21}). As $V_\eff(x)$ has the same non-logarithmic singularities as $V(x)$ by definition \eqref{eq:89}, we can use it to define a basis $\g_i$ as follows. For any saddle point $m$ such that $V_\eff '(m)=0$ (including the ends of spectrum), we define an ascending path $\g(m)$ emanating from $m$ such that $\Im V_\eff(x)=\Im V_\eff(m)$ and $\Re V_\eff(x)$ is strictly increasing for $x$ away from $m$. Such a path will end on a singularity of $V(x)$ or a point $p$ on $D_\r$. In the latter case, we further extend $\g(m)$ along $D_\r$ (on which $V_\eff$ is constant) to reach another end $m'$ of $D_\r$ and then continue to a singularity of $V(x)$ along an ascending path $\g(m')$. Define a descent-ascent path \cite{Eynard:2015aea} as the union of two different ascending paths starting at the same saddle point $m$. See Fig. \ref{fig:veff-illus} for illustration. Our $\g_i$ is chosen as a complete and independent set from all descent-ascent paths.  Each such contour $\g_i$ has a unique minimum value of $\Re V_\eff (x)$ and extends to singularities of $V(x)$ monotonically.

Let us apply above construction of the basis $\g_i$ to the matrix model dual to JT gravity. The effective potential away from $D_{\r_{JT}}=\R_{+}$
is \cite{Saad:2019lba}
\begin{equation}
V_{\text{eff}}^{JT}(E)=\f{e^{S_{0}}}{4\pi^{3}}\left(\sin(2\pi\sqrt{2\phi_{b}(-E)})-2\pi\sqrt{2\phi_{b}(-E)}\cos(2\pi\sqrt{2\phi_{b}(-E)})\right)\label{eq:118}
\end{equation}
which has infinitely many saddles at $E=-n^2/(8\phi_b)$ for $n\in\Z$  (see Fig. \ref{fig:5a}). Using the construction of descent-ascent paths, we can choose the basis $\g_i$ as plotted in Fig. \ref{fig:5b}, where four nearby basis contours are separated around the local maximum
of $V_{\text{eff}}^{JT}(E)$ at $E=-(2n-1)^{2}/(8\phi_{b})$ for $n\in\N$.
As there are infinite many $n$, the dimension of the basis is infinite.

Note that the interval around the saddles at $-k^2/(2\phi_b)$ for $k\in\Z$ has lower effective
potential than that on $D_{\r_{JT}}$ (see Fig. \ref{fig:5a}). 
This means that the integral contour $\G$ for the spectrum of JT gravity cannot be along
the real axis, otherwise there would be infinitely many eigenvalues filling
these energy wells on $\R_{-}$. This is a  known  non-perturbative
instability \cite{Saad:2019lba, Johnson:2019eik, Johnson:2020lns}. Indeed, in order to have $\r_{JT}(E)$ supported only on the positive
real axis, we must choose the defining contour $\G_i$ for every eigenvalue $\lambda_{i}$ in (\ref{eq:117}) to be one of the rightmost two contours in Fig. \ref{fig:5b} that are
labeled by bright red and green respectively. Each of these two contours
consists of two pieces (one along the real axis and the other extending
into upper/lower half complex $E$ plane) joined at the first saddle
of $V_{\text{eff}}^{JT}(E)$ at $E=E_{saddle}=-1/(8\phi_{b})$. Other
choices of contours will lead to nonzero support of spectrum
in some regions of $\R_{-}$. Given these two contours, there are still many ways to choose how many eigenvalues are assigned to each contour. Each choice defines a nonperturbative completion of the model. The most natural choice of $\G$ is to assign half of the eigenvalues to the
 upper contour and the other half to the lower contour, as this leads to
a real partition function. This is the prescription we will use to
study the phase transition.

Adding the extra potential from EOW branes changes the spectrum as seen in the gravitational computation. Increasing $K$ in \eqref{eq:88} from zero will also smoothly deform $\G$ as the basis defined as descent-ascent roads also smoothly moves. Let us analyze the effective potential near the critical point $K_{cr}^>$ when $\lambda>0$. Using the near edge spectral density \eqref{eq:114-1}, the effective potential for $E<E_{0}$ is
\begin{equation}
V_{\text{eff}}(E)\sim\f{4a_{1}}3(E_{0}-E)^{3/2}-\f{16a_{2}}{15}(E_{0}-E)^{5/2}\label{eq:115-1}
\end{equation}
Before reaching the critical point, $V_{\text{eff}}(E)$ is positive in a small
neighborhood of $E_{0}$ but it becomes negative at the critical point when
$a_{1}=0$. For $a_{1}>0$ but close to zero, we can solve for the rightmost
saddle of $V_{\text{eff}}(E)$ in the above approximation as
\begin{equation}
E_{saddle}\app E_{0}-\f{3a_{1}}{4a_{2}}
\end{equation}
At critical point, $a_{1}$ vanishes and $E_{saddle}$ coincides with
$E_{0}$. Note that increasing $K$ does not change the direction of left-pointing extra force from $\d V$ and the two integration contours in $\G$ each has
two pieces joined at $E_{saddle}$. Given the fact that the end points of the tree-level spectrum of matrix models must deform continuously (not necessarily smoothly at phase transition point),\footnote{Even for Stokes' phenomenon in matrix models that have two or more competitive saddles, the end points of the tree-level spectrum deforms continuously at the critical point in generic cases. See \cite{Marino:2009dp} as an example. } this means that past the critical
point, the spectrum cannot extend along the real axis into the region with negative effective potential but rather it goes into the
upper or lower complex $E$ plane along $\G$. 
Therefore, the support of spectrum after phase transition must be a ``Y'' shape
with two complex conjugate pieces joined with a piece along real axis
(see Fig. \ref{fig:The-support-of}).\footnote{As contour is defined only up to deformations that preserve the asymptotes, one may suspect another scenario where the spectrum splits into two disjoint conjugate curves in the upper and lower half plane respectively. However, this cannot be a saddle of the matrix model immediately after the critical point because the Coulomb repulsive force between eigenvalues on these two curves would be arbitrarily large if they have infinitesimal separation. Hence, the two curves must join somewhere and this is the ``Y" shaped spectrum.}

Such transition to ``Y'' shaped phases have been studied in other complex matrix models (for example in  \cite{bleher2016mother}). For readers not familiar with this topic, we examine a simple matrix model in Appendix \ref{app:cubic} with cubic potential $V(x)=x^{3}/3-tx^{2}/2$ and an even distribution
of eigenvalues on complex conjugate contours $\mC_{+}$ and $\mC_{-}$,
where $\mC_{\pm}$ is a contour connecting $\pm e^{2i\pi/3}\infty$
to $+\infty$. In Appendix \ref{app:cubic}, we show that there are two critical values $\pm t_{cr}=\pm2^{2/3}\cdot3^{1/2}\app\pm2.749$
such that for $t>t_{cr}$ the spectrum is one-cut; for $-t_{cr}<t<t_{cr}$
the spectrum is ``Y'' shaped; for $t<-t_{cr}$ the spectrum
becomes to one-cut again.

\subsection{``Y'' shaped solution of the spectral density} \label{sec:4.2}

\begin{figure}
\begin{centering}
\includegraphics[width=5cm]{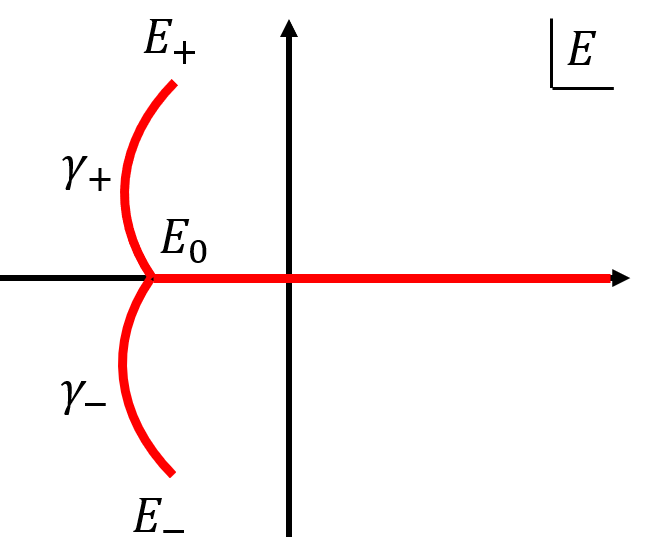}
\par\end{centering}
\caption{The support of ``Y'' shaped spectrum. The two complex conjugate ends
are $E_{\pm}$ and the joint point is $E_{0}$. $\protect\g_{\pm}$
are two arcs connecting $E_{0}$ to $E_{\pm}$ respectively. \label{fig:The-support-of}}
\end{figure}
Let us denote the two complex ends of the spectrum as $E_{\pm}$, the
intersection on real axis as $E_{0}$ and the two arcs connecting $E_{0}$
to $E_{\pm}$ as $\g_{\pm}$ (see Fig. \ref{fig:The-support-of}).
As we assume eigenvalues are evenly distributed on the upper and lower contour,
 $E_{\pm}$ and $\g_{\pm}$ are complex conjugates.\footnote{If we choose to put eigenvalues on these two contours with different fractions, we will still have a ``Y" shaped solution but with $E_+$ not conjugate to $E_-$. In this case, we have an additional parameter, which is fixed by the extra equation from the difference of normalization on two contours that depends on the filling fraction.} Though eigenvalues can be complex, the spectral density $\r(E)dE$
must be real and nonnegative along $\g_{\pm}$. This is a strong condition
that determines the shape of $\g_{\pm}$. Assuming $E_{0}<0$, it
is easy to see that same equation (\ref{eq:101-dR}) and the contour
integral techniques of the one-cut case in Section \ref{sec:3.3-one-cut} are both applicable to solve for $\d R(E)$.
As usual, $\d R(E)$ is the sum of a universal term and a potential-dependent term.

The computation is similar to the one-cut case, and we relegate it to Appendix
\ref{sec:Y-shape-spectrum}. The result is as follows. We need to
place the branch cut of $\d R(E)$ along $\g_{+}\cup\g_{-}\cup[E_{0},+\infty)$.
For $E\in[E_{0},+\infty)$, we have 
\begin{align}
\r_{U}(E) & =\f{e^{S_{0}}\phi_{b}^{2}\sqrt{(E-E_{+})(E-E_{-})(E-E_{0})}}{\pi}\int_{V_{3}}\f{I_{1}\left(2\pi\sqrt{2\phi_{b}[E+\sum_{i=0,\pm}t_{i}(E_{i}-E)]}\right)}{\sqrt{E+\sum_{i=0,\pm}t_{i}(E_{i}-E)}t_{0}^{1/2}t_{+}^{1/2}t_{-}^{1/2}}dt_{0}dt_{+}dt_{-}\label{eq:rhou}\\
\r_{K}(E) & =\f{K\phi_{b}}{\pi}\int_{\mD}d\a \, m(\a) \, \f{\sqrt{(E-E_{0})(E-E_{+})(E-E_{-})}}{(\a^{2}+2\phi_{b}E)\sqrt{(\a^{2}/2\phi_{b}+E_{0})(\a^{2}/2\phi_{b}+E_{+})(\a^{2}/2\phi_{b}+E_{-})}}\label{eq:rhok}
\end{align}
where
\begin{equation}
V_{3}\equiv\{(t_{0},t_{+},t_{-})|t_{i}\geq0,\sum_{i=0,\pm}t_{i}\leq1\}
\end{equation}
For $E\in\g_{\pm}$, $\r(E)$ has the same formula as \eqref{eq:rhou} and \eqref{eq:rhok} but with nontrivial branch-cut choice of $\sqrt{(E-E_{0})(E-E_{+})(E-E_{-})}$, which is determined by the locus of $\g_{\pm}$ by
requiring $\r(E)dE$ to be real and nonnegative along $\g_{\pm}$. 

We need three equations to determine three real parameters of the spectrum
$E_{0}$, $\Re E_{+}$ and $\Im E_{+}$. By a similar argument to Section \ref{sec:determinE0}, there are two ways to impose appropriate UV data to determine these three IR parameters. One can either use a regulated measure and impose normalizability of $\d\r(E)$, or use the unregualted measure \eqref{eq:malpha} and specify the same large $E$ asymptotic behavior in \eqref{eq:uvdefn}. They are equivalent in the limit that the regulator is removed just as in the one-cut case. Let us take the large $E$ behavior defined by \eqref{eq:uvdefn} as an example. Indeed, it is very easy to see that (\ref{eq:rhok})
only has three non-normalizable pieces at large $E$, namely $E^{1/2}$, $E^{-1/2}\log E$ and $E^{-1/2}$. Including the contribution from $\d\r_U(E)$ at large $E$, the first piece should vanish and the remaining two pieces should obey \eqref{eq:uvdefn}. As computed in
Appendix \ref{sec:Y-shape-spectrum}, matching these three orders gives two equations
\begin{equation}
2\pi Ke^{-S_{0}}\int_{\mD}d\a\f{m(\a)}{\eta(\a)}=\int_{0}^{1}\f{dw}{(1-w)^{1/2}}\int_{0}^{1}du\f{I_{0}(2\pi\sqrt{2\phi_{b}(E_{0}+w(E_{+-}u-E_{0-}))})}{u^{1/2}(1-u)^{1/2}}\label{eq:cond1}
\end{equation}
\begin{align}
 &\int_{0}^{1}\f{dw}{(1-w)^{1/2}}\int_{0}^{1}du\f{I_{0}(2\pi\sqrt{2\phi_{b}(E_{0}+w(E_{+-}u-E_{0-}))})}{(\phi_{b}E_{0}+\phi_{b}E_{+-}(1-2u))^{-1}u^{1/2}(1-u)^{1/2}}-\sqrt{2\phi_{b}E_{0}}I_{1}(2\pi\sqrt{2\phi_{b}E_{0}})\nonumber \\
= & 2\pi Ke^{-S_{0}}\left[\lambda+\int_{\mD}d\a m(\a)\left(\f{\a^{2}+(E_{0}+E_{+}+E_{-})\phi_{b}}{\eta(\a)}-\f 1 \a\right)\right]\label{eq:cond2}
\end{align}
where 
\begin{equation}
\eta(\a)\equiv\sqrt{(\a^{2}+2\phi_{b}E_{0})(\a^{2}+2\phi_{b}E_{+})(\a^{2}+2\phi_{b}E_{-})}
\end{equation}
It is a little surprising that three orders give us only two equations. This means that the $E^{-1/2}\log E$ piece is universal and independent of the phase transition in the IR.

The last equation comes from requiring $\r(E)dE$ to be a
real non-negative measure on $\g_{\pm}$. From the general expression
for $\r(E)$ given in (\ref{eq:rhou}) and (\ref{eq:rhok}), we have
\begin{equation}
\r(E)dE=\r_{r}(E)dE_{r}-\r_{i}(E)dE_{i}+i\left(\r_{r}(E)dE_{i}+\r_{i}(E)dE_{r}\right)
\end{equation}
where $r$ and $i$ subscripts represent real and imaginary parts
respectively. Even though we do not know where the cut is {\it a priori},
the above decomposition holds up to an overall $\pm$ sign because
$\r(E)$ is a two fold map. Being a real measure gives an ordinary differential
equation
\begin{equation}
\r_{r}(E)dE_{i}+\r_{i}(E)dE_{r}=0,\qquad E\in\g_{\pm}
\end{equation}
which has a unique solution given an initial point $E=E_{\pm}$. On
the other hand, we require $\g_{\pm}$ to join $E_{\pm}$ and $E_{0}$,
which gives another initial condition for this differential equation,
leading to the final condition for determining the three parameters
of the spectrum. On the other hand, since $\rho(E)$ is analytic except at the
branching point, the integral from $E_{0}$ to $E_{\pm}$ is invariant
under path deformation. Therefore, we can pick a simple one, for example
the straight line from $E_{0}$ to $E_{+}$, and require the integral
to be real
\begin{equation}
\Im\int_{E_{0}}^{E_{+}}dE\r(E)=0\label{eq:132}
\end{equation}
This condition is independent on the choice of branch cut as $\r(E)$
is a two fold map \cite{Eynard:2015aea} which only leads to  a sign ambiguity that
is irrelevant in (\ref{eq:132}). The same applies to the path from $E_{0}$
to $E_{-}$. This leads to the third equation
required to determine the three parameters. In Appendix \ref{sec:Y-shape-spectrum},
we rewrite it explicitly as
\begin{align}
\Im\left[E_{0+}^{2}E_{-0}^{1/2}\left(\int_{0}^{1}dy\int_{-y}^{y}dx\f{(x+y)^{1/2}\mF(E_{0}-E_{0r}y+iE_{i}x)}{(y-x)^{1/2}(2+x-y)^{1/2}}\int_{\f{x+y}{2+x-y}}^{1}dq\f{_{2}F_{1}(-\f 12,\f 32,2,q\f{E_{0+}}{E_{0-}})}{(q-\f{x+y}{2+x-y})^{1/2}}\right.\right.\nonumber \\
\left.\left.+Ke^{-S_{0}}\phi_{b}^{1/2}\int_{\mD}d\a m(\a)\f{F_{1}(\f 32;-\f 12,-\f 12;3;\f{E_{0+}}{E_{0-}},\f{E_{0+}}{\a^{2}/2\phi_{b}+E_{0}})}{(\a^{2}+2\phi_{b}E_{0})\eta(\a)}\right)\right]  =0\label{eq:cond3}
\end{align}
where $\mF(x)\equiv I_{1}(2\pi\sqrt{2\phi_{b}x})/\sqrt{x}$, $F_{1}$
is the Appell function and $E_{\pm}=E_{r}\pm iE_{i},E_{r,i}\in\R$.

\subsection{Phase transition to the ``Y'' shape} \label{sec:4.3yshape}

\begin{figure}
\begin{centering}

\begin{subfigure}{.49\textwidth}
\includegraphics[height=5.2cm]{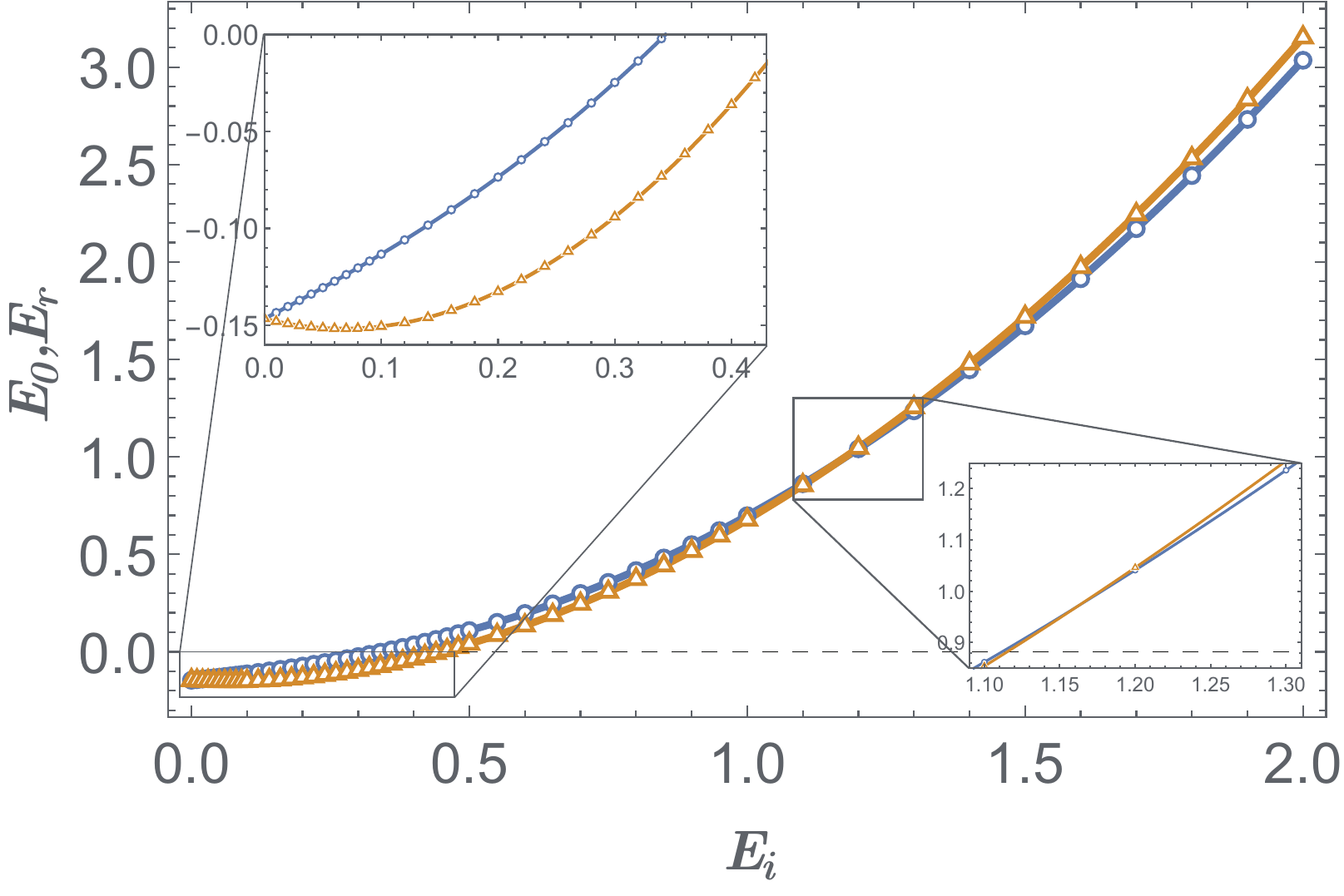}
\caption{}
\label{fig:8a}
\end{subfigure}
\begin{subfigure}{.49\textwidth}
\includegraphics[height=5.2cm]{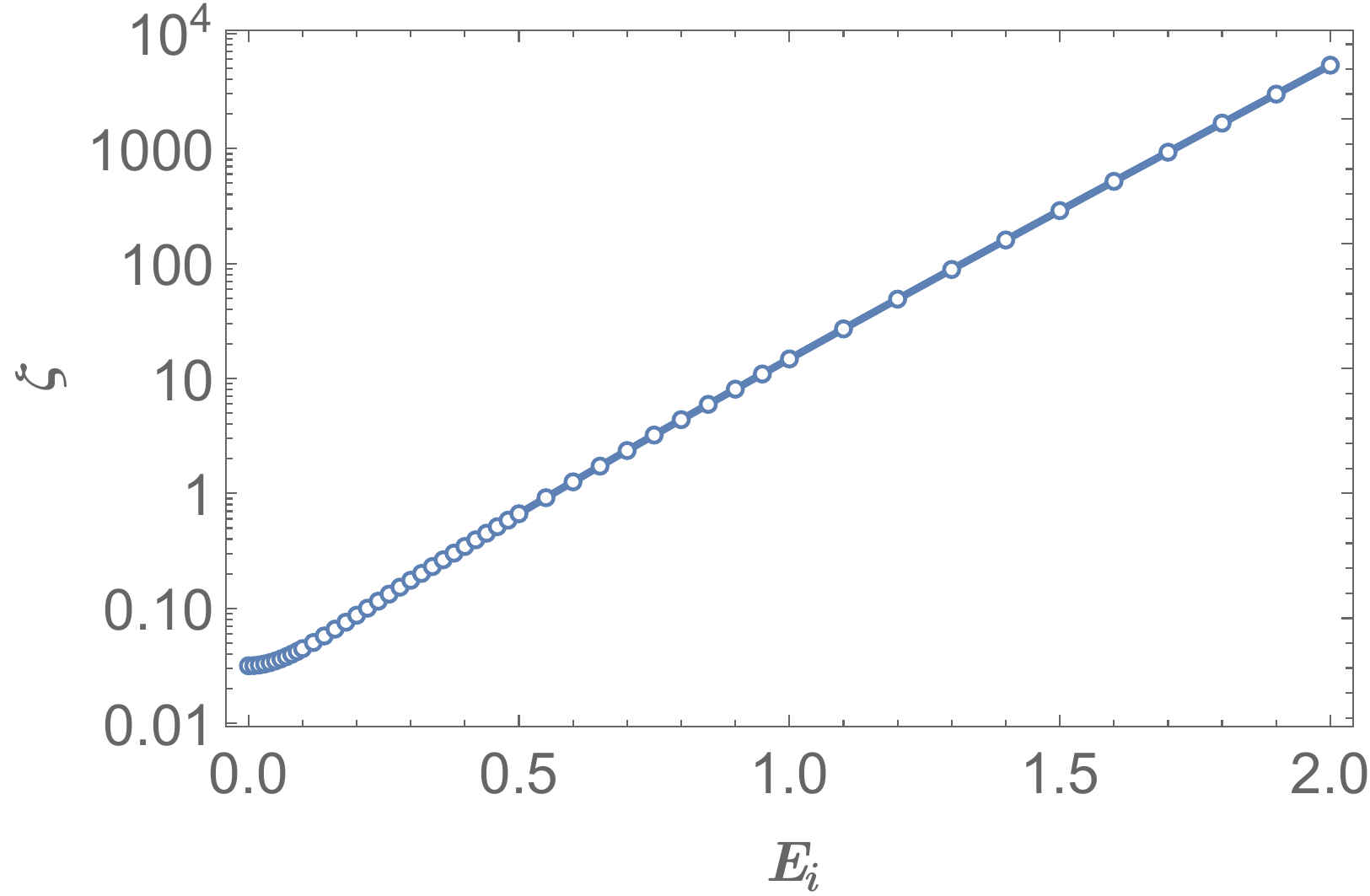}
\caption{}
\label{fig:8b}
\end{subfigure}

\par\end{centering}
\caption{(a) The numerical solution of $E_{0}$ (blue) and $E_{r}$ (yellow)
as a function of $E_{i}$. As $E_{i}$ increases, $E_{0}$ increases
monotonically and $E_{r}$ decreases in the beginning then increases.
At around $E_{i}\protect\app1.2$, $E_{r}$ becomes larger than $E_{0}$.
(b) The numerical solution of $\zeta$ as a function of $E_{i}$.
As $E_{i}$ increases, $\zeta$ increases exponentially. In both pictures,
we set $2\phi_{b}=1$. \label{fig:The-support-of-1}}
\end{figure}
Before we solve for the ``Y'' shaped configuration numerically, let us
first show that this ``Y'' shaped solution continuously connects to one-cut solution
at critical point $E_{0}=E_{+}=E_{-}$. The last condition
for solving the ODE is trivial and we only need to check the first two
conditions (\ref{eq:cond1}) and (\ref{eq:cond2}). Taking $E_{0}=E_{+}=E_{-}$
in (\ref{eq:cond1}), we have
\begin{align}
0 & =-2\pi Ke^{-S_{0}}\int_{\mD}d\a\f{m(\a)}{(\a^{2}+2\phi_{b}E_{0})^{3/2}}+2\pi I_{0}(2\pi\sqrt{2\phi_{b}E_{0}})\nonumber \\
 & =4\pi Ke^{-S_{0}}f'_\lambda (2\phi_{b}E_{0})+2\pi I_{0}(2\pi\sqrt{2\phi_{b}E_{0}})\label{eq:95}
\end{align}
where we used (\ref{eq:60}). Taking $E_{0}=E_{+}=E_{-}$ in (\ref{eq:cond2}),
we have
\begin{align}
&2\pi\phi_{b}E_{0}I_{0}(2\pi\sqrt{2\phi_{b}E_{0}})-\sqrt{2\phi_{b}E_{0}}I_{1}(2\pi\sqrt{2\phi_{b}E_{0}}) \nn\\
=&2\pi Ke^{-S_{0}}\left[\lambda+\int_{\mD}d\a m(\a)\left(\f{\a^{2}+3E_{0}\phi_{b}}{(\a^{2}+2\phi_{b}E_{0})^{3/2}}-\f 1 \a\right)\right]\nonumber \\
 =& 2\pi Ke^{-S_{0}}\left(f_\lambda(2\phi_{b}E_{0})-2\phi_{b}E_{0}f'_\lambda(2\phi_{b}E_{0})\right)\label{eq:96}
\end{align}
Plugging (\ref{eq:95}) into (\ref{eq:96}), we recover (\ref{eq:23-1}).
(\ref{eq:95}) is exactly the critical condition that $G'(2\phi_{b}E_{0})=0$
where $G(x)$ is defined in (\ref{eq:119}). This shows that the phase transition to ``Y''
shape is second or higher order.

To do numerics, we will take $K\sim O(e^{S_0})$, $\lambda\sim O(1)$ and $\mu^2 \gg Ke^{-S_0}$  to simplify computation. This corresponds to a large number of heavy EOW branes. Let us first look at one-cut case. In large $\mu$ limit, we have $Ke^{-S_0}f'_\lambda(x)\sim Ke^{-S_0}\mu^{-2}\ra 0$ in \eqref{eq:23}. The only contribution to $\r(E)$ is from the universal piece. For the zero point equation \eqref{eq:23-1}, using \eqref{eq:regflam} we see $f_\lambda(x)= \lambda$ is just a constant. In this limit, we can find the critical point for $\lambda>0$ in Section \ref{sec:gravspectrum} easily. Define $\zeta\equiv K e^{-{S_0}}\lambda>0$. As we increase $\zeta$ from zero to the critical value, the solution
of $E_{0}$ from (\ref{eq:23-1}) moves from zero to some negative
$E_{cr}$ that is the largest solution to
\begin{equation}
I_{0}(2\pi\sqrt{2\phi_{b}E_{cr}})=0\implies E_{cr}\app-\f{0.146}{2\phi_{b}},\quad\zeta_{cr}\app0.0316 \label{eq:cr-exact}
\end{equation}

\begin{figure}
\begin{centering}

\begin{subfigure}{.49\textwidth}
\includegraphics[height=6cm]{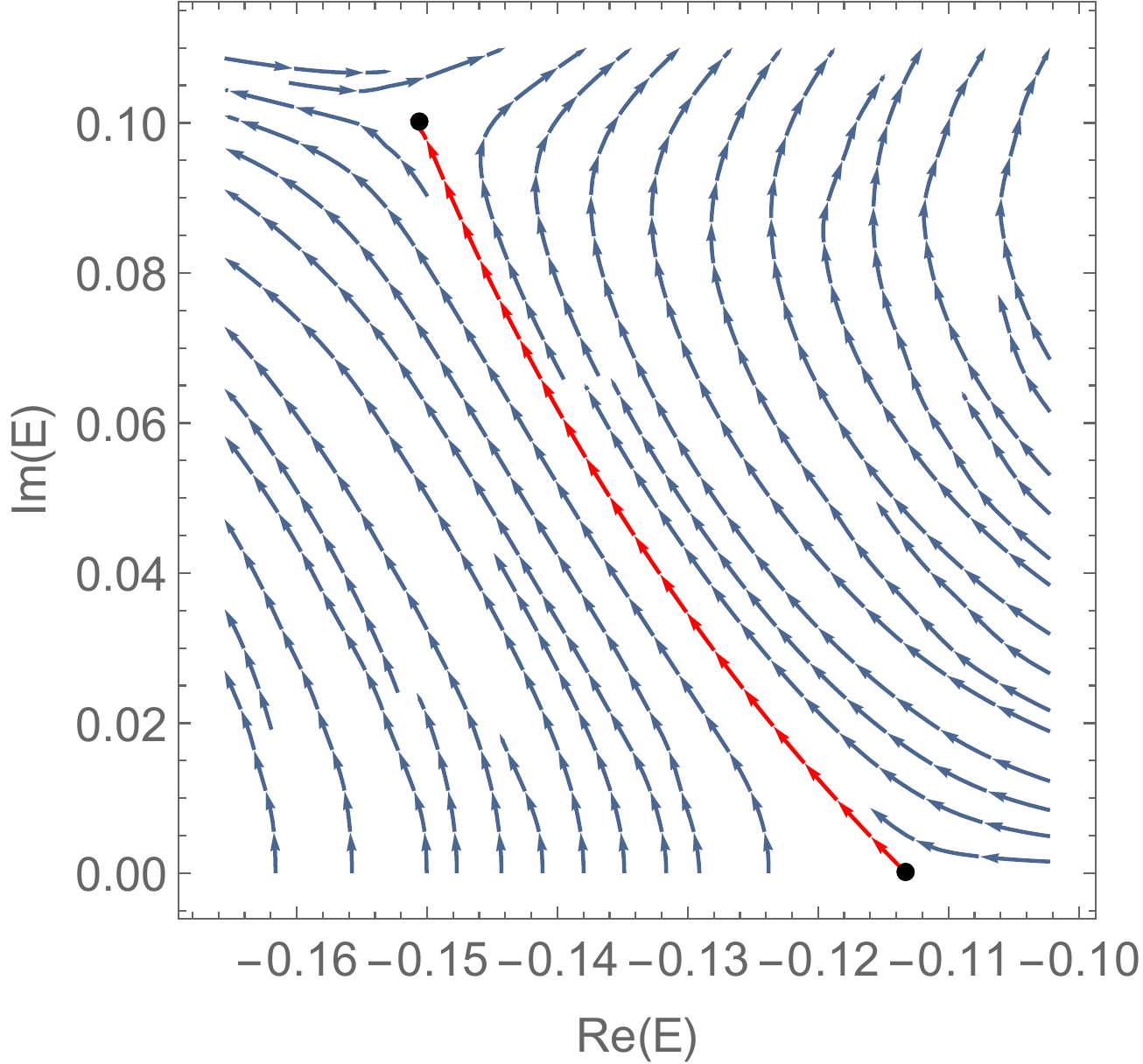}
\caption{$E_{0}=-0.113,\;E_{r}=-0.150,\;E_{i}=1/10$\label{fig:9a}}
\end{subfigure}
\begin{subfigure}{.49\textwidth}
\includegraphics[height=6cm]{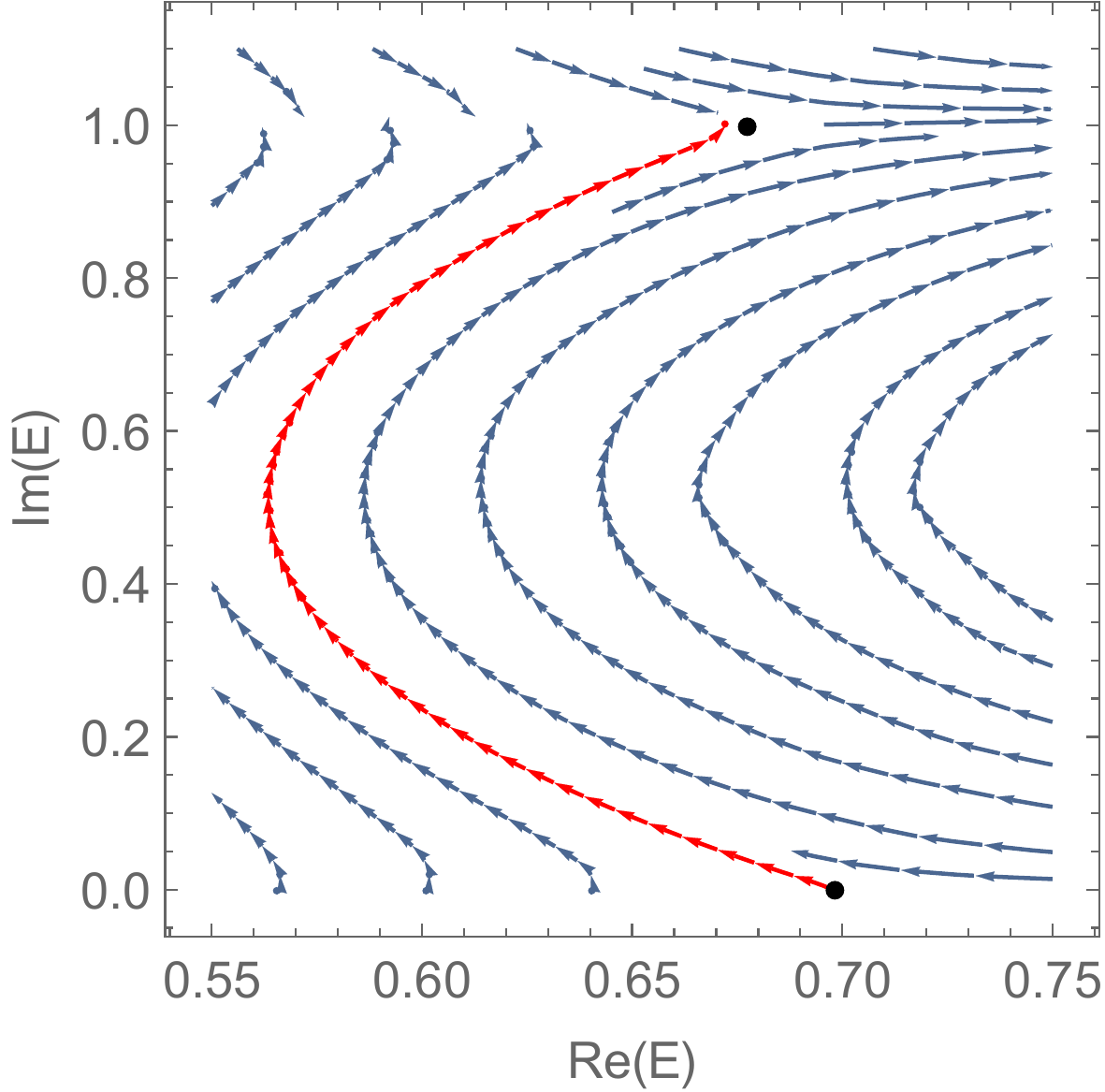}
\caption{$E_{0}=0.698,\;E_{r}=0.678,\;E_{i}=1$\label{fig:9b}}
\end{subfigure}

\par\end{centering}
\caption{The stream plot of vector field $(\protect\r_{r}(E),-\protect\r_{i}(E))$
on upper half complex $E$ plane. The black dot on real axis is $E_{0}$
and the black dot with complex value is $E_{+}$. The red curve connecting
two black dots is the support of density $\protect\g_{+}$ on which
$\protect\r(E)dE$ is nonnegative and real. $\protect\g_{-}$ is just
mirror symmetric curve with $\Im E\protect\ra-\Im E$. In both pictures,
we set $2\phi_{b}=1$. \label{fig:The-support-of-1-1}}
\end{figure}

The simplification is similar in the ``Y" shaped case. In \eqref{eq:rhok} we have $\r_K\sim K\mu^{-4}\ll \r_U\sim O(e^{S_0})$ and thus can be neglected. The three conditions \eqref{eq:cond1}, \eqref{eq:cond2} and \eqref{eq:cond3} simply to
\begin{equation}
\int_{0}^{1}\f{dw}{(1-w)^{1/2}}\int_{0}^{1}du\f{I_{0}(2\pi\sqrt{2\phi_{b}(E_{0}+w(E_{+-}u-E_{0-}))})}{u^{1/2}(1-u)^{1/2}}=0\label{eq:cd1}
\end{equation}
\begin{equation}
\int_{0}^{1}\f{dw}{(1-w)^{1/2}}\int_{0}^{1}du\f{u^{1/2}I_{0}(2\pi\sqrt{2\phi_{b}(E_{0}+w(E_{+-}u-E_{0-}))})}{(1-u)^{1/2}}+\f{2\pi\zeta+\sqrt{2\phi_{b}E_{0}}I_{1}(2\pi\sqrt{2\phi_{b}E_{0}})}{2\phi_{b}E_{+-}}=0\label{eq:cd2}
\end{equation}
\begin{equation}
\Im\int_{0}^{1}dy\int_{-y}^{y}dx\f{E_{0+}^{2}E_{-0}^{1/2}(x+y)^{1/2}I_{1}(2\pi\sqrt{2\phi_{b}(E_{0}-E_{0r}y+iE_{i}x)})}{(E_{0}-E_{0r}y+iE_{i}x)^{1/2}(y-x)^{1/2}(2+x-y)^{1/2}}\int_{\f{x+y}{2+x-y}}^{1}dq\f{_{2}F_{1}(-\f 12,\f 32,2,q\f{E_{0+}}{E_{0-}})}{(q-\f{x+y}{2+x-y})^{1/2}}=0\label{eq:cd3}
\end{equation}
One may naively think that each of (\ref{eq:cd1}) and (\ref{eq:cd2})
contain two real equations as $E_{\pm}$ are complex variables. However,
one can show that (\ref{eq:cd1}) is always real given the symmetry
$u\ra1-u$ and the integral in (\ref{eq:cd2}) is always imaginary
using (\ref{eq:cd1}) (still under transformation $u\ra1-u$). Therefore,
they only give two real equations. 

Our numerical strategy is to solve (\ref{eq:cd1}) and (\ref{eq:cd3})
for a given $E_{i}$ and use $E_{i}$ to compute $\zeta$ using (\ref{eq:cd2}).
This reduces the numerical work to solving only two integral equations.
The result is plotted in Fig. \ref{fig:8a} and \ref{fig:8b}, where
we set $2\phi_{b}=1$ for simplicity. At $E_{i}=0$, both $E_{0}$
and $E_{r}$ start at $E_{cr}.$ As $E_{i}$ increases, $E_{0}$
increases monotonically and $E_{r}$ first decreases, then
increases. Before reaching $E_{i}\app1.2$, $E_{r}$ is less than $E_{0}$,
and it becomes larger for $E_{i}\gtrsim1.2$. While $E_{0,r}$ seems
to increase in roughly the same order as $E_{i}$ in the numerical range,
$\zeta$ increases exponentially with $E_{i}$. It is interesting
that as $E_{i}$ increases (corresponding to increasing $\zeta$),
both $E_{0}$ and $E_{r}$ become positive. As we will see later,
this implies that the canonical ensemble partition function for fixed
temperature will decrease for growing $\zeta$.

Given the solution for $E_{0,r,i}$, we can find the support of the density
$\g_{\pm}$ by solving ODE (\ref{eq:132}). We illustrate the result
for two parameter choices in Fig. \ref{fig:9a} and \ref{fig:9b}.
In these two pictures, we plot the vector field $(\r_{r}(E),-\r_{i}(E))$
on the upper half complex $E$ plane. The red curve connecting two black
dots is the support of the eigenvalues $\g_{+}$ on which $\r(E)dE$ is nonnegative
and real. $\g_{-}$ is the reflected curve with $\Im E\ra-\Im E$.
In Fig. \ref{fig:9a}, $E_{i}$ is small and the whole $\g_{+}$ has
negative real part, for which particularly $E_{r}<E_{cr}$ and $E_{0}>E_{cr}$;
in Fig. \ref{fig:9b}, $E_{i}$ is large and the whole $\g_{+}$ has
positive real part. This is consistent with the behavior in Fig. \ref{fig:8a}.
As $E_{0}$ is a triple branch point, near $E_{0}$ we have $\r(E)dE\sim d(E-E_{0})^{3/2}$,
which implies that the angle between any two branches is $2\pi/3$. 

\begin{figure}
\begin{centering}

\begin{subfigure}{.49\textwidth}
\includegraphics[height=5cm]{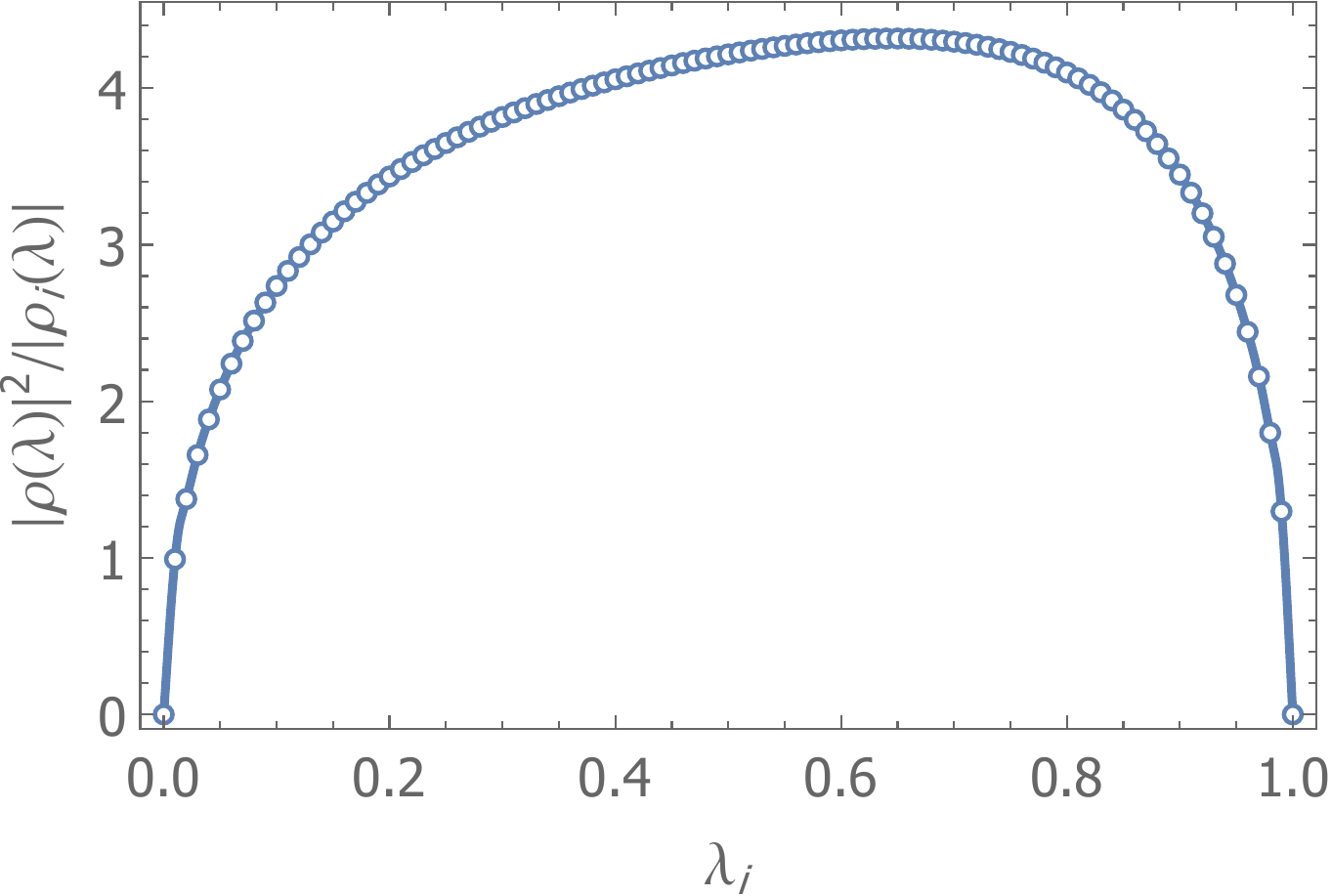}
\caption{\label{fig:10a}}
\end{subfigure}
\begin{subfigure}{.49\textwidth}
\includegraphics[height=5cm]{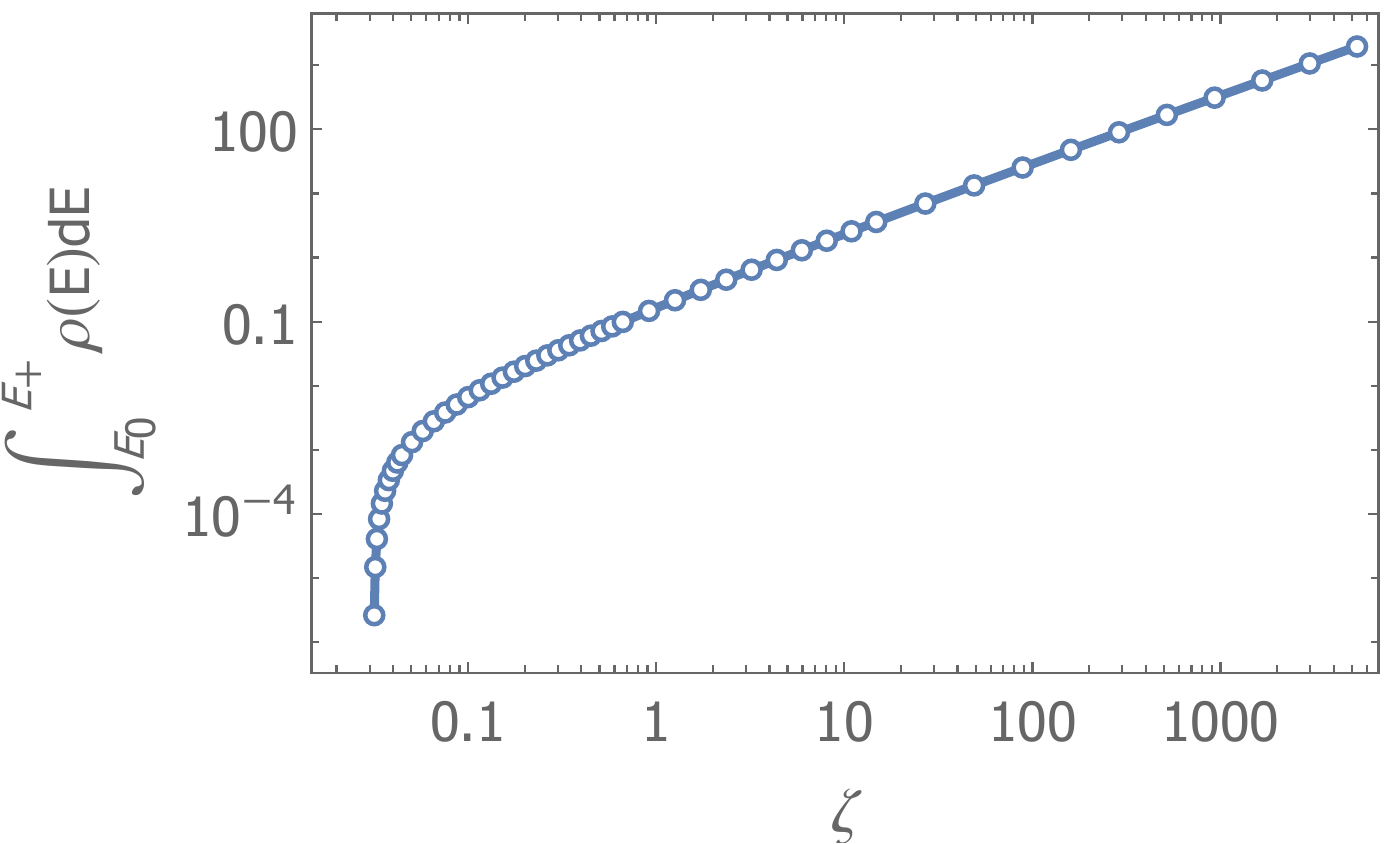}
\caption{\label{fig:10b}}
\end{subfigure}

\par\end{centering}
\caption{(a) The plot of density $|\protect\r(\lambda(\lambda_{i}))|^{2}/|\protect\r_{i}(\lambda(\lambda_{i}))|$
along $\protect\g_{+}$ as a function of $\lambda_{i}$, the imaginary
part of $\lambda$, with $E_{i}=1$. (b) The amount of eigenvalues
on $\protect\g_{+}$ as a function of $\zeta$. In both pictures,
we set $2\phi_{b}=1$ and $e^{S_{0}}=1$.}
\end{figure}
To have an intuitive impression of $\r(\lambda)d\lambda$ on $\g_{\pm}$,
using $d\lambda\propto\r_{r}(\lambda)-i\rho_{i}(\lambda)$ for $\lambda\in\g_{\pm}$,
we have
\begin{equation}
\r(\lambda)d\lambda=\pm\f{|\r(\lambda(\lambda_{i}))|^{2}}{\r_{i}(\lambda(\lambda_{i}))}d\lambda_{i},\qquad\lambda(\lambda_{i})\equiv\lambda_{r}(\lambda_{i})+i\lambda_{i}\in\g_{\pm}
\end{equation}
where $\lambda(\lambda_{i})$ is the locus of $\g_{\pm}$ with affine
parameter $\lambda_{i}$. Note that we need to choose the right sign
of $\pm$ such that $\r(\lambda)d\lambda$ is nonnegative. In Fig.
\ref{fig:10a}, we plot $|\r(\lambda(\lambda_{i}))|^{2}/|\r_{i}(\lambda(\lambda_{i}))|$
along $\g_{+}$ for $E_{i}=1$. The total number of eigenvalues on $\g_{+}$
(which is the same as that on $\g_{-}$) is computed in (\ref{eq:184})
by integration of $\r(E)dE$ along $\g_{+}$ (or equivalently the
straight line) from $E_{0}$ to $E_{+}$. As a function of $\zeta$,
we plot it in Fig. \ref{fig:10b}. From the plot, it seems that the
number of eigenvalues on $\g_{+}$ quickly becomes a power law growth
in $\zeta$ (note that Fig. \ref{fig:10b} is log-log plot).

\begin{figure}
\begin{centering}
\includegraphics[height=5.2cm]{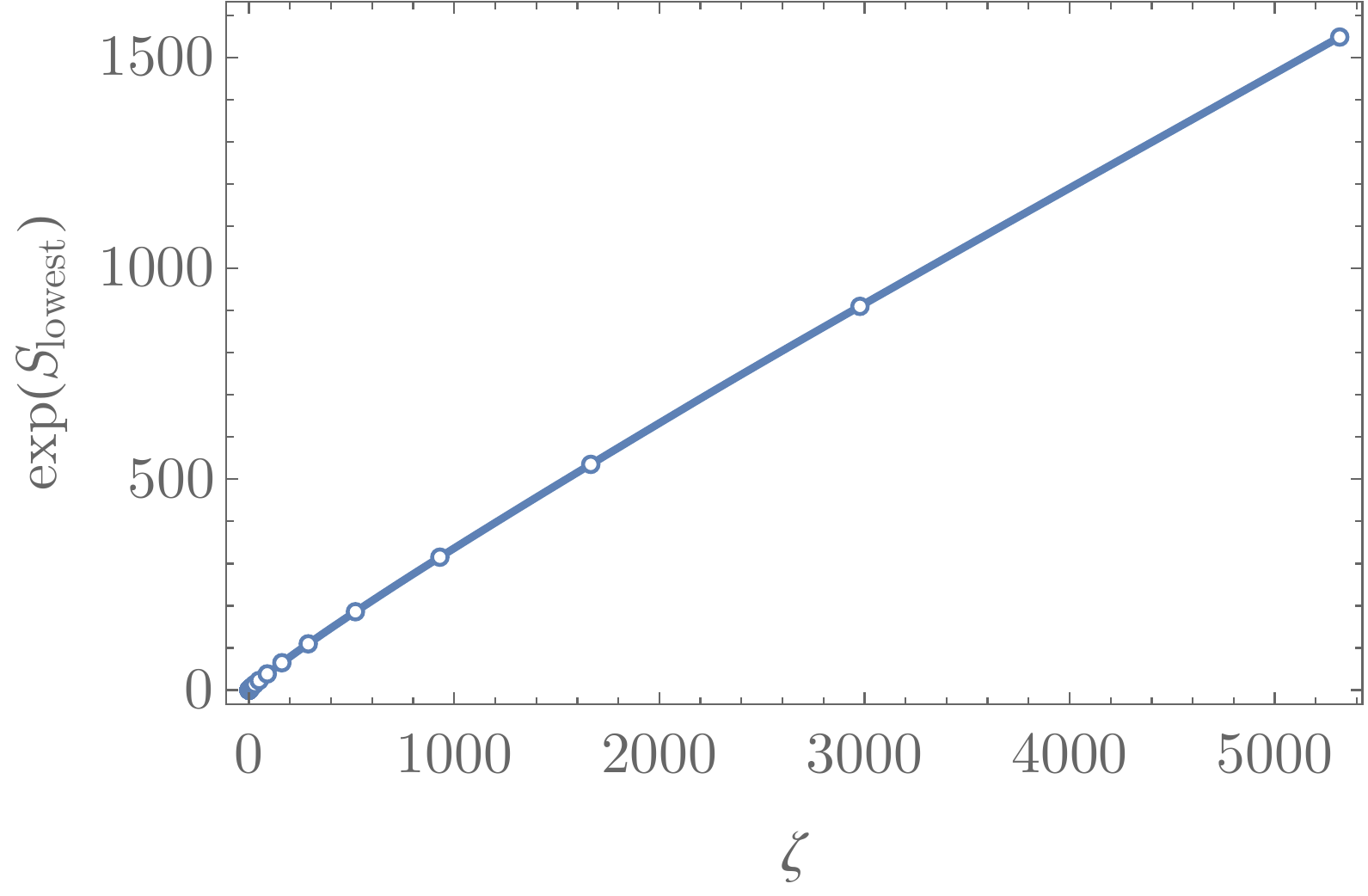}
\par\end{centering}
\caption{$\exp(S_{lowest})$ as a function of $\zeta$. In this plot, we set
$\protect\f 23\protect\d E^{3/2}=e^{S_{0}}=1$ and $2\phi_{b}=1$.
It is clear that $\exp(S_{lowest})$ scales linearly in $\zeta$ for
large $\zeta$.\label{fig:-as-a}}
\end{figure}

We now turn to the interpretation of the complex energy spectrum resulting from the matrix model analysis in terms of black hole physics. In next section, we will use the effective $W(\phi)$ dilaton
JT gravity to illustrate a similar ``Y'' shaped phase transition
using semiclassical analysis of Euclidean black holes. Based on that
observation, we will interpret the complex energy states in spectrum
as unstable black holes that decay to eigenbranes, which are not included
in current matrix model. See section \ref{subsec:Hawking-Page-phase-transition}
for more details. The bottom line is that we can treat the real energy
part of spectrum as stable black hole states. For this piece of the spectrum,
we would like to define the lowest energy black hole entropy (similar to zero-temperature entropy in Section \ref{sec:2.5intEOW}) as
\begin{equation}
S_{lowest}=\lim_{E\ra E_{0}}\f{\r(E)}{\sqrt{E-E_{0}}}
\end{equation}
This $S_{lowest}$ could also be understood as the effective extremal entropy $S_{0,\eff}$ of stable black holes due to integrating out EOW branes along the lines of the discussion in Section \ref{sec:2.5intEOW}, although the full complex spectrum does not start at $E_0$. In the large $\mu$ limit, the leading contribution to $\r(E)$ is the universal piece $\r_U(E)$. Using
(\ref{eq:rhou}), we have
\begin{equation}
S_{lowest}=\f{e^{S_{0}}\phi_{b}^{2}\prod_{i=\pm}(E_{0}-E_{i})^{\f 12}}{\pi}\int_{V_{2}}\f{2dt_{+}dt_{-}I_{1}\left(2\pi\sqrt{2\phi_{b}[E_{0}+\sum_{i=\pm}t_{i}(E_{i}-E_{0})]}\right)}{\sqrt{E_{0}+\sum_{i=\pm}t_{i}(E_{i}-E_{0})}(1-t_{+}-t_{-})^{-1/2}t_{+}^{1/2}t_{-}^{1/2}}
\end{equation}
We plot $\exp(S_{lowest})$ as a function of $\zeta$ in Fig. \ref{fig:-as-a}.
From this plot, it is clear that $\exp(S_{lowest})$ scales linearly
in $\zeta$ for large $\zeta$, which is relevant for  $K\gg e^{S_0}$. In other words, numerics suggest the lowest energy
black hole entropy scales as
\begin{equation}
S_{lowest}\sim\log(Ke^{-S_0} \lambda)=\log K-S_{0}+o(\log K,S_0,1/\mu)
\end{equation}
As mentioned in Section \ref{sec:2.5intEOW}, it is very natural to interpret this as a state counting result, where
$K$ flavors of EOW branes behind the horizon are microstates of the lowest energy black holes.

\section{Effective $W(\Phi)$ dilaton gravity} \label{sec:5}


\subsection{Gas of cusps for heavy EOW branes} \label{subsec:-dilaton-JT}

In Section \ref{sec:2.5intEOW}, we derived an effective $W(\Phi)$ dilaton gravity for the $\lambda=0$ case by comparing spectral densities. In this section, we will study the effective $W(\Phi)$ related to heavy EOW branes with $\lambda\neq 0$.

The large $\mu$ limit  basically receives
only contributions from small $b$ of order $1/\mu$ in the EOW brane measure $\mM(b)$.
On the other hand, the $b\ra0$ limit shrinks the EOW loops into cusps as discussed in Section \ref{sec:EOWbranequantization}. Therefore, adding heavy EOW branes to JT gravity can be understood as deformation by cusps, the $2\pi$ deficit angle, which is described by a new potential for the dilaton $\Phi$  \cite{Witten:2020wvy, Maxfield:2020ale}
\begin{equation}
W(\Phi)=2\Phi+2\chi e^{-2\pi\Phi}\label{eq:wphi}
\end{equation}
with total Euclidean action
\begin{equation}
I=-S_{0}-\f 12\int d^{2}x\sqrt{g}(\Phi R+W(\Phi))
\end{equation}
On the other hand, large $\mu$ leads to $f_\lambda(x)=\lambda$. Comparing \eqref{eq:23-1} with (8.4) in \cite{Witten:2020wvy}, we can determine
\begin{equation}
\chi=K e^{-S_{0}} \lambda \label{eq:lambda}
\end{equation}



\begin{figure}
\begin{centering}
\subfloat[$\zeta<\zeta_{cr}$\label{fig:13a}]{\begin{centering}
\includegraphics[height=3.4cm]{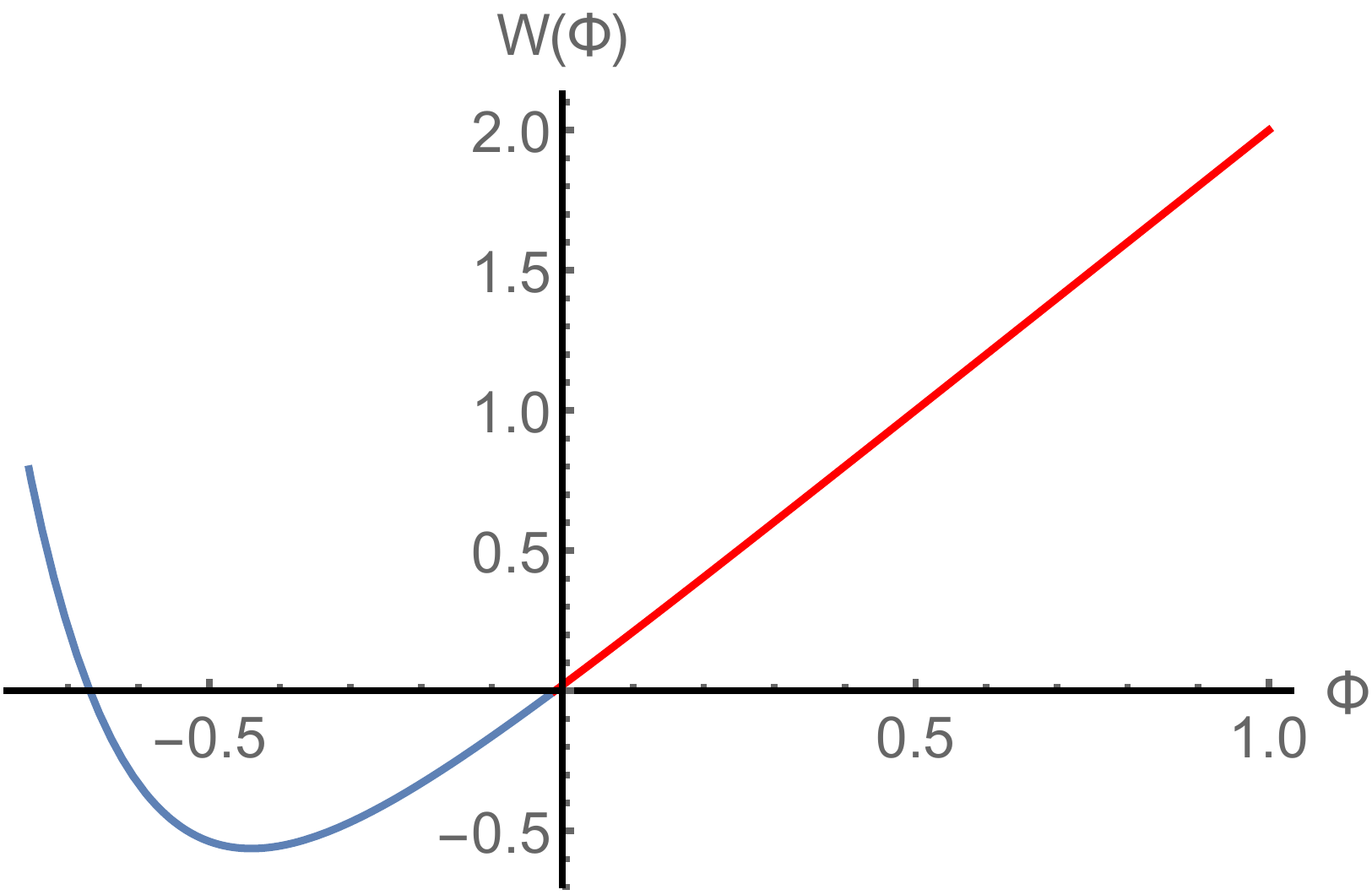}
\par\end{centering}
}\subfloat[$\zeta=\zeta_{cr}$\label{fig:13b}]{\begin{centering}
\includegraphics[height=3.4cm]{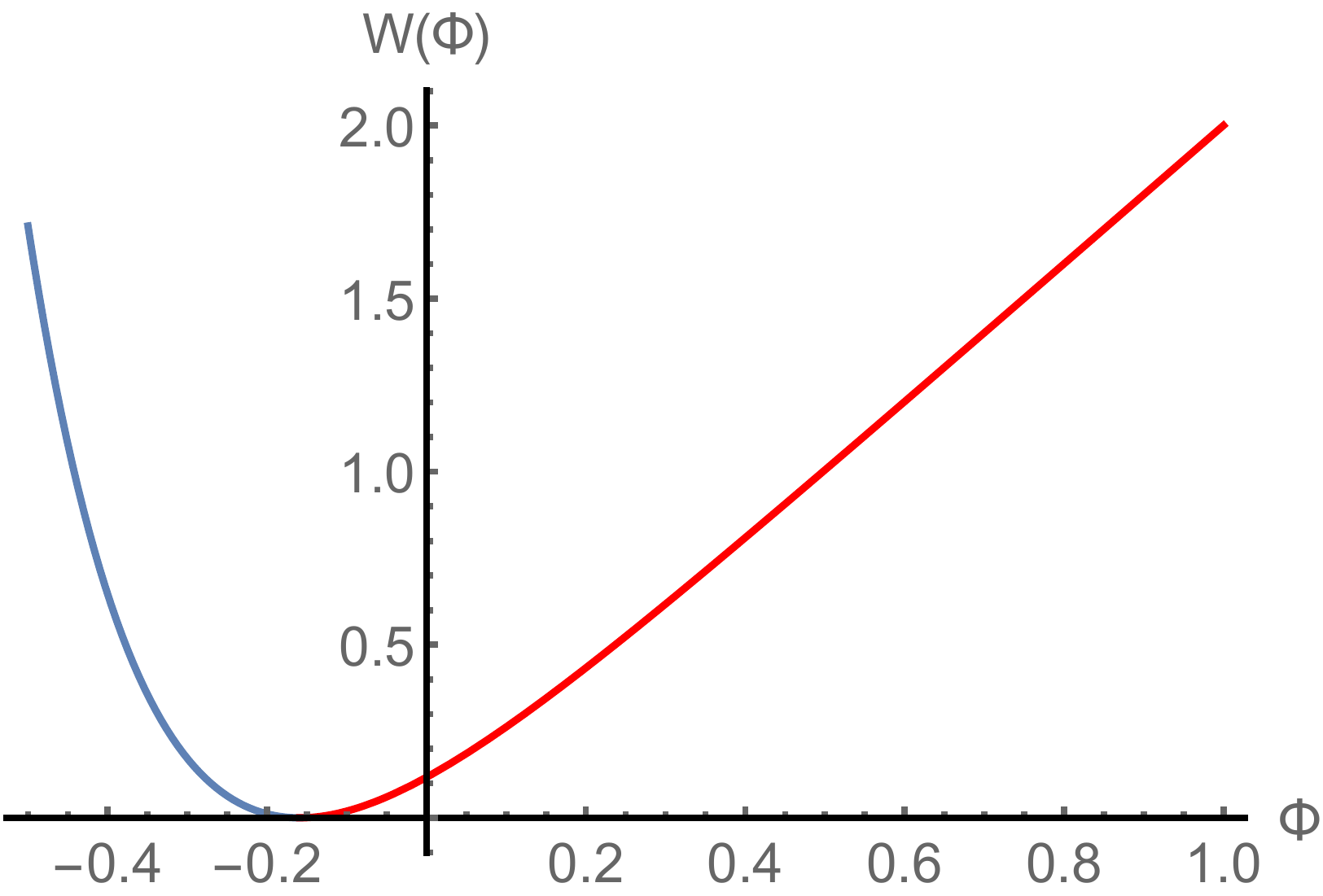}
\par\end{centering}
}\subfloat[$\zeta>\zeta_{cr}$\label{fig:13c}]{\begin{centering}
\includegraphics[height=3.4cm]{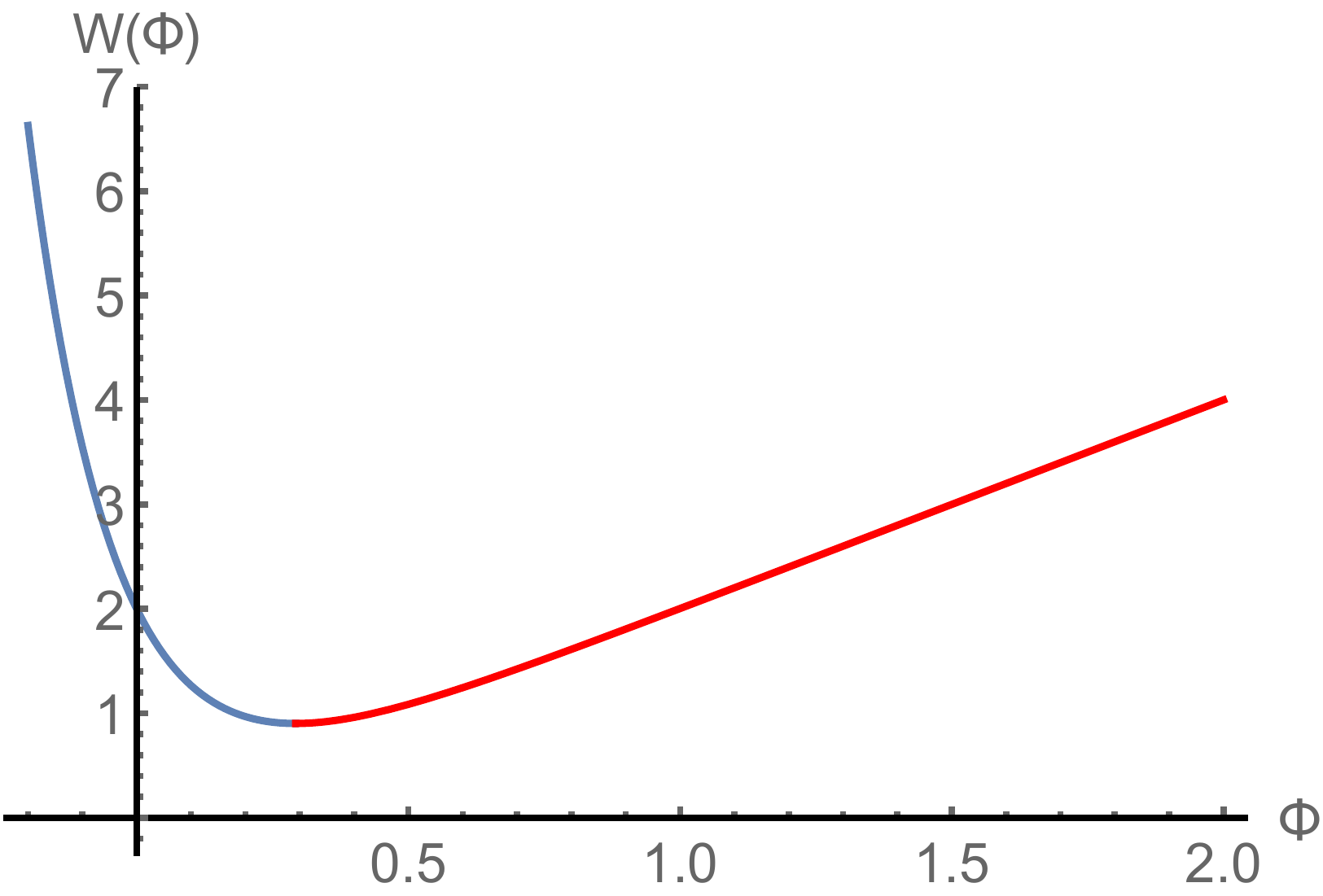}
\par\end{centering}
}
\par\end{centering}
\caption{$W(\Phi)$ for different values of $\zeta$. Red part corresponds
to the spectrum of thermodynamical stable black holes. \label{fig:Black-hole-microcanonical-1}}
\end{figure}

Given a generic dilaton potential $W(\Phi)$ with asymptotic behavior
$W(\Phi)\sim2\Phi$ for $\Phi\ra+\infty$, the classical solutions
of the equations of motion were studied in \cite{PhysRevD.49.5227} and recently in \cite{Witten:2020ert}.
Choosing the diffeomorphism gauge $\Phi=\phi_h+r$, the solution for the 2d metric is 
\begin{equation}
ds^{2}=A(r)dt^{2}+\f 1{A(r)}dr^{2},\qquad A(r)=\int_{0}^{r}dxW(\phi_{h}+x)\label{eq:154}
\end{equation}
where $\phi_{h}$ is the horizon value of the dilaton and $r\in[0,+\infty)$. Matching with the
AdS boundary condition (\ref{eq:adsbdc}) at $r=r_{\infty}$, we have
\begin{equation}
t=\phi_{b}^{-1}du_{E},\quad r_{\infty}=\phi_{b}/\e
\end{equation}
The ADM energy of such a black hole is 
\begin{equation}
E(\phi_{h})=\f 1{2\phi_{b}}\left((\phi_h+r_{\infty})^{2}-\int_{0}^{r_{\infty}}dxW(\phi_{h}+x)\right)\label{eq:admenergy}
\end{equation}
Taking $r_{\infty}\ra\infty$ limit, and using (\ref{eq:wphi}) and
(\ref{eq:lambda}), we have
\begin{equation}
E(\phi_{h})=\f 1{2\phi_{b}}\left(\phi_{h}^{2}-\f{\zeta}{\pi}e^{-2\pi\phi_{h}}\right),\label{eq:156}
\end{equation}
where $\zeta= K e^{-{S_0}}\lambda$ as defined in Section \ref{sec:4.3yshape}. The black hole temperature is determined by requiring smoothness of the metric at $r=\phi_{h}$.
Setting $y=\sqrt{r}$, and $A(r)=W(\phi_{h})r+o(r)$,
we have for $y\ll1$
\begin{equation}
ds^{2}=\f 4{W(\phi_{h})}\left(dy^{2}+\f{W(\phi_{h})^{2}}4y^{2}dt^{2}\right)\label{eq:nhorizon}
\end{equation}
For $t\in\R$, this metric has a conical singularity at $y=0$ unless
\begin{equation}
t\simeq t+\phi_{b}^{-1}\b,\quad\b=\f{4\pi\phi_{b}}{W(\phi_{h})},\quad T(\phi_{h})=\f{W(\phi_{h})}{4\pi\phi_{b}}\label{eq:159}
\end{equation}
In the following, we will mainly consider $\lambda>0$ (namely $\zeta>0$).
 
For small $\zeta$, $W(\Phi)=0$ has two real roots (see Fig. \ref{fig:13a}).
Based on \cite{Witten:2020ert}, we should take the larger root as the lower
bound for $\phi_{h}$ because we require $A(r)>0$ for all $r>0$.
The red part of Fig. \ref{fig:13a} shows the allowed range of $\phi_{h}$,
which by (\ref{eq:156}) determines the spectrum. In particular, the black
hole temperature can take any nonnegative value. There is a critical
value of $\zeta_{cr}$ where these two roots coincide at $\phi_{cr}$
(see Fig. \ref{fig:13b})
\begin{equation}
W'(\phi_{cr})=W(\phi_{cr})=0\implies\zeta_{cr}=\f 1{2\pi e}\app0.0585,\quad\phi_{cr}=-\f 1{2\pi}
\end{equation}
This gives a critical zero point energy 
\begin{equation}
E_{cr}=E(\phi_{cr})=-\f 1{8\pi^{2}\phi_{b}}\app-\f{0.0253}{2\phi_{b}}
\end{equation}
If we compare with the exact result (\ref{eq:cr-exact}), we see they
are off to some extent but not too much. As this is only a semiclassical
analysis, we should not expect an exact match with the full quantum computation.
They match qualitatively, especially in that for some positive $\zeta$
we get negative critical energy $E_{cr}$. 

\begin{figure}
\begin{centering}
\subfloat[\label{fig:14a}]{\begin{centering}
\includegraphics[height=4cm]{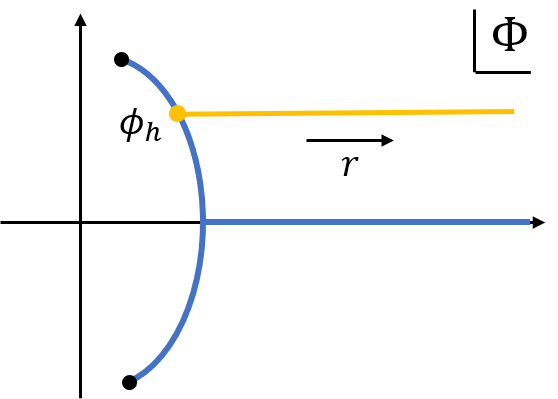}
\par\end{centering}
}\subfloat[\label{fig:14b}]{\begin{centering}
\includegraphics[height=4cm]{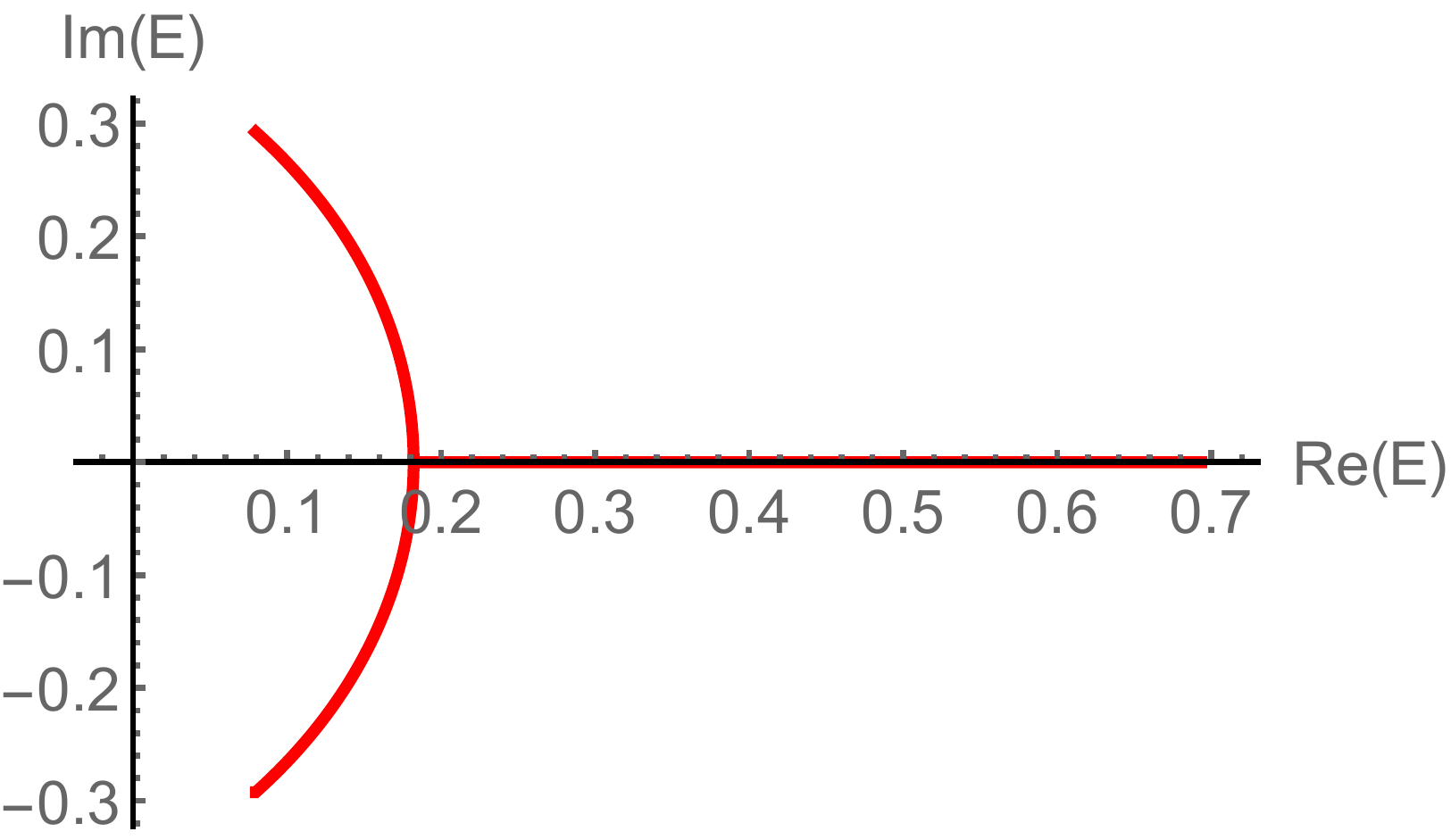}
\par\end{centering}
}
\par\end{centering}
\caption{(a) On the $\Phi$ complex plane, the blue curve is the locus for which
$W(\Phi)\in\protect\R$, and $\phi_{h}$ is valued on this curve. For a given solution, $\Phi$ takes value on the yellow straight line emanating from $\phi_h$, which is parameterized by radial coordinate $r\in[0,+\infty)$. (b) The support $\protect\g_{*}$
of energy spectrum on complex $E$ plane. It has two branches $\protect\g_{*,\pm}$
conjugate to each other. \label{fig:(a)-On-}}
\end{figure}
If we increase $\zeta$ past the critical value, we will have all
$\phi_{h}\in\R$ allowed in the spectrum by the condition $A(r)>0$ for
$r>0$. In this case, as $W(\phi)$ is lower bounded by a positive
value, the minimum temperature of black holes is nonzero and given by
\begin{equation}
T_*=T(\phi_{*})=\f{1+\log2\pi\zeta}{4\pi^{2}\phi_{b}}
\end{equation}
where $\phi_{h}=\phi_{*}=(2\pi)^{-1}\log2\pi\zeta$ with $W'(\phi_{*})=0$.
For a given temperature $T>T_*$, there are two solutions for
black holes with different $\phi_{h}=\phi_{h,\pm}$. It has been shown
in \cite{Witten:2020ert} that the free energy difference between these two
solutions is
\begin{equation}
\D F=2\pi\int_{\phi_{h,-}}^{\phi_{h,+}}(T(x)-T)dx>0
\end{equation}
which implies that the larger black hole with $\phi_{h,+}$ is thermodynamically
stable. Equivalently, one can also show that the specific heat $dE/dT$
is negative for the smaller black hole with $\phi_{h,-}$. 

Note that we
have set $\kappa=(8\pi G)^{-1}$ equal to 1 throughout this paper. Restoring the $G_{N}$ dependence, 
the contribution to the partition function from smaller black holes is
relatively $O(e^{-1/G_N})$ suppressed in the small $G_N$ limit. Therefore, if we only look
at the spectrum of canonically stable black holes, the energy will be lower
bounded by 
\begin{equation}
E_{*}=E(\phi_{*})=\f{(\log2\pi\zeta)^{2}-2}{8\pi^{2}\phi_{b}}
\end{equation}
This physically corresponds to the lower bound $E_{0}$ of the real
energy spectrum in the ``Y'' shaped phase of the matrix model computation.

The entropy of this lowest energy stable black hole is
\begin{equation}
S_{*}=2\pi\phi_{*}=\log2\pi\zeta
\end{equation}
which scales logarithmically with $K$. This also matches with the
numerics for $S_{lowest}$ in the large $K$ limit as
shown in Fig. \ref{fig:-as-a}. 

Similarly, one can easily show that for $\lambda<0$, the zero point energy is given by $W(\phi_*)=0$, which leads to same logarithmic scaling $S_*\sim \log K$ for $Ke^{-S_0}\gg 1$. This, at the semiclassical level, justifies the induced gravity interpretation mentioned at the end of Section \ref{sec:2.5intEOW}.

\begin{figure}
\begin{centering}
\subfloat[\label{fig:15a}]{\begin{centering}
\includegraphics[height=4cm]{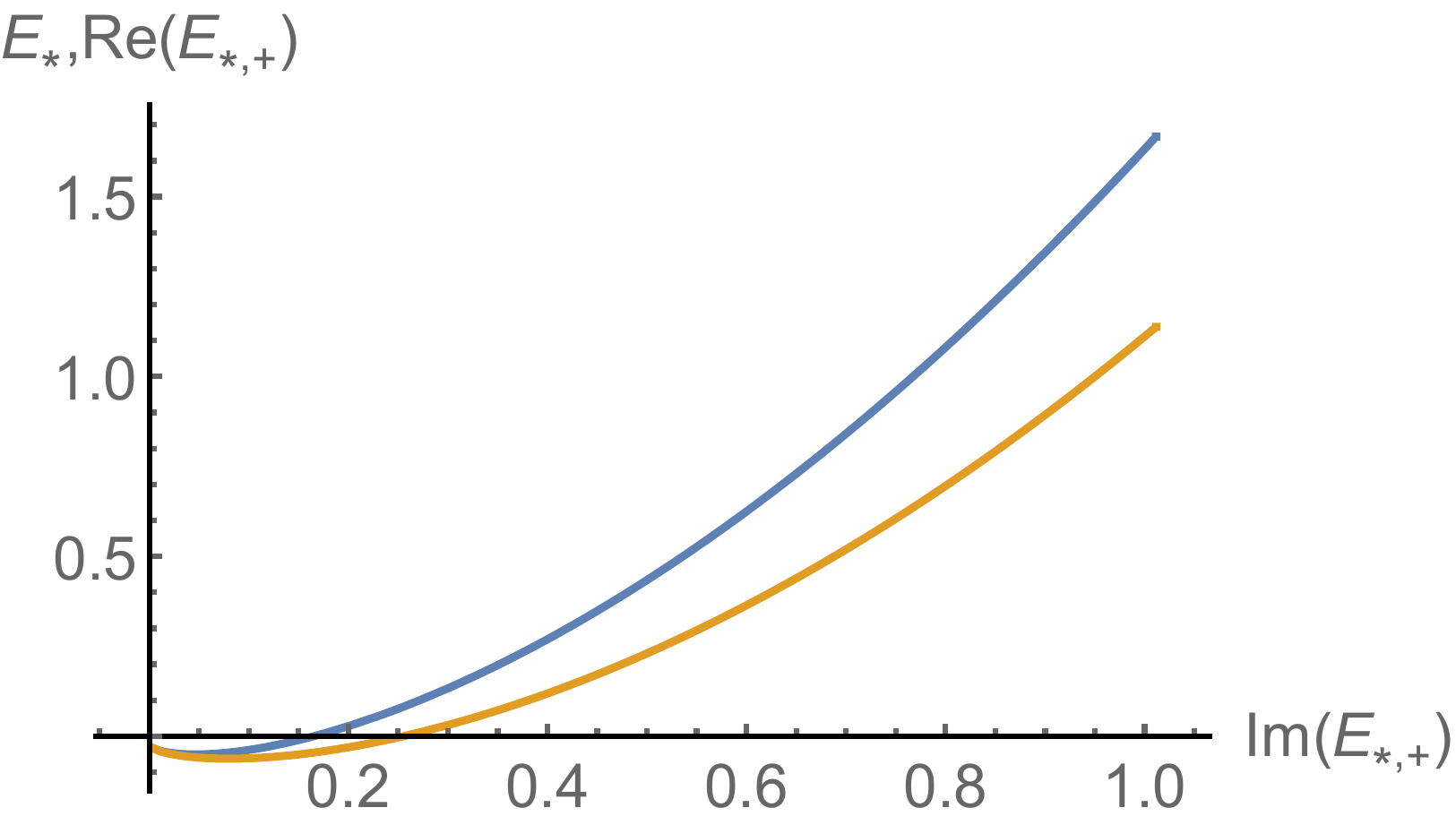}
\par\end{centering}
}\subfloat[\label{fig:15b}]{\begin{centering}
\includegraphics[height=4cm]{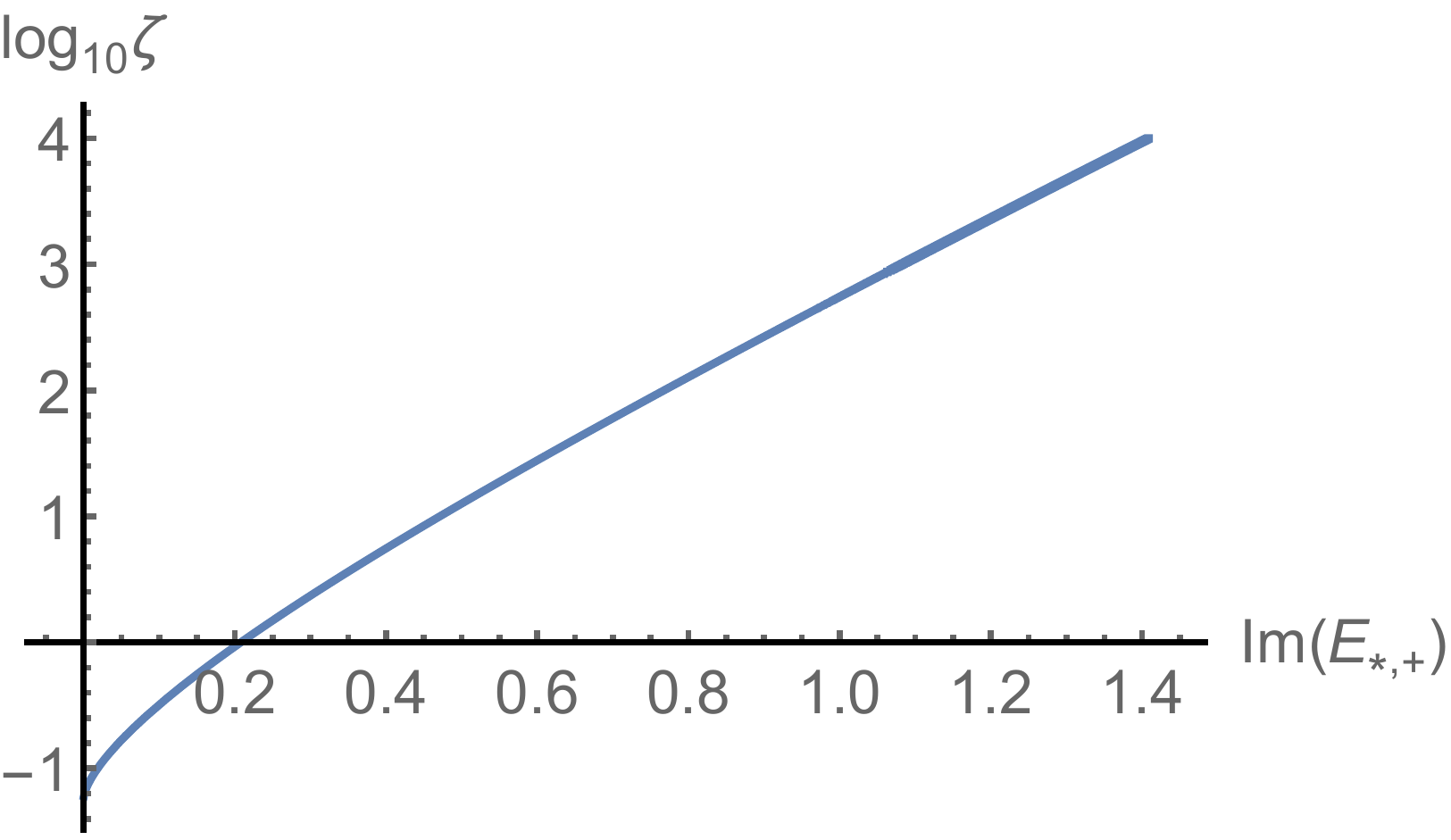}
\par\end{centering}
}
\par\end{centering}
\caption{$\Re(E_{*,+})$, $E_{*}$ and $\log_{10}\zeta$ as a function of $\Im(E_{*,+})$.
We set $2\phi_{b}=1$ in both plots. \label{fig:,--and}}
\end{figure}
One can continue the condition $W(\phi_{h})=0$ that defined the endpoint of the spectrum 
for $\zeta<\zeta_{cr}$ past the critical value, where $\phi_h$ becomes complex. This corresponds to
the complex part of spectrum in the matrix model.
We
could consider complex saddles of Euclidean action simply by analytic continuation of $\phi_h$ to complex numbers in \eqref{eq:154}. As $\phi_h$ is a free parameter, it is obvious that such a complex metric and dilaton solve the equations of motion (see Fig. \ref{fig:14a}). 
For this complex solution to be a saddle, we still need to require the complex metric
is smooth around $\phi_{h}$. Expanding in small $r$,
(\ref{eq:nhorizon}) still holds. 
For $t\in\R$, the metric has conical singularity at $y=0$ unless
$W(\phi_{h})\in\R$ and $t$ have the same periodicity as (\ref{eq:159}).
The condition  $W(\phi_{h})\in\R$ gives the allowed complex $\phi_{h}$,
which defines the support of the complex part of the spectrum $\g_{*}$ via
$E(\phi_{h})$ (see Fig. \ref{fig:14b}). The endpoint of $\g_{*}$ 
corresponds to $W(\phi_{h})=0$ as it is the singularity of metric \eqref{eq:nhorizon}.

Let us check what follows from this complex saddle. For a given $\zeta$,
the solution of $W(\phi_{h})=w$ is given by
\begin{equation}
\phi_{h}=\f w2+\f 1{2\pi}\mW(-2\pi\zeta e^{-\pi w})\label{eq:20-1}
\end{equation}
where $\mW(z)$ is product logrithmic function (or Lambert W function) defined implicitly as $\mW e^\mW=z$. This is a multivalued
function but we will only choose branches that connect to real $\phi_{h}$
for $\zeta<\zeta_{cr}$. This leads two solutions $\phi_{h,\pm}(w)$
conjugate to each other and thus splits $\g_{*}$ into two pieces
$\g_{*,\pm}$ joint at $E_{*}$, similar to $\g_{\pm}$ in matrix
model (see Fig. \ref{fig:14b}). Setting $w=0$ and using in (\ref{eq:156}),
we can solve the two ends of $\g_{*,\pm}$
\begin{equation}
E_{*,\pm}=\f{\mW(-2\pi\zeta)(2+\mW(-2\pi\zeta))}{8\pi^{2}\phi_{b}}\label{eq:estar-pm}
\end{equation}
To have a comparison with matrix model result, we plot $\Re(E_{*,+})$,
$E_{*}$ and $\zeta$ as a function of $\Im(E_{*,+})$ in Fig. \ref{fig:,--and}.
Qualitatively, this matches with the exact result in Fig. \ref{fig:The-support-of-1}.

\subsection{Dirichlet-Neumann EOW branes}

We found that for $\zeta>\zeta_{cr}$, no real solutions
for $W(\phi_{h})=0$ exist. This means that there is no zero temperature black hole. It is natural to ask what happens if we decrease
the temperature below $T_*$. This phenomenon is not unique to JT gravity. There is also a minimum temperature $T_{\min}$ for $d > 3$ dimensional AdS-Schwarzschild black holes. For any fixed temperature above $T_{\min}$, two black holes
solutions exist and only the bigger one is thermodynamically stable.

Hawking and Page studied this property in their well-known paper \cite{Hawking:1982dh} and found a first order phase transition from black holes
to a thermal gas at some higher temperature $T_{HP}>T_{\min}$. At
$T=T_{HP}$, if we decrease the total energy, black holes will be in equilibrium with a thermal gas in the microcanonical ensemble. For $T<T_{HP}$, one transitions completely to the thermal gas phase. 

By analogy with higher dimensional black holes, we consider how a ``thermal gas'' could exist in $W(\Phi)$ dilaton gravity for temperatures below a Hawking-Page temperature $T_{HP}>T_*$. The simplest example
is to consider another type of EOW brane that leads to spacetimes
different from black holes. All types of EOW branes in pure JT gravity
are classified in \cite{Goel:2020yxl} based on their boundary conditions.
We can perform a similar classification for $W(\Phi)$ dilaton gravity.  The Dirichlet-Neumann (DN) EOW
brane (in the language of \cite{Goel:2020yxl}), corresponding to a location where the dilaton reaches $-\Phi_0$, is the dimensional reduction of the smooth origin of polar coordinates in the center of empty AdS. Thus it is the natural candidate for states that behave like the thermal gas. 

Let us consider the Euclidean action
\begin{equation}
I=-\f 12\int drdt\sqrt{g}(\Phi R+W(\Phi))-\int_{AdS}du\sqrt{g_{uu}}\Phi(K-1)-\int_{\del}dv\sqrt{g_{vv}}(\Phi K+d)
\end{equation}
where $d$ is a constant defined on the DN brane world line $\del$.
The variation of action gives equations of motion as well as boundary
conditions (see Appendix \ref{sec:Variation-of-JT}), which on DN
brane are
\begin{equation}
(n^{a}\del_{a}\Phi+d)h_{ab}\d g^{ab}=K\d\Phi=0
\end{equation}
We fix $\Phi=\phi_{DN}$ (Dirichlet) and also $n^{a}\del_{a}\Phi+d=0$ (Neumann). The bulk
solution has the same form as \eqref{eq:154} but with a different ``horizon" parameter $\phi_{\bar h}$. Note that this spacetime does not have any real horizon at $\phi_{\bar h}$ because we truncate the spacetime at a new boundary (DN brane) before it, which restricts $\phi_{DN}>\phi_{\bar{h}}$. 

Using solution (\ref{eq:154}), we
have
\begin{equation}
n^{a}\del_{a}\Phi=-\sqrt{A(\phi_{DN})}\implies d=\sqrt{A(\phi_{DN})}\label{eq:169-x}
\end{equation}
The Hamiltonian defined in (\ref{eq:30H}) only depends on the asymptotic
behavior of $\Phi$ and thus is given by the same formula as (\ref{eq:admenergy})
with $\phi_{h}\ra\phi_{\bar{h}}$.\footnote{No Hamiltonian arises on the DN brane due to the boundary conditions.}
As $d$ is a fixed constant, (\ref{eq:169-x}) determines $\phi_{\bar{h}}$
and completely fixes the Hamiltonian to be
\begin{equation}
H=\f 1{2\phi_{b}}\left(r_{\infty}^{2}-\int_{\phi_{DN}}^{r_{\infty}}dxW(x)-d^{2}\right)
\end{equation}
This corresponds to the eigenbrane first introduced in \cite{Blommaert:2019wfy}
that has fixed energy. In the dual matrix model, it gives delta function
in spectrum \cite{Goel:2020yxl}. 

To discuss dominance for a given temperature, we will evalute the
Euclidean on-shell action for DN brane and black hole respectively.
In both cases, Euclidean time is periodic 
\begin{equation}
t\sim t+\phi_{b}^{-1}\b
\end{equation}
For the DN brane, extrinsic curvature is 
\begin{equation}
K=-\f{A'(\phi_{DN})}{2\sqrt{A(\phi_{DN})}}=-\f{W(\phi_{DN})}{2\sqrt{A(\phi_{DN})}}
\end{equation}
The action on the DN brane is
\begin{equation}
I_{\del}=-\f{\b}{2\phi_{b}}(2d^{2}-\phi_{DN}W(\phi_{DN}))
\end{equation}
Similarly, the extrinsic curvature on the AdS boundary is
\begin{equation}
K=\f{W(r_{\infty})}{2\sqrt{A(r_{\infty})}}
\end{equation}
The action on the AdS boundary is
\begin{equation}
I_{AdS}=-\b\f{\phi_{b}}{\e^{2}}\left[\f{W(r_{\infty})}{2\sqrt{A(r_{\infty})}}-1\right]\ra-\f{\b}{2\phi_{b}}\left(r_{\infty}^{2}-\int_{\phi_{DN}}^{r_{\infty}}dxW(x)-d^{2}\right)
\end{equation}
The bulk action is
\begin{align}
I_{bulk} & =-\f{\b}{2\phi_{b}}\int_{\phi_{DN}}^{r_{\infty}}dr(A'(r)-rA''(r))=-\f{\b}{2\phi_{b}}\int_{\phi_{DN}}^{r_{\infty}}dr(2A(r)-rA'(r))\nonumber \\
 & =\f{\b}{\phi_{b}}\left(r_{\infty}^{2}-\int_{\phi_{DN}}^{r_{\infty}}dxW(x)-\f{\phi_{DN}W(\phi_{DN})}2\right)
\end{align}
Altogether we find that the Euclidean on-shell action for DN branes is
\begin{equation}
I_{DN}=\f{\b}{2\phi_{b}}\left(r_{\infty}^{2}-\int_{\phi_{DN}}^{r_{\infty}}dxW(x)-d^{2}\right)=\b E_{DN}
\end{equation}
which is as expected because the DN brane is an energy eigen state and thus
has zero entropy $S=\b E-I=0$. Note that there is no topological
term linear in $S_{0}$ in $I_{DN}$ because the topology is a cylinder. 

On the other hand, the Euclidean action for black holes only has two
terms, $I_{AdS}$ and $I_{bulk}$, because the geometry is smooth
at the horizon $\phi=\phi_{h}$. For black holes, $\phi_h$
is not a free parameter and is determined by $\b$ using (\ref{eq:159}).  We have
\begin{equation}
I_{AdS}=-\f{\b}{2\phi_{b}}(r_{\infty}^{2}-A(r_{\infty})),\quad I_{bulk}=\f{\b}{\phi_{b}}(r_{\infty}^{2}-A(r_{\infty})-\phi_{h}W(\phi_{h})/2)
\end{equation}
As the topology is a disk, we do have the topological term $-S_{0}=-2\pi\phi_{0}$.
This leads to
\begin{equation}
I_{bh}=\f{\b}{2\phi_{b}}(r_{\infty}^{2}-A(r_{\infty}))-2\pi(\phi_{h}+\phi_{0})
\end{equation}
where we used (\ref{eq:159}) to write $I_{bh}$ in the form of $\b E-S$.

\subsection{Hawking-Page phase transition\label{subsec:Hawking-Page-phase-transition}}

\begin{figure}
\begin{centering}
\subfloat[\label{fig:16a}]{\begin{centering}
\includegraphics[height=3cm]{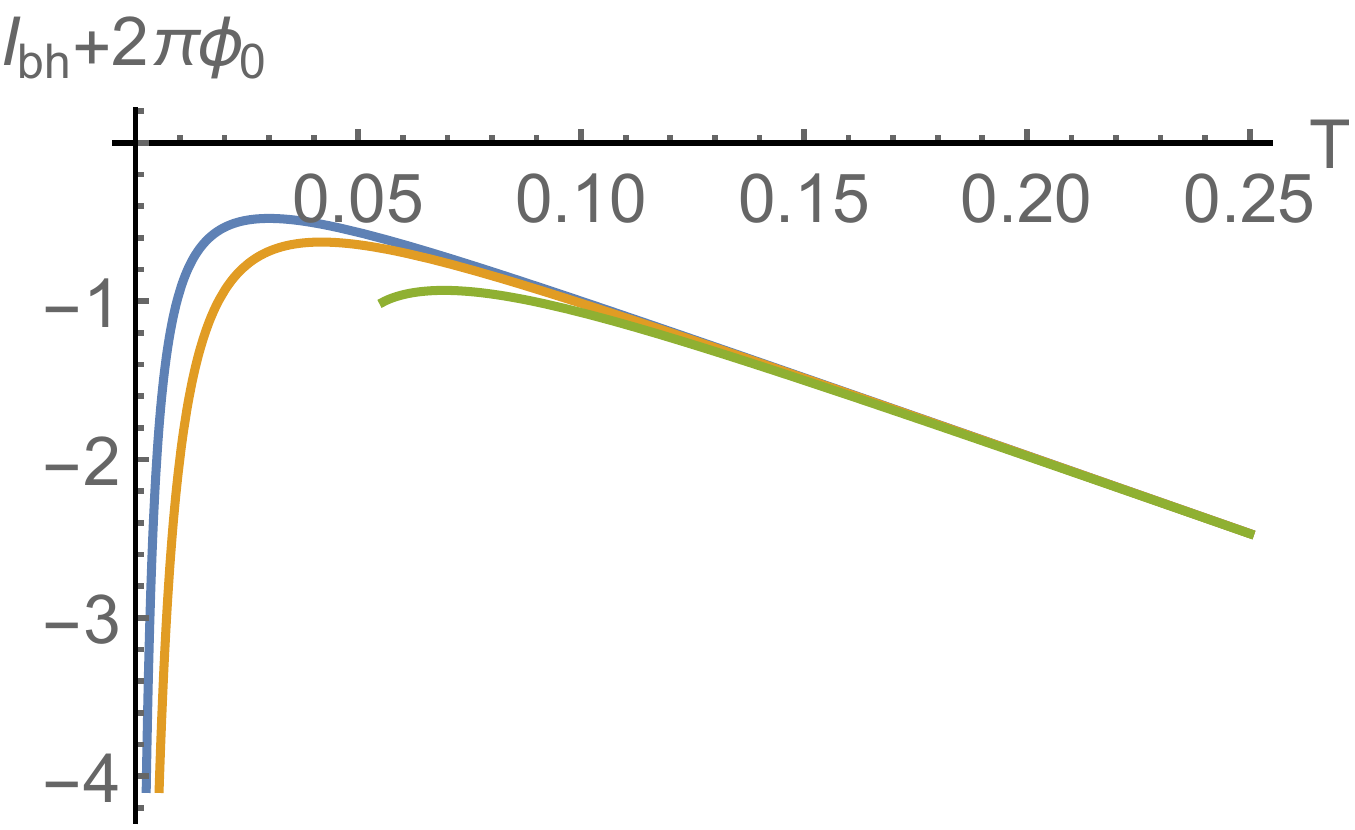}
\par\end{centering}
}\subfloat[$\zeta<\zeta_{cr}$]{\begin{centering}
\includegraphics[height=3cm]{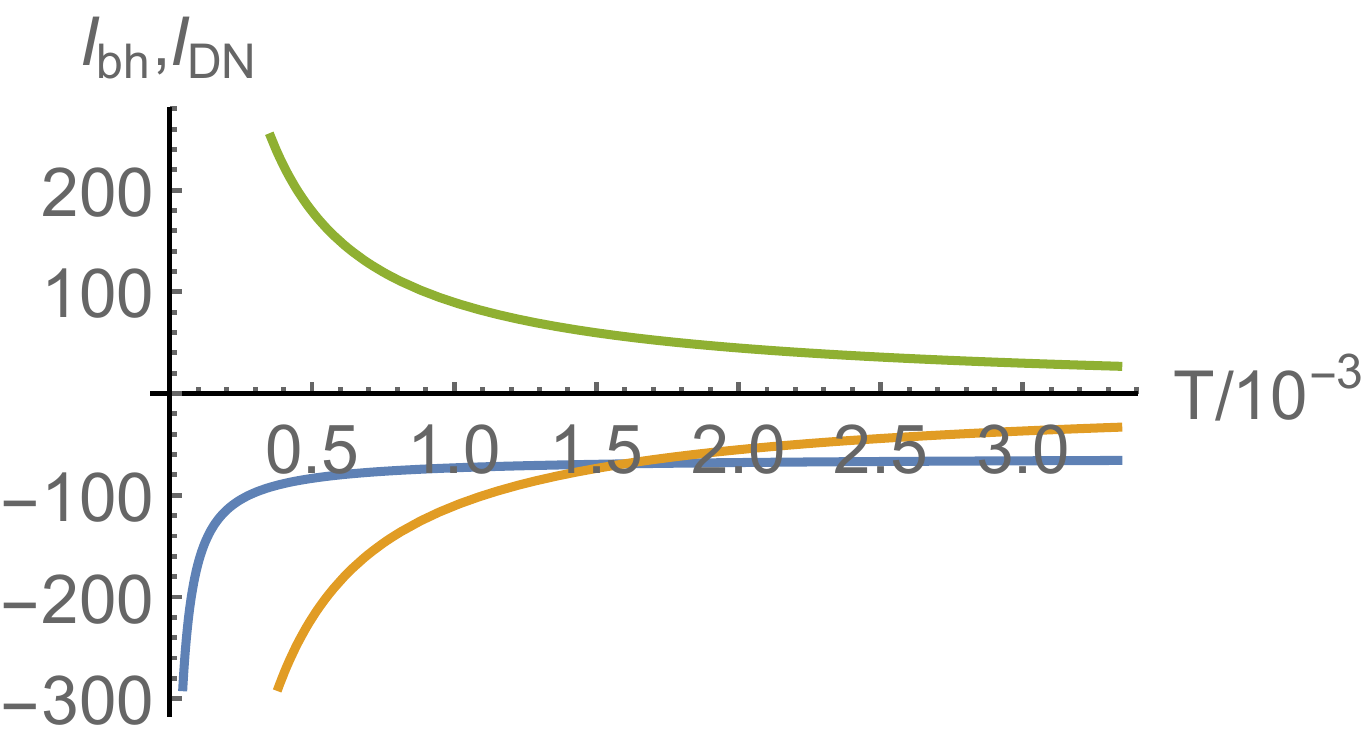}
\par\end{centering}
}\subfloat[$\zeta>\zeta_{cr}$]{\begin{centering}
\includegraphics[height=3cm]{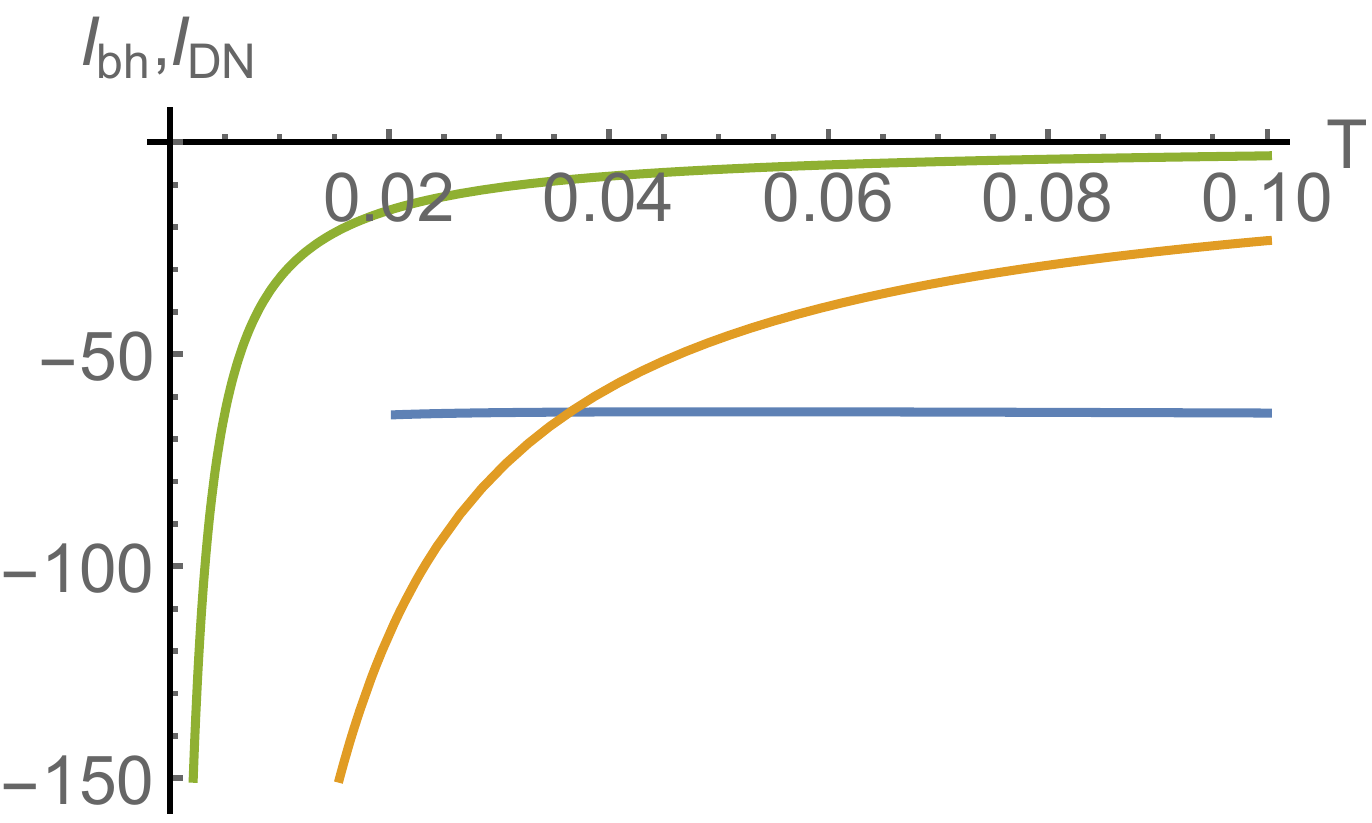}
\par\end{centering}
}
\par\end{centering}
\caption{(a) Plot of $I_{bh}+2\pi\phi_{0}$ as a function of $T$ for different $\zeta$, where blue is for $\zeta<\zeta_{cr}$, yellow is for $\zeta=\zeta_{cr}$, and green is for $\zeta>\zeta_{cr}$; (b) and
(c) are both plots of $I_{bh}$ and $I_{DN}$ as a function of $T$,
where blue curve is for $I_{bh}$, yellow curve is for $I_{DN}$ with
$E_{DN}<E_{DN,\max}$ and green is for $I_{DN}$ with $E_{DN}>E_{DN,\max}$.
We set $2\phi_{b}=1$ and $S_{0}=10$ in these plots. \label{fig:(a)-Plot-of}}
\end{figure}
The Hawking-Page phase transition is determined by dominance of $e^{-I_{DN}}$
versus $e^{-I_{bh}}$. A quick observation is that for any temperature,
for large enough $d$, the DN brane is always dominant. This
corresponds to very low energy eigenbranes. 

There is no principle
that determines $d$ \textit{a priori}, thus the Hawking-Page
phase transition is model dependent. However, the upper bound of eigenbrane
energies is for $d=0$ and we can compute a lower bound for the phase transition
temperature corresponding to this case. 

For $W(\Phi)$ potential (\ref{eq:wphi}), we have 
\begin{align}
I_{DN} & =\f{\b}{2\phi_{b}}\left(\phi_{DN}^{2}-\f{\zeta}{\pi}e^{-2\pi\phi_{DN}}\right)\\
I_{bh} & =\f{\b}{2\phi_{b}}\left(\phi_{h}(\b)^{2}-\f{\zeta}{\pi}e^{-2\pi\phi_{h}(\b)}\right)-2\pi(\phi_{h}(\b)+\phi_{0})
\end{align}
where $\phi_{h}(\b)$ is found by solving (\ref{eq:159}) 
\begin{equation}
\phi_{h}(\b)=\f{2\pi\phi_{b}}{\b}+\f 1{2\pi}\mW(-2\pi\zeta e^{-\f{4\pi^{2}\phi_{b}}{\b}})
\end{equation}
The plot of $I_{bh}+2\pi\phi_{0}$ as a function of $T$ is shown
in Fig. \ref{fig:16a} for a few different $\zeta$. For $\zeta\leq\zeta_{cr}$,
$I_{bh}+2\pi\phi_{0}$ has a maximum at finite temperature and tends
to negative infinity for both zero and infinite temperature. For $\zeta>\zeta_{cr}$,
$I_{bh}+2\pi\phi_{0}$ still tends to negative infinity for infinite
temperature but ends at a finite value at $T_*$. The temperature
dependence of $I_{DN}$ is very simple, namely inversely proportional
to $T$. 

For large $T$, it is clear that $I_{DN}>I_{bh}$ and black holes
dominate at high temperature. For small $T$, in order to have $I_{DN}<I_{bh}$,
we must require $\phi_{DN}^{2}-\f{\zeta}{\pi}e^{-2\pi\phi_{DN}}<\phi_{h}(\b)^{2}-\f{\zeta}{\pi}e^{-2\pi\phi_{h}(\b)}$
for $\b=1/T_*$ ($\zeta>\zeta_{cr}$) or $\b\ra+\infty$ ($\zeta\leq\zeta_{cr}$).
Taking the $T$ derivative of $I_{DN}-I_{bh}$, we have
\begin{equation}
\del_{T}(I_{DN}-I_{bh})=\f{8\pi^{2}\phi_{b}}{W(\phi_{h}(\b))^{2}}\left[\left(\phi_{h}(\b)^{2}-\f{\zeta}{\pi}e^{-2\pi\phi_{h}(\b)}\right)-\left(\phi_{DN}^{2}-\f{\zeta}{\pi}e^{-2\pi\phi_{DN}}\right)\right]>0
\end{equation}
because $\phi_{h}(\b)$ is a monotonically increasing function of $T$
in the allowed range of temperature. This means that there is only
one Hawking-Page phase transition point if it exists. In other words,
a necessary condition for a Hawking-Page phase transition is that the
energy of DN brane must be below the stable black hole spectrum.

For $\zeta\leq\zeta_{cr}$, the lowest energy of a stable black hole
is $E_{*,\pm}$ in (\ref{eq:estar-pm}) (it takes real values in this
case). For any $E_{DN}$ below it, the Hawking-Page phase transition
always exists because $I_{DN}-I_{bh}\leq\b(E_{DN}-E_{*,\pm})+2\pi(\phi_{h}(\b)+\phi_{0})<0$
for large enough $\beta$ since $\phi_{h}(\b)$ is an increasing but
bounded function of $\b$. In other words, the upper
bound of $E_{DN}$ is $E_{DN,\max}=E_{*,\pm}$ for $\zeta\leq\zeta_{cr}$.
For $\zeta>\zeta_{cr}$, the upper bound $E_{DN,\max}$
depends on the value of $\phi_{0}$ because Hawking-Page phase transition
must occur above $T_*$. To find it, we need to solve $I_{DN}=I_{bh}$
at $\b=1/T_*$. It turns out that
\begin{equation}
E_{DN,\max}=-\f 1{8\pi^{2}\phi_{b}}\left(1+(1+\log2\pi\zeta)^{2}+4\pi\phi_{0}(1+\log2\pi\zeta)\right)
\end{equation}
We plot with a few examples in Fig. \ref{fig:(a)-Plot-of} to show
that Hawking-Page phase transition occurs only when $E_{DN}<E_{DN,\max}$.

The lesson of above discussion is that including eigenbranes
with low enough energy in the black hole system is necessary to have a canonical
ensemble description at all temperatures. If we isolate the black
holes, zero temperature (defined by the periodicity of Euclidean time) can
only be achieved by complex geometries.
Therefore, it is natural to interpret these complex black holes as
metastable states that decay to eigenbranes in finite time.

On the matrix model side,
adding eigenbranes corresponds to fixing some eigenvalues $\lambda_{i}$ by
inserting $\sum_{i}\d(\lambda_{i}-\lambda_{DN,i})$ with a set of
fixed numbers $\lambda_{DN,i}$ in the matrix integral \cite{Blommaert:2019wfy}.
If we take $\lambda_{DN,i}$ to be real numbers to the left of the
support of spectrum, they will exert a right-pointing repulsive force
on all other eigenvalues, which can balance the
left-point force exerted by the potential of EOW branes in (\ref{subsec:Hawking-Page-phase-transition}).
It would be interesting to see how this would change the solution
of the  matrix model. In particular, we expect to see no complex part of
spectrum if we assign eigenbranes properly. This question is beyond
the scope of this paper and we will leave it for future works.

\section{Discussion and conclusion} \label{sec:disc}

\subsection*{Matrix model dual to JT gravity with deficit angles}
The matrix model dual to JT gravity with deficit angles has been studied in \cite{Witten:2020wvy, Maxfield:2020ale}. For a single type of deficit angle $2\pi-\t$ ($0\leq\theta<\pi$) with weight $\varepsilon$, the genus zero one-cut spectral density is
\be\label{eq:6.2}
\r(E)=\f{e^{S_{0}}\phi_{b}}{2\pi}\int_{2\phi_{b}E_{0}}^{2\phi_{b}E}\f{d\xi}{\sqrt{2\phi_{b}E-\xi}}\left(I_{0}(2\pi\sqrt{\xi})+\varepsilon \f{\t}{\sqrt{\xi}}I_1(\t\sqrt{\xi})\right)
\ee
where the zero-point energy $E_0$ is determined by the largest real solution to
\be
\sqrt{2\phi_b E_0}I_1(2\pi\sqrt{2\phi_b E_0})+2\pi \varepsilon I_0(\t \sqrt{2\phi_b E_0})=0 \label{eq:6.2norm}
\ee
If $\t\neq 0$, the second term in brackets of \eqref{eq:6.2} scales like $\sim e^{\t \sqrt{\xi}}\xi^{-3/4}$ in the large $\xi$ limit. This term leads to a spectral density that differs by an exponential amount at large $E$ from that of pure JT gravity, $\r_{JT}(E)$, in contrast to the behavior of our matrix model dual to EOW branes. In matrix model language, this corresponds to a huge left-pointing extra force $\d F_\Lambda$ that pushes a large number of eigenvalues into the double-scaled IR regime. In the following, we will study this matrix model by deriving the corresponding potential deformation $\d V$.

For JT gravity with deficit angles, the measure is formally an analytic continuation of a delta function, $\d (b-i\t)$. The way to define this analytic continuation in the matrix model is as follows. First, we assume the measure $\mM(b)$ to be $\d(b-b_0)$ for $b_0>0$, which leads to
\be\label{eq:defmalpha}
m(\a)=\f 1 {2\pi}\int_{-\infty}^{+\infty}dk \d(\a-(\d-ik)) e^{(\d-ik)b_0}
\ee
where $b_0>0$ and infinitesimal $\d>0$.\footnote{Here the complex delta function is a formal device that just means $\a$ is taken to a specific value when integrated over $\mD$. A mathematically rigorous formulation of $\delta(\a-\a_0)$ could be $\f{1}{2\pi i (\a-\a_0)}$ with $\mD$ defined as an anticlockwise closed contour encircling $\a_0$. However, this subtlety does not affect computations in this paper, and we will use the formal but simpler notation.} For any rational function $F(\a)$ that is integrated against $m(\a)$, adding a lower half plane infinite arc for $k$ does not change the result because $b_0>0$. Therefore, we can define the analytic continuation of $b_0$ based on a slightly different $m(\a)$ with this extra lower half plane infinite arc
\be\label{eq:fulldefmalpha}
m(\a)=\f 1 {2\pi}\oint_{\mC^-_\infty}dk \d(\a-(\d-ik)) e^{(\d-ik)b_0}
\ee
where $\mC^-_\infty$ is the clockwise contour consisting of the real axis and the lower half plane infinite arc. Using this definition, the integral of $F(\a)$ against $m(\a)$ becomes contour integrals of $k$ around the singularities of  $F(\d-i k)$ and thus $b_0$ can be safely continued to complex numbers. Taking this definition in \eqref{eq:102}, we have
\be
-\d V'(E)=-\sqrt{\f{2\phi_b}{E}}\sin\left(\sqrt{2\phi_b E}b_0\right) \label{eq:6.5}
\ee
For generic complex $b_0$, we will have a complex potential along the real axis. Putting \eqref{eq:fulldefmalpha} into the first line of \eqref{eq:32}, where we need to assign the contour of $\lambda$ at infinity inside of the contour of $-\a^2/(2\phi_b)$,\footnote{This is justified as one could imagine that in the non-double-scaled matrix model the right end of spectrum is finite and we take the double scaling limit afterward.} \eqref{eq:36} follows and using the first line of \eqref{eq:36-1}, we have (assuming $\xi>0$)
\begin{align}
\r(E)\supset& -\f{\varepsilon e^{S_0}\phi_b}{(2\pi)^2}\int_{2\phi_b E_0}^{2\phi_b E}\f {d\xi}{\sqrt{2\phi_b E-\xi}}\oint_{\mC^-_\infty}dk \f{e^{(\d-ik)b_0}}{((\d-ik)^2+\xi)^{3/2}}\nn\\
=& -\f{\varepsilon e^{S_0}\phi_b}{(2\pi)^2}\int_{2\phi_b E_0}^{2\phi_b E}\f {d\xi}{\sqrt{2\phi_b E-\xi}} \f{2 \pi b_0}{\sqrt{\xi}}J_1(b_0\sqrt{\xi}) \label{eq:6.4rho}
\end{align}
For $\xi<0$, we simply deform the contour of $k$ in the first line appropriately such that the second line holds. It is clear that \eqref{eq:6.4rho} matches with the second term in \eqref{eq:6.2} after continuing $b_0\ra i \t$. From \eqref{eq:6.5}, it is also clear that the potential exponentially grows for large $E$ for $b_0\ra i\t$, and this huge potential deformation explains the deviation of $\r$ from $\r_{JT}$ as mentioned below \eqref{eq:6.2norm}.

As discussed in Section \ref{sec:determinE0}, to find the equation determining the zero point energy $E_0$ in the matrix model, we need to input UV data of either a regulated measure or asymptotic large $E$ behavior of $\d\r(E)$. While it is unclear how to define a regulated measure for a deficit angle, figuring out required asymptotic large $E$ behavior is straightforward. Indeed, we can rewrite \eqref{eq:6.4rho} as
\begin{align}
\r_\varepsilon(E)& =\f{\varepsilon e^{S_0}\phi_b}{\pi}\int_{2\phi_b E_0}^{2\phi_b E}\f {d\xi}{\sqrt{2\phi_b E-\xi}} \del_\xi J_0(b_0\sqrt{\xi})\nn\\
& =\f{\varepsilon e^{S_0}\phi_b^{1/2}}{\sqrt{2E}\pi}(1- J_0(b_0\sqrt{2\phi_b E_0}))+\f{\varepsilon e^{S_0}\phi_b}{\pi}\int_{0}^{2\phi_b E}\f {d\xi}{\sqrt{2\phi_b E-\xi}} \del_\xi J_0(b_0\sqrt{\xi}) +o(E^{-1})
\end{align}
Comparing with the first term in \eqref{eq:3.40}, it obvious that we need to require that for $E\ra +\infty$
\be\label{eq:6.8asymp}
\d\r(E)\ra \f{\varepsilon e^{S_0}\phi_b}{\pi}\left[\int_{0}^{2\phi_b E}\f {d\xi}{\sqrt{2\phi_b E-\xi}} \del_\xi J_0(b_0\sqrt{\xi}) +\f 1 {\sqrt{2\phi_b E}}\right]+o(E^{-1})
\ee
which leads to
\be \label{eq:6.9norm}
\sqrt{2\phi_b E}I_1(2\pi\sqrt{2\phi_b E_0})+2\pi \varepsilon J_0( b_0 \sqrt{2\phi_b E_0})=0
\ee
Continuing $b_0\ra i \t$, it matches with \eqref{eq:6.2norm}. Note that for generic complex $b_0$, the zero point energy $E_0$ can take multiple complex values because \eqref{eq:6.9norm} has multiple solutions. We will pick the one smoothly connecting to the $E_0$ of $b=i \theta$ case, which is the largest real solution to \eqref{eq:6.2norm}. Given a complex $E_0$, the support of the spectrum $D_\r$ is defined by requiring $\r(E)dE$ to be a real nonnegative measure along $D_\r$ asymptotically extending to infinity. This ODE completely fixes the one-cut spectrum of the matrix models. Indeed, the above construction defines a class of matrix models labelled by the complex number $b_0$, which contains two special cases with geometric duals in deformed JT gravity: with deficit angles ($b_0= i\theta$) and with fixed length EOW branes ($b_0>0$).

At first glance, specifying asymptotic behavior \eqref{eq:6.8asymp} looks like a fine-tuning starting with the SSS matrix model because a deviation at any $E^p~(p>-1/2)$ order will strongly violate the zero point equation although it seems negligible compared to the exponentially large behavior at large $E$ for $\Im b_0\neq 0$. However, the SSS matrix model is equally  fine-tuned in this sense, since its spectral density also grows exponentially at large $E$ in double scaling limit.

The method used in \cite{Witten:2020wvy, Maxfield:2020ale} to solve JT gravity with deficit angles by summing over topologies only applies to one-cut case. Given the dual matrix model defined above, it would be straightforward to study the spectrum beyond the critical point in this model requiring the same asymptotic behavior \eqref{eq:6.8asymp} of the spectral density. With this definition, we expect to see a phase transition similar to our ``Y" shape because heavy EOW branes with $\lambda>0$ can be regarded as the special case $\t=0$ for deficit angles as discussed in Section \ref{subsec:-dilaton-JT}. Our method should thus offer a reasonable solution to the negative spectrum (and non-perturbative instability) problem in \cite{Johnson:2020lns, Witten:2020wvy, Maxfield:2020ale}.



\subsection*{Comparison with previous work}

Recently, a different non-pertubative completion for JT gravity and JT supergravity has been discussed \cite{Johnson:2020heh, Johnson:2020exp, Johnson:2020mwi, Johnson:2019eik} that in some cases avoids non-pertubative instability. In their method, the Hermitian matrices are promoted to squares of complex matrices and the spectrum is nonperturbatively bounded from below. Alternatively, one could use the original Hermitian matrix but introduce a hard wall in the potential. This completion was further generalized to JT gravity with deficit angles \cite{Johnson:2020lns}. 

There, all eigenvalues are still integrated along the real axis but the matrix model potential is modified (in particular, by fixing a specific asymptotic behavior for the function $u(x)$ that appears in the string equation as $x\ra \pm\infty$). The computational method uses an auxiliary Schr\"{o}dinger equation that is associated with the orthogonal polynomial formalism of double-scaled matrix models with even potentials \cite{Banks:1989df}. The standard orthogonal polynomial technique only applies to the case where all eigenvalues are on the same integration contour. Generalizing to the case of multiple distinct contours may require multi-matrix techniques, in which the  coupling among eigenvalues on different contours from the Vandermonde could be treated as interactions among different matrices.

Our method to nonperturbatively define a double-scaled matrix model starts with the observation that in a non-double-scaled matrix model with analytic potential, the tree-level spectral density plus a choice of contours for the eigenvalues completely specify the model. The double-scaled matrix models in this paper are  defined such that this property still holds.\footnote{This property rules out the existence of a hard wall at $E_0$. It is important to distinguish between double-scaled and non-double-scaled models here because a double-scaled model with a hard wall may be obtained as a limit of non-double-scaled models with analytic potentials. One could create a hard wall in the double-scaling limit by adding a $e^{- \Lambda (E - E_0)}$ term to the potential, where $\Lambda \rightarrow \infty$ in the double-scaling limit.}
We promote the Hermitian matrices to normal matrices along some contour for each eigenvalue, and the effective potential is defined everywhere by analytically continuing the spectral curve. 
The nonperturbative definition is thus supplied by the spectral curve and the choice of eigenvalue contours. This is along the same lines of \cite{Saad:2019lba}. 

More importantly, one of our results is this completion naturally applies to the multi-cut case and avoids the non-perturbative instability of \cite{Johnson:2020lns}. Our completion allows order $e^{S_0}$ eigenvalues to be complex  in the ``Y" shaped phase, and it seems impossible to interpret each matrix in the ensemble as an instantiation of a Hermitian Hamiltonian. As argued in Section \ref{sec:5}, we can interpret the complex eigenvalues as metastable black hole states that decay in finite time. Indeed, this implies that JT gravity with EOW branes is not a complete theory and one must include other objects with lower energy, {\it e.g.} DN branes, to have an unitary theory with Hermitian Hamiltonian.

On the other hand, the formalism of \cite{Johnson:2020heh, Johnson:2020exp, Johnson:2020mwi, Johnson:2019eik}  can easily compute higher genus contributions. We only considered the tree-level (genus zero) spectral density that solves the saddle equation for the matrix model potential. It would be interesting to study higher genus contribution to our matrix model in future works.

\begin{figure}
\begin{centering}
\subfloat[\label{fig:17a}]{\begin{centering}
\includegraphics[height=4cm]{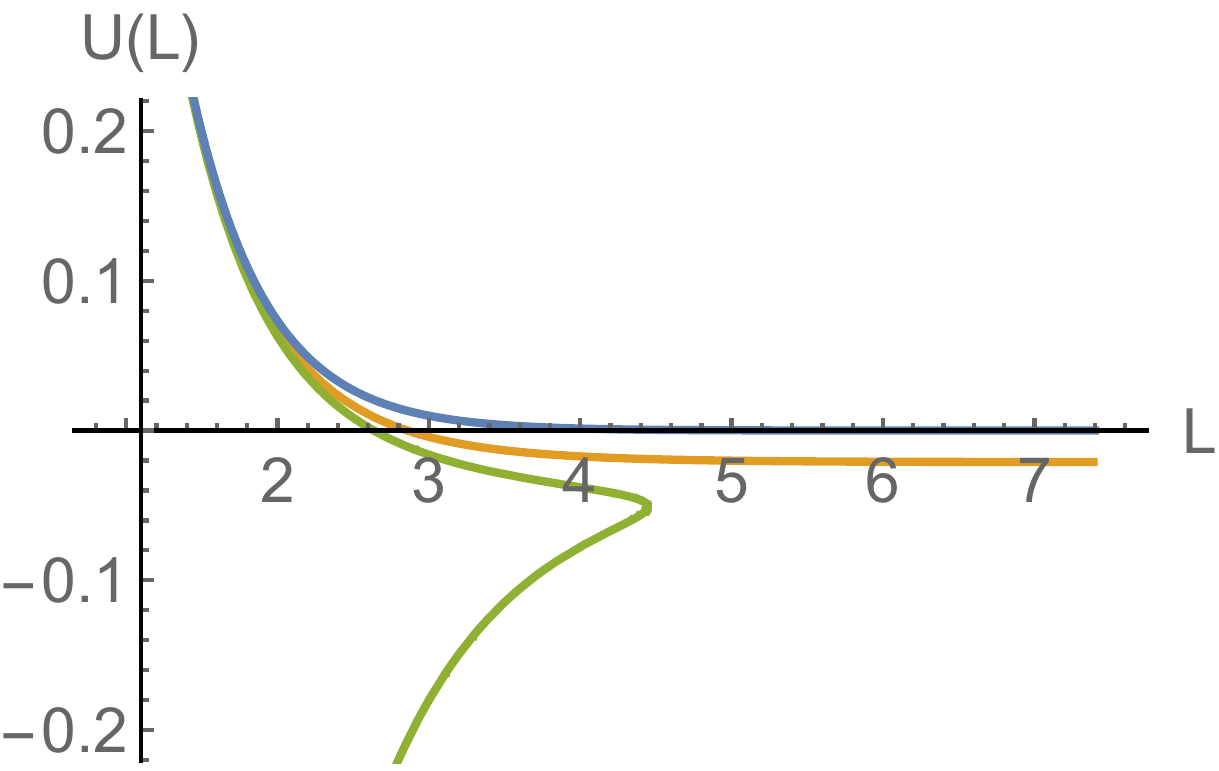}
\par\end{centering}
}\subfloat[\label{fig:17b}]{\begin{centering}
\includegraphics[height=4cm]{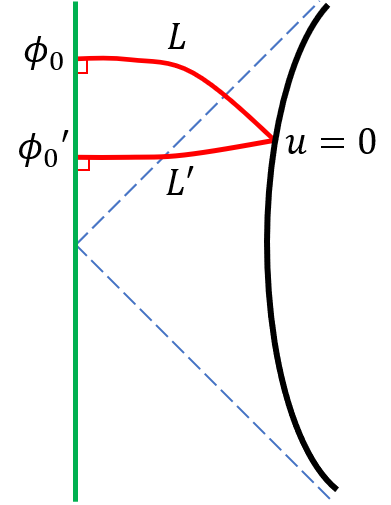}
\par\end{centering}
}
\par\end{centering}
\caption{(a) $U(L)$ for different $\zeta$, where blue is for $\zeta=0$, yellow is for $0<\zeta<\zeta_{cr}$ and green is for $\zeta>\zeta_{cr}$. (b) $\zeta>\zeta_{cr}$. Black curve is AdS boundary, green line is $\mu=0$ EOW brane, dashed lines are horizon and red curves are two geodesics. In this case, there is an ambiguity of choice of geodesics emanating from the same boundary point at $u=0$. In both plots, $2\phi_b=1$ \label{fig:17}}
\end{figure}

\subsection*{Phase space of $W(\Phi)$ dilaton gravity}

Given that heavy EOW branes with $\lambda>0$ lead to an effective $W(\Phi)$ dilaton gravity, it is also interesting to study the phase space for this theory along the lines of Section \ref{sec:2} and \cite{Harlow:2018tqv}. Let us consider the case of one-sided black holes in $W(\Phi)$ dilaton JT gravity with one AdS boundary and one $\mu=0$ EOW brane boundary, for simplicity. This is equivalent to half of a single sided black hole. The Hamiltonian is given by \eqref{eq:admenergy}. Using the solution from \cite{PhysRevD.49.5227}, one can show that the regularized geodesic length connecting the AdS boundary at $u=0$ to the EOW brane normally is given by
\be\label{eq:6.4}
L=\int_{\phi_0}^{r_\infty}\f{d\phi}{\sqrt{\int_{\phi_0}^{\phi}W(r)dr}}-\log r_\infty
\ee
where $\phi_0$ is the dilaton value at the end of geodesic on the EOW brane, which depends on the location of the other end at the AdS boundary in a complicated way. After some algebra, one can show that the Hamiltonian can be written in a canonical form in terms of $L$ and its conjugate momentum $P$ as
\be
H= \f {1}{2\phi_b}\left(P^2+r_\infty^2-\int_{\phi_0(L)}^{r_\infty}W(r)dr\right)\equiv \f{P^2}{2\phi_b}+U(L)
\ee
where $\phi_0(L)$ is defined implicitly by \eqref{eq:6.4}. One can check that for $W(r)=2r$, it reduces to the pure JT result \eqref{eq:2.37} with $\mu=0$.

For $W(\Phi)$ given by \eqref{eq:wphi}, we plot the potential in Fig. \ref{fig:17a} for different $\zeta$. It turns out that for $0<\zeta<\zeta_{cr}$, the potential is similar to Liouville potential ($\zeta=0$) but with a negative shift of the ground energy. For $\zeta>\zeta_{cr}$, the potential is double valued for some $L$ and thus ill-defined. Also, the potential is unbounded from below and the allowed range of $L$ is finite. This matches with the expectation from Euclidean computations \eqref{eq:156} that the spectrum is unbounded from below for $\zeta>\zeta_{cr}$ because $\phi_h$ can take any real value.

In the Lorentzian picture, this ill-defined potential reflects the fact that $L$ is no long an appropriate phase space variable because of the ambiguity of two geodesics emanating from same AdS boundary point but ending on two different points on the $\mu=0$ EOW brane normally (see Fig. \ref{fig:17b}). In other words, using $L$ as a radial coordinate is no longer a good gauge for the metric. It would be interesting to find an appropriate gauge in this case and see how the quantization leads to results consistent with the semiclassical analysis of Section \ref{sec:5}.

\subsection*{Modified inner product from geodesic separable objects}

\begin{figure}
\begin{centering}
\subfloat[\label{fig:18a}]{\begin{centering}
\includegraphics[height=1.5cm]{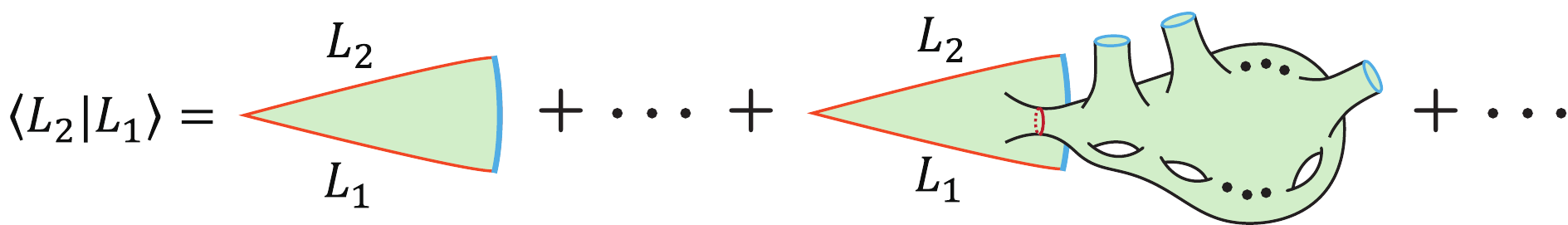}
\par\end{centering}
}\\
\subfloat[\label{fig:18b}]{\begin{centering}
\includegraphics[height=1.65cm]{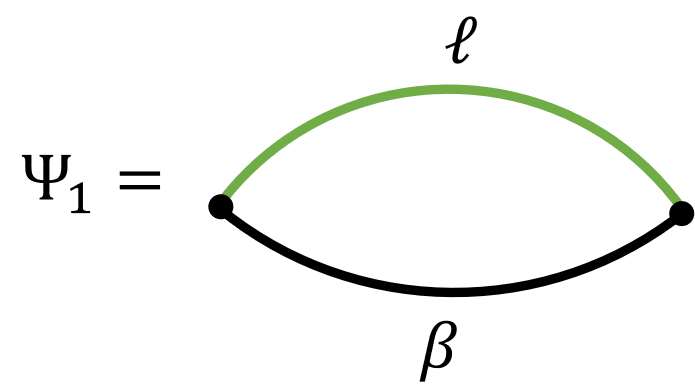}
\par\end{centering}
}\subfloat[\label{fig:18c}]{\begin{centering}
\includegraphics[height=3.3cm]{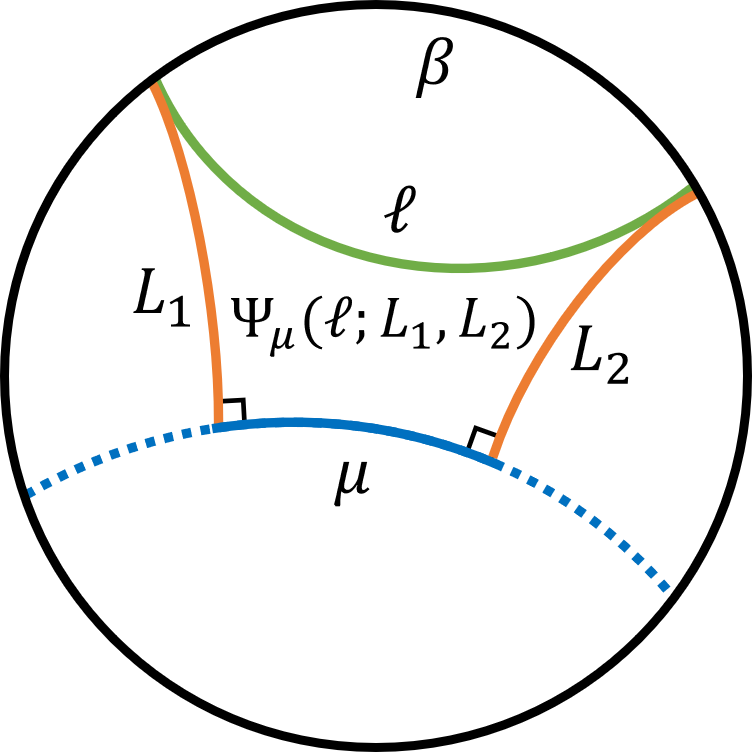}
\par\end{centering}
}\subfloat[\label{fig:18d}]{\begin{centering}
\includegraphics[height=1.7cm]{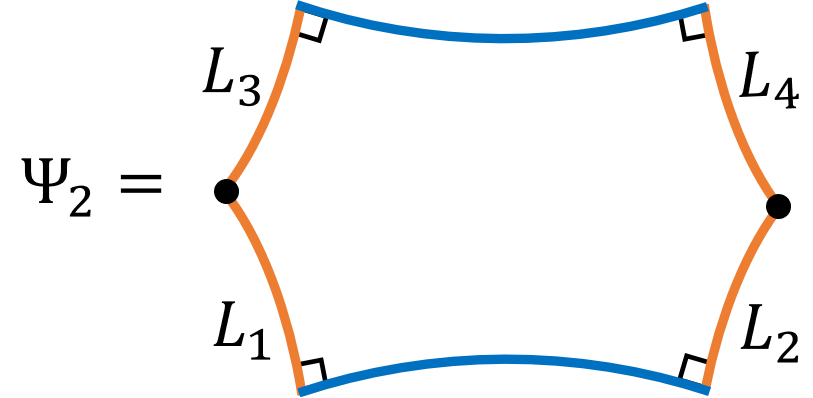}
\par\end{centering}
}\subfloat[\label{fig:18e}]{\begin{centering}
\includegraphics[height=1.65cm]{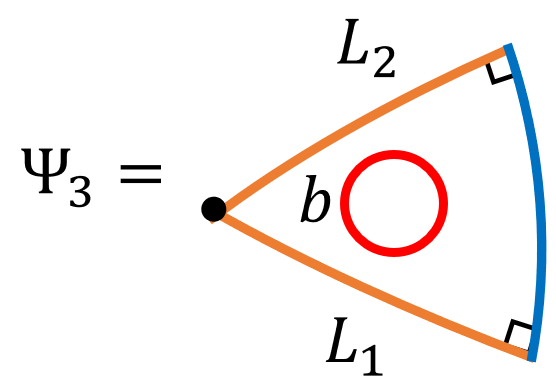}
\par\end{centering}
}
\par\end{centering}
\caption{(a) Modified inner product of $\avg{L_2|L_1}$ by including handles and EOW brane loops in Euclidean path integral. Blue curve is EOW branes and orange curve is geodesic connecting AdS boundary and EOW brane. The red loop is a closed geodesic which bounds all nontrivial topologies. (b) to (e) are a few geometric building blocks to compute $\avg{L_2|L_1}$, where black curves are AdS boundaries, green curves are geodesics connecting two points on AdS boundary, and red/blue/orange curves represent the same objects as (a). (b) $\Psi_1$ is the wavefunction of the Hartle-Hawking state of two-sided AdS system in JT gravity. (c) $\Psi_\mu(\l;L_1,L_2)$ is a propagator $G_{\mu,\b}(L_2,L_1)$ with a $\Psi_1$ pinched off. (d) $\Psi_2$ is gluing of two $\Psi_\mu$ along $l$. (e) $\Psi_3$ is gluing of $L_3$ with $L_4$ in $\Psi_2$ with one EOW brane measure $\mM(b)$ pinched off. Topological contribution to modified inner product $\avg{L_2|L_1}$ is $\Psi_3$ gluing with all other geometric objects (handles and EOW brane loops) along the geodesic $b$. \label{fig:18}}
\end{figure}

As formulated in Section \ref{sec:2}, the Hilbert space for gravity in spacetimes with one AdS boundary and one EOW brane can be found by quantizing in the regularized geodesic length $\bra{L}$ basis. Canonical quantization requires orthogonality of this basis $\avg{L_2|L_1}=\d(L_1-L_2)$. Such Lorentzian analysis is based on a fixed spacetime topology whereas full quantum gravity allows for contributions from all topologies. To take this into account, we can define a modified inner product between $\bra{L}$ states using Euclidean path integrals including handles and loops of EOW branes between the two geodesic slices as shown in Fig. \ref{fig:18a}.

Similarly to the gluing of partition function in Fig. \ref{fig:The-zero-genus}, the idea is to find the geometric building block $\Psi_3$ in Fig. \ref{fig:18e} that contains a geodesic loop with length $b$ between two geodesics $L_1$ and $L_2$ that connect the AdS boundary with the EOW brane. The contribution of other topologies to modified inner product $\avg{L_2|L_1}$ is obtained from $\Psi_3$ by gluing with all possible geometric objects (handles and EOW brane loops) along the geodesic loop of length $b$. In other words, $\Psi_3$ plays a role similar to the trumpet in the computation of partition functions. 

To evaluate $\Psi_3$, we need some gluing and pinch-off surgery of the Euclidean propagator  $G_{\mu,\b}(L_2,L_1)$  in \eqref{eq:11}. Diagrammatically in Fig. \ref{fig:18c}, $G_{\mu,\b}(L_2,L_1)$ is the same as the Euclidean path integral over the region bounded by an AdS boundary arc with length $\b$ (black curve), two geodesics $L_1$ and $L_2$ (orange curves) and an EOW brane (blue curve). We will first pinch off the region ($\Psi_1$ in Fig. \ref{fig:18b}) bounded by the AdS boundary arc and the geodesic $\l$ connecting its two ends (green curve). Indeed, $\Psi_1$ is the Hartle-Hawking state in the two-sided AdS system in JT gravity \cite{Jafferis:2019wkd}
\be
\Psi_1(\b,\l)=\f{e^{S_{0}/2}}{4\pi^{1/2}}\int_0^{\infty} dke^{-\b k^{2}/(8\phi_{b})}\r_0(k)K_{ik}(y)
\ee
where we defined 
\be
\rho_0(k)\equiv\f{2k\sinh\pi k}{\pi},~~y\equiv 4e^{-\l/2}
\ee
Using normalization of Bessel function
\be
\int_{-\infty}^{+\infty}d\l K_{ik}(y)K_{ik'}(y)=\d(k-k')/\rho_0(k)
\ee
one can show that gluing two Hartle-Hawking states along geodesic $\l$ leads to disk partition function
\be
Z_{\text{disk}}(\b_{1}+\b_{2})=\int_{-\infty}^{+\infty} d\l\Psi_1(\b_{1},\l)\Psi_1(\b_{2},\l)
\ee
Pinching off $\Psi_1$ from $G_{\mu,\b}$ leads to a piece $\Psi_\mu(\l;L_1,L_2)$ bounded by four geodesics as labelled in Fig. \ref{fig:18c}. It can be written as
\be
\Psi_{\mu}(\l;L_{1},L_{2})=4\pi^{1/2}e^{-S_{0}/2}\int_0^\infty dk\p(k)(z_{1}z_{2})^{-1/2}W_{-\mu,ik}(z_{1})W_{-\mu,ik}(z_{2})K_{2ik}(y)
\ee
where $z\equiv 4 e^{-L}$ and $\p(k)$ is given in \eqref{eq:63}, because one can check easily that
\be
\int_{-\infty}^{+\infty}d\l \Psi_{\mu}(\l;L_{1},L_{2}) \Psi_1(\b,\l)=G_{\mu,\b}(L_2,L_1)
\ee
Then, we will glue two $\Psi_\mu$ along the geodesic $\l$ to get $\Psi_2$ in Fig. \ref{fig:18d}
\begin{align}
\Psi_{2}(L_1,L_2;L_3,L_4)&=\int_{-\infty}^{+\infty}d\l \Psi_{\mu}(\l;L_{1},L_{2})\Psi_{\mu}(\l;L_{3},L_{4})\nn\\
&=\f 2{\pi^{2}}e^{-S_{0}}\int_0^\infty dk|\G(\f 12+\mu-ik)|^{4} k\sinh2\pi k\prod_{s=1,2,3,4}z_{s}^{-1/2}W_{-\mu,ik}(z_{s})
\end{align}
Lastly, we glue $L_3$ and $L_4$ of $\Psi_2$, which forms an EOW brane loop
\be\label{eq:6.19}
\int_0^\infty db\Psi_{3}(L_{1},L_{2},b)\mM(b)=\int_{-\infty}^{+\infty} dL\Psi_{2}(L_1,L_2;L,L),
\ee
where $\Psi_3$ is depicted in Fig. \ref{fig:18e}, and pinch off the EOW brane measure $\mM(b)$ on that loop to get
\be\label{eq:phi3}
\Psi_{3}(L_1,L_2,b)=8e^{-S_{0}}\int_0^\infty dk\f{\cos kb}{\rho_0(2k)}\Psi_k(z_{1})\Psi_k(z_{2})
\ee
where $\Psi_k(z)$ is the normalized eigen function from \eqref{eq:63}, and we used the technique from \eqref{eq:11} to \eqref{eq:50} to separate $\mM(b)$ from RHS of \eqref{eq:6.19}.

It follows that the modified inner product in Fig. \ref{fig:18a} is
\be\label{eq:6.21}
\avg{L_2|L_1}=\d (L_1-L_2)+8e^{-S_0}\int_0^\infty db X(b)\int_0^\infty dk  \f{\cos kb}{\rho_0(2k)} \Psi_k(z_1)\Psi_k(z_2) 
\ee
where $X(b)$ represents all topologies of handles and EOW brane loops that are separated by the geodesic loop of length $b$. Indeed, \eqref{eq:6.21} can be applied to any geometric objects that are separable by a geodesic loop as they are all accountable by some $X(b)$. We can rewrite $\cos k b$ as the inverse Laplace transformation of the trumpet using \eqref{eq:48}. Note that the partition function can be written in a similar form as
\be
Z(\b)=Z_{\text{disk}}(\b)+\int_0^\infty db Z_{\text{trumpet}}(\b,b) X(b)
\ee
It follows that
\begin{align}
\avg{L_2|L_1}&=\d (L_1-L_2)+8e^{-S_0}\int_0^\infty db X(b)\int_0^\infty dk  k \int_{\e-i\infty}^{\e+i\infty}d\b \f{Z_{\trumpet}(\b,b)e^{\b k^2/(2\phi_b)}}{2\phi_b i\rho_0(2k)} \Psi_k(z_1)\Psi_k(z_2) \nn\\
&=\d (L_1-L_2)+8e^{-S_0}\int_0^\infty dk  k \int_{\e-i\infty}^{\e+i\infty}d\b \f{(Z(\b)-Z_{\text{disk}}(\b))e^{\b k^2/(2\phi_b)}}{2\phi_b i\rho_0(2k)} \Psi_k(z_1)\Psi_k(z_2)\nn\\
&=\int_0^\infty dk \f{\rho(k^2/2\phi_b)}{\rho_{JT}(k^2/2\phi_b)}\Psi_k(z_1)\Psi_k(z_2)
\end{align}
where in the second line we assumed that the $b$ integral is interchangeable with the other two, and in the last line $\rho(E)$ is the full spectral density for $E=k^2/2\phi_b$ and inverse Laplace transformed from the partition function $Z(\b)$. If $\rho=\rho_{JT}$, we get back to $\avg{L_2|L_1}=\d (L_1-L_2)$ as expected.

This result shows that the modified inner product among $\bra{L}$ basis can still be diagonalized in terms of the positive energy basis of the Morse potential quantum mechanics even after taking all nontrivial topologies into account. In particular, the contribution from each energy is proportional to its spectral density. 

On the other hand, if the full spectral density allows negative energy, like the tree-level spectrum with $\lambda>0$ in this paper, the states $\bra{L}$ are blind to it. Indeed, if we naively extend $\Psi_k(z)$ to imaginary $k$, this wave function becomes non-normalizable. 

Importantly, if the full spectral density is discrete \cite{Blommaert:2019wfy}, $\bra{L}$ becomes overcomplete and there are infinitely many null states, namely the energy eigenstates that are not on the support of the spectrum. It would be interesting to understand what geometric objects are required to have such discrete spectra and how they could lead to a unique $\a$ state of baby universes in such a modified JT gravity, in the language similar to pure topological 2D gravity \cite{Marolf:2020xie}.

\section*{Acknowledgements} We would like to thank Daniel Harlow, Hong Liu, Baur Mukhametzhanov, Julian Sonner, and Douglas Stanford for 
stimulating and helpful discussions. The work of DLJ and DK was supported in
part by the US Department of Energy grant DE-SC0021528. PG is supported by the US Department of Energy grants DE-SC0018944 and DE-SC0019127, and also the Simons foundation as a member of the {\it It from Qubit} collaboration.

\appendix

\section{Variation of JT action\label{sec:Variation-of-JT}}

Here we review the variation of the JT action. The spacetime $M$ has the topology
of a strip with two timelike boundaries, $\del M$. The outward
normal unit vector on $\del M$ is $n^{a}$. The metric can be decomposed
as $g_{ab}=n_{a}n_{b}+h_{ab}$ and $g^{ab}=n^{a}n^{b}+h^{ab}$. As
we are in two dimensions, $h_{ab}$ can be further
written as $-t_{a}t_{b}$, where $t^{a}$ is a tangent unit vector on
$\del M$. The volume form on $M$ (resp. $\del M$) is $\e$ (resp. $\hat{\e}$). 
We define $\d g_{ab}\equiv\d(g_{ab})$ , $\d g^{ab}\equiv-\d(g^{ab})$
and $\d g\equiv g_{ab}\d g^{ab}=g^{ab}\d g_{ab}$. This unconventional notation is designed so that $\delta g^{ab} = g^{ac} g^{bd} \delta g_{c d}$.

The action is as follows
\begin{equation}
I=I_{M}+I_{\del M}=\int_{M}\f 12(R+2)\Phi\e+\int_{\del M}(\nabla_{a}n^{a}\Phi-C\Phi-D)\hat{\e},
\end{equation}
where $C$ and $D$ are constants. The variation of the bulk action is
\begin{align}
\d I_{M}= & \f 12\int_{M}\left[\d g^{ab}(\nabla_{a}\nabla_{b}\Phi-g_{ab}\nabla^{2}\Phi)+(R+2)\d\Phi-(R_{ab}-(\f 12R+1)g_{ab})\d g^{ab}\Phi\right]\e\nonumber \\
 & +\f 12\int_{\del M}\left[(\nabla_{a}\d g^{ac}-\nabla^{c}\d g)\Phi\e_{c}-\d g^{ab}\nabla_{b}\Phi\e_{a}+\d g\nabla_{a}\Phi\e^{a}\right],
\end{align}
where $\e_{a}=n_{a}\hat{\e}$, and we used Stokes' theorem twice to
obtain the boundary terms. In two dimensions, any metric is locally conformally
flat and thus the Einstein tensor $R_{ab}-\f 12Rg_{ab}=0$ trivially.

Next, we consider the variation of the boundary action. Let the boundary be the locus that obeys $f(x)=\const$ for some smooth function $f(x)$. Then the normal
1-form is
\begin{equation}
n_{a}=\f{\del_{a}f}{\sqrt{\del_{a}f\del_{b}fg^{ab}}}.
\end{equation}
The variation of the normal 1-form becomes
\begin{equation}
\d n_{a}=\f 12n_{a}n_{b}n_{c}\d g^{bc}\implies\d n^{a}=-\f 12(h_{\phantom{a}c}^{a}+\d_{c}^{a})n_{b}\d g^{cb}.
\end{equation}
Furthermore, note that 
\begin{align}
\d\G_{ab}^{a} & =\f 12\left(\nabla_{b}\d g+g_{db}\nabla_{a}\d g^{da}-\nabla^{a}\d g_{ab}\right),\\
\d\e_{a} & =\f 12\e_{a}\d g\implies\d\hat{\e}=\d(n^{a}\e_{a})=\f 12h_{ab}\d g^{ab}\hat{\e}.
\end{align}
The variation of the boundary action becomes
\begin{align}
\label{eq:boundaryactionvariation}
\d I_{\del M}= & -\f 12\int_{\del M}\left[h_{\phantom{a}c}^{a}\nabla_{a}(n_{b}\d g^{db}h_{\phantom{a}d}^{c})+\left(\nabla_{b}\d g^{ab}-\nabla^{a}\d g\right)n_{a}-(\nabla_{a}n^{a}h_{bc}-\nabla_{b}n_{c}+n^{a}\nabla_{a}n_{c}n_{b})\d g^{bc}\right]\Phi\hat{\e}\nonumber \\
 & +\int_{\del M}\left[(\nabla_{a}n^{a}-C)\d\Phi-\f 12(C\Phi+D)h_{ab}\d g^{ab}\right]\hat{\e}.
\end{align}
Using the fact that 
\begin{equation}
h_{\phantom{a}b}^{a}\nabla_{a}V^{b}=\hat{\nabla}_{a}V^{a},\quad\text{if }V^{a}n_{a}=0
\end{equation}
where $\hat{\nabla}_{a}$ is the covariant derivative constructed
from the induced metric on $\del M$, we can write the first term in \eqref{eq:boundaryactionvariation} as
\begin{equation}
\f 12\int_{\del M}(n_{b}\d g^{db}h_{\phantom{a}d}^{c})\hat{\nabla}_{c}\Phi\hat{\e}=\f 12\int_{\del M}(n_{b}\d g^{db}h_{\phantom{a}d}^{c})\del_{c}\Phi\hat{\e}.
\end{equation}
The third term in $\delta I_{\del M}$ is nontrivial in general dimensions
but vanishes in 2d. This can be easily seen using $h_{ab}=-t_{a}t_{b}$.
Putting the bulk and boundary variations together, we have
\begin{align}
\d I= & \f 12\int_{M}\left[\d g^{ab}(\nabla_{a}\nabla_{b}\Phi-g_{ab}\nabla^{2}\Phi+g_{ab}\Phi)+(R+2)\d\Phi\right]\e\nonumber \\
 & +\f 12\int_{\del M}\left[n_{b}h_{\phantom{a}a}^{c}\del_{c}\Phi-n_{a}\del_{b}\Phi+g_{ab}n^{c}\del_{c}\Phi-(C\Phi+D)h_{ab}\right]\d g^{ab}\hat{\e}+\int_{\del M}(\nabla_{a}n^{a}-C)\d\Phi\hat{\e}\nonumber \\
= & \f 12\int_{M}\left[\d g^{ab}(\nabla_{a}\nabla_{b}\Phi-g_{ab}\nabla^{2}\Phi+g_{ab}\Phi)+(R+2)\d\Phi\right]\e\nonumber \\
 & +\f 12\int_{\del M}\left[n^{c}\del_{c}\Phi-C\Phi-D\right]h_{ab}\d g^{ab}\hat{\e}+\int_{\del M}(\nabla_{a}n^{a}-C)\d\Phi\hat{\e}\label{eq:41}
\end{align}

The boundary condition for a well defined phase space must be one of the
following two types
\begin{equation}
\begin{cases}
n^{c}\del_{c}\Phi=C\Phi+D & \text{or}\quad h^{ab}\d g_{ab}=0\\
\nabla_{a}n^{a}=C & \text{or}\quad\d\Phi=0
\end{cases}
\end{equation}

\section{Lagrange inversion theorem\label{sec:Lagrange-inversion-theorem}}

Consider two smooth functions $g(x)$ and $f(x)$ that satisfy $f(0)=g(0)=0$ and $f(g(x))=x$. The Lagrange inversion
theorem is a formula for the Taylor series coefficients of $g(x)$ in terms of the Taylor series coefficients of its inverse, $f(x)$. The formula is given by
\begin{equation}
g(y)=\sum_{n=1}^{\infty}\f {y^{n}}{n!}\left.\del_{x}^{n-1}\left(\f{x^{n}}{f(x)^{n}}\right)\right|_{x=0}.
\end{equation}
This formula can be slightly generalized to compute $h(g(x))$ for any
smooth function $h$ defined in a neighborhood of $0$. The result is
\begin{equation}
h(g(y))=h(0)+\sum_{n=1}^{\infty}\f {y^{n}}{n!}\left.\del_{x}^{n-1}\left(\f{x^{n}h'(x)}{f(x)^{n}}\right)\right|_{x=0}
\end{equation}
We could choose $f(x)=x/\phi(x)$ for $\phi(0)\neq0$, which leads
to
\begin{equation}
h(g(y))=h(0)+\sum_{n=1}^{\infty}\f 1{n!}\left.\del_{x}^{n-1}(h'(x)\phi(x)^{n})\right|_{x=0}y^{n}
\end{equation}
Let $f_{\lambda}(x)$, $g_{\lambda}(x)$ be a one-parameter family of pairs of functions obeying the same properties as $f(x)$ and $g(x)$ above. Let $f_{\lambda}(x)=x/\phi(x+\lambda)$.
We have
\begin{equation}
h(\lambda+g_{\lambda}(y))=h(\lambda)+\sum_{n=1}^{\infty}\f 1{n!}\del_{\lambda}^{n-1}(h'(\lambda)\phi(\lambda)^{n})y^{n}.
\end{equation}
We may define $g_{\lambda}$ implicitly in terms of $\phi$ as follows:
\begin{equation}
g_{\lambda}(x)=x \, \phi(g_{\lambda}(x)+\lambda).
\end{equation}
We now write
\begin{equation}
h(\lambda+g_{\lambda}(y))=h(\lambda)+\sum_{n=1}^{\infty}\f 1{n!}\del_{\lambda}^{n-1}(h'(\lambda)\Phi(\lambda,y)^{n}),
\end{equation}
where
\begin{equation}
g_{\lambda}(y)=\Phi(g_{\lambda}(y)+\lambda,y),\qquad\Phi(\lambda,y)=y\phi(\lambda).
\end{equation}
Consider a family of variables $y_i$ and functions $\phi_i(x)$, indexed by $i$. We can generalize the above as follows:
\begin{equation}
h(\lambda+g_{\lambda}(\vec{y}))=h(\lambda)+\sum_{n=1}^{\infty}\f 1{n!}\del_{\lambda}^{n-1}(h'(\lambda)\Phi(\lambda,\vec{y})^{n})\label{eq:19}
\end{equation}
where
\begin{equation}
g_{\lambda}(\vec{y})=\Phi(g_{\lambda}(\vec{y})+\lambda,\vec{y}),\qquad\Phi(\lambda,\vec{y})=\sum_{i=1}^{K}y_{i}\phi_{i}(\lambda).
\end{equation}

If we integrate $\lambda$ in (\ref{eq:19}) from $a$ to $b$, we get
\begin{equation}
\int_{a}^{b}d\lambda h(\lambda+g_{\lambda}(\vec{y}))=\int_{a}^{b}d\lambda h(\lambda)+\int_{a}^{b}d\lambda h'(\lambda)\Phi(\lambda,\vec{y})+\sum_{n=2}^{\infty}\f 1{n!}\left.\del_{\lambda}^{n-2}(h'(\lambda)\Phi(\lambda,\vec{y})^{n})\right|_{\lambda=a}^{\lambda=b}.
\end{equation}
If there exists an $a$ such that $\del_{a}^{n-2}(h'(a)\Phi(a,\vec{y})^{n})=0$
for all $n$, we may derive a new series
\begin{equation}
\sum_{n=2}^{\infty}\f 1{n!}\del_{b}^{n-2}(h'(b)\Phi(b,\vec{y})^{n})=\int_{a}^{b}d\lambda\left[h(\lambda+g_{\lambda}(\vec{y}))-h(\lambda)-h'(\lambda)\Phi(\lambda,\vec{y})\right]\label{eq:22-1}
\end{equation}
We have not discussed the radius of convergence of the above series. In
practice, we need to sum the series within its radius of convergence and then analytically continue $\vec{y}$ as needed.

\section{Phase transition in matrix model with cubic potential}\label{app:cubic}

\begin{figure}
\begin{centering}
\includegraphics[height=4cm]{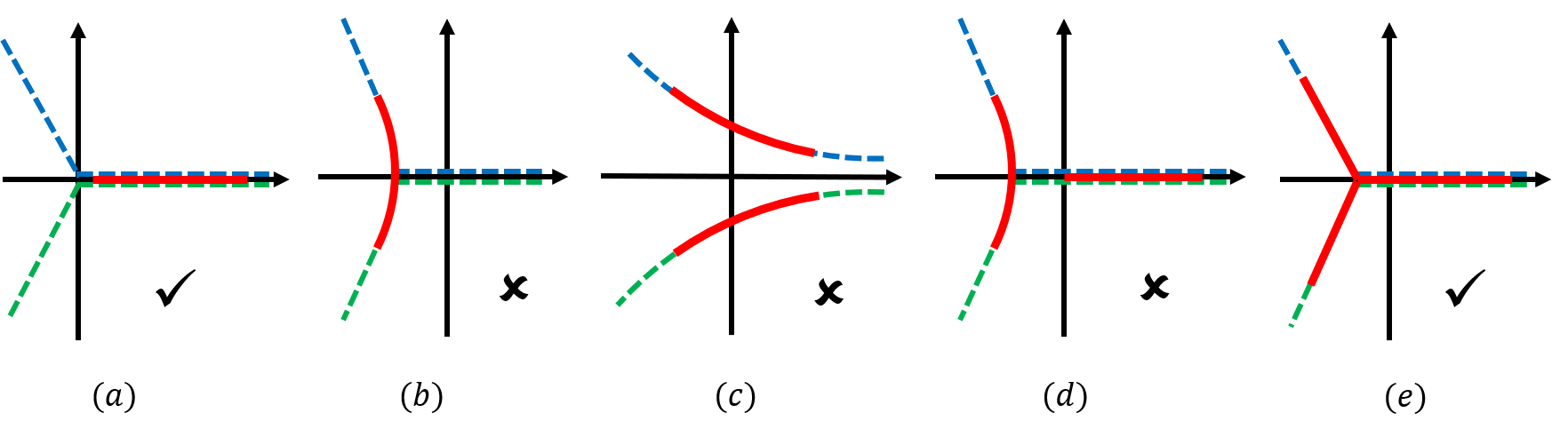}
\par\end{centering}
\caption{The red curves specify the possible cut shapes for $D_{\protect\r}$ in the complex
$E$ plane. The blue dashed line is $\protect\mC_{+}$, which connects $e^{2\pi i/3}\infty$
to $+\infty$, and the green dashed line is $\protect\mC_{-}$, which connects
$e^{-2\pi i/3}\infty$ to $+\infty$ . (a)(b) are one-cut configurations,
and (c)(d)(e) are two-cut configurations. Only (a) and (e) are solutions
to our cubic potential matrix model.\label{fig:All-possible-of}}
\end{figure}

Let us consider the matrix model with cubic potential $V(x)=x^{3}/3-tx^{2}/2$.
For $x\ra\infty$, there are three directions where $\Re V(x)\ra+\infty$,
namely $x\ra+\infty$ and $x \rightarrow \pm e^{2i\pi/3}\infty$. As discussed in
Section \ref{sec:4.1}, there are three contour choices for each eigenvalue.
We will assign half of the eigenvalues to the contour $\mC_{+}$ and the other half to the contour $\mC_{-}$, where $\mC_{\pm}$ is defined to be the contour that connects $\pm e^{2i\pi/3}\infty$ to $+\infty$ (see Fig. \ref{fig:All-possible-of}).
In this appendix, we use  a slightly different convention and divide both
$\r(E)$ and $R(E)$ in \eqref{eq:3.20} and \eqref{eq:3.21} by $N$ such that the density of states is normalized to unity.

\begin{figure}
\begin{centering}
\subfloat[$t>t_{cr}$]{\begin{centering}
\includegraphics[height=2.45cm]{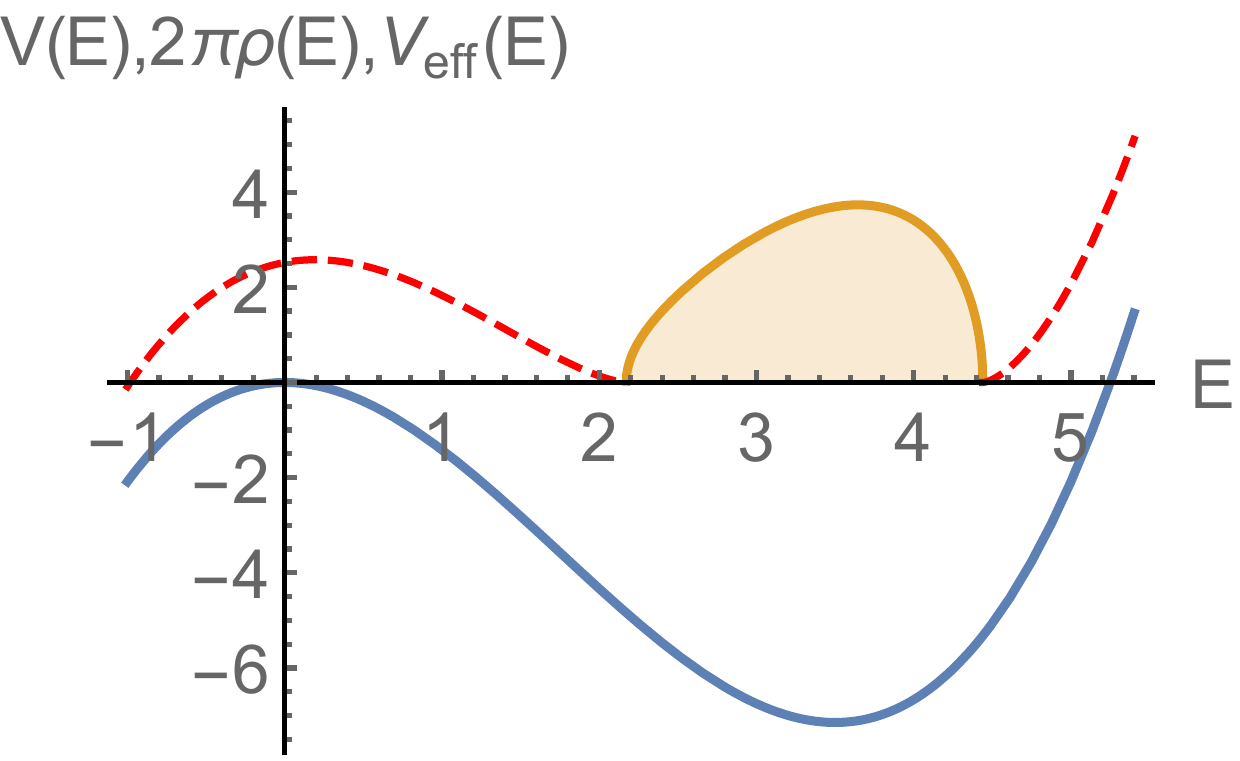}
\par\end{centering}
}\subfloat[$t=t_{cr}$]{\begin{centering}
\includegraphics[height=2.45cm]{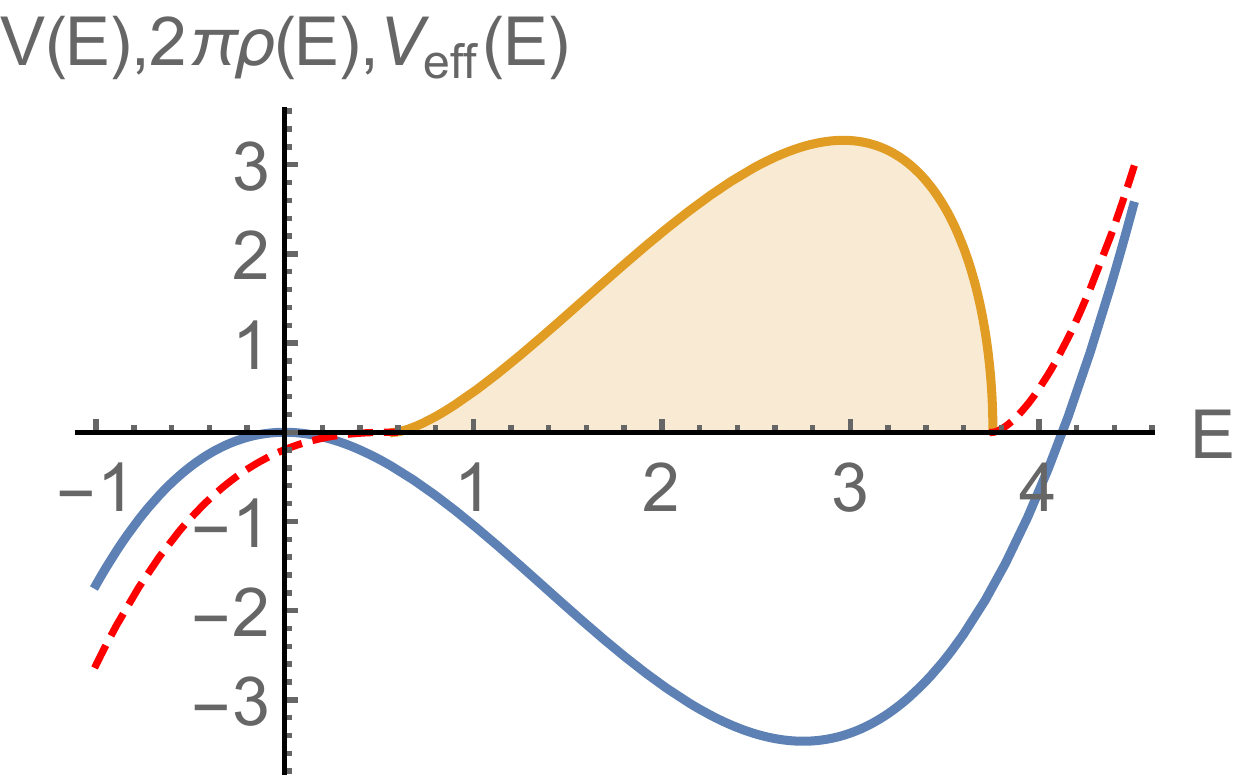}
\par\end{centering}
}\subfloat[$t=-t_{cr}$]{\begin{centering}
\includegraphics[height=2.45cm]{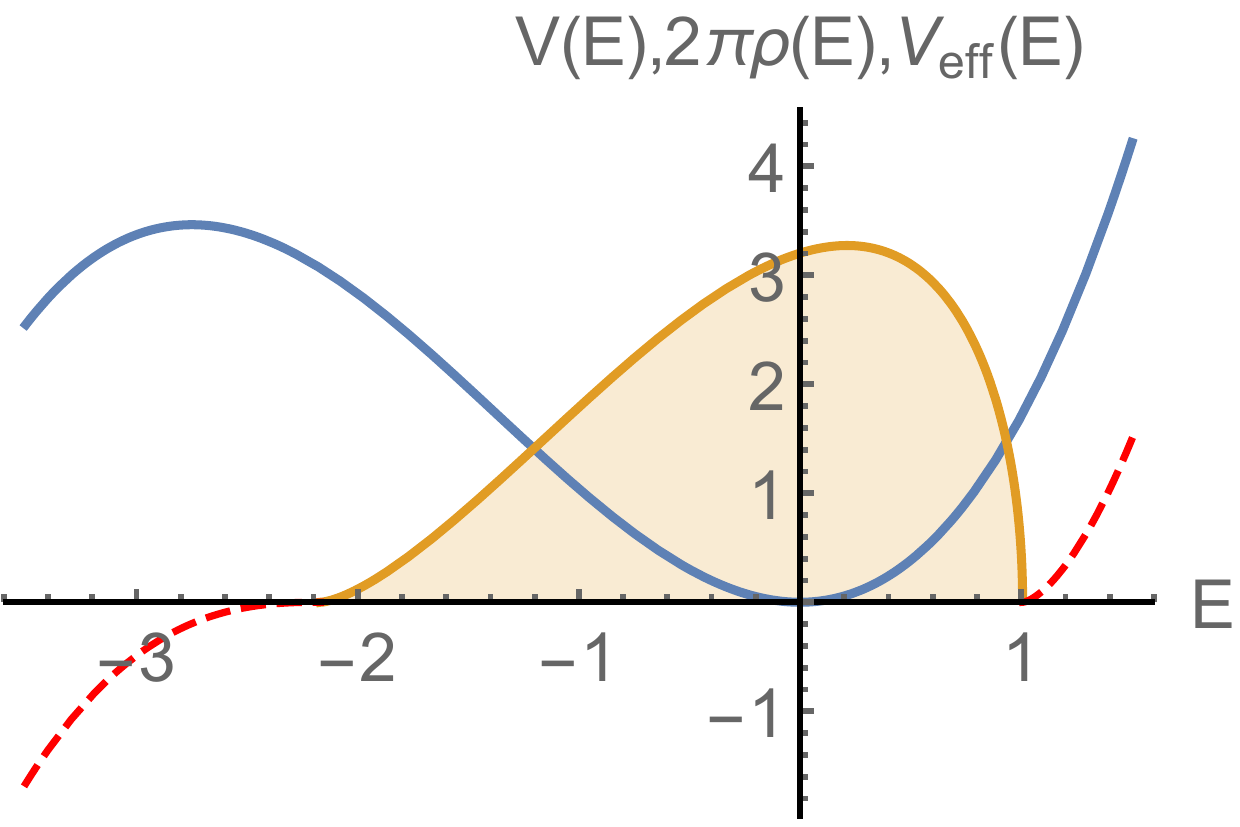}
\par\end{centering}
}\subfloat[$t<-t_{cr}$]{\begin{centering}
\includegraphics[height=2.45cm]{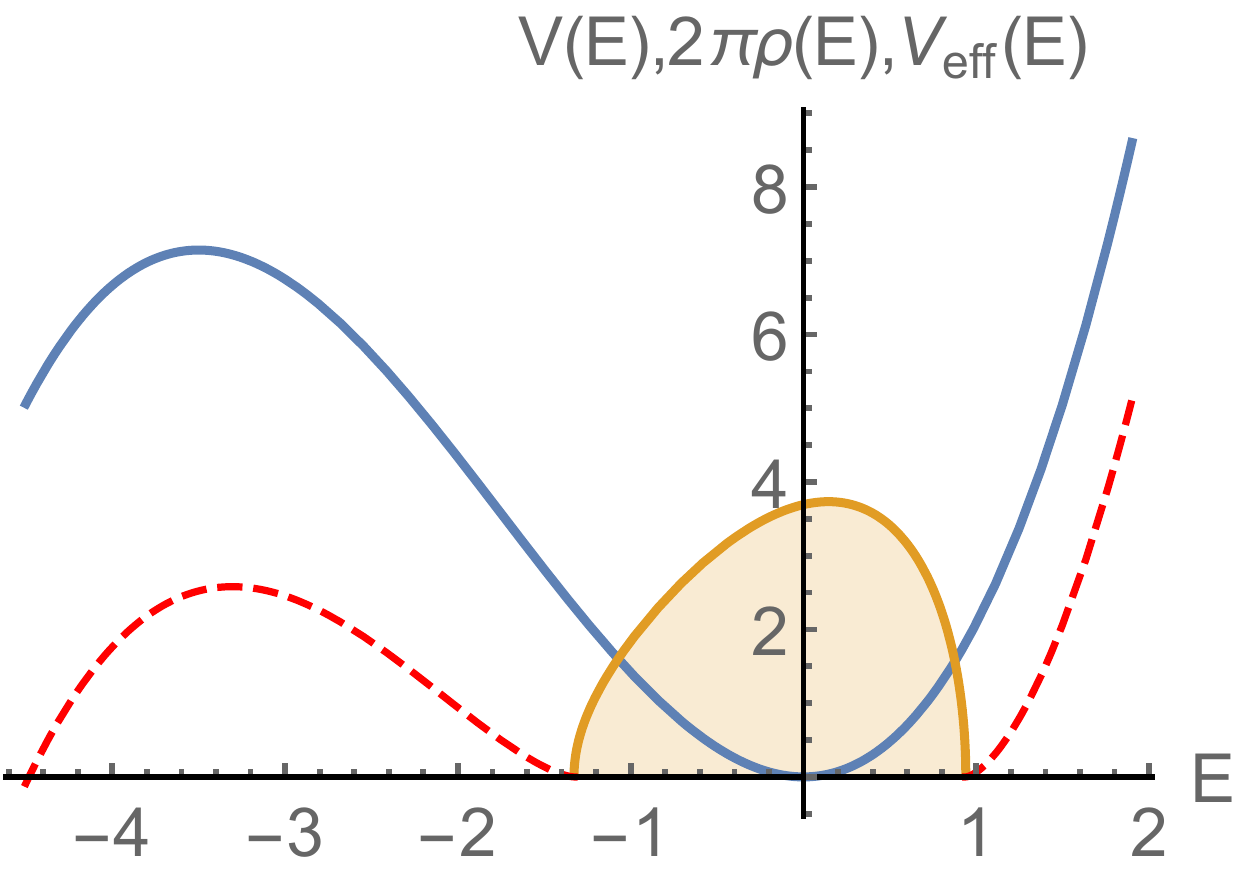}
\par\end{centering}
}
\par\end{centering}
\centering{}\caption{The potential $V(E)$ (blue), (rescaled) spectral density $2\pi\protect\r(E)$
(shaded yellow), and effective potential $V_\text{eff}(E)$ (dashed red) in the one-cut case for different $t$. \label{fig:Potential--(blue)}}
\end{figure}

For large $|t|$, we see that the potential has a deep well along
 the real axis (see Fig. \ref{fig:Potential--(blue)}), and we expect
 the eigenvalues to stabilize in the well and form a one-cut solution along the real axis. Because we have evenly assigned the eigenvalues to the two contours $\mC_{\pm}$, the spectrum must be invariant under a reflection across the real axis. Aside from a single cut along the real axis, another possibility that respects this symmetry is a single cut along a complex curve that is invariant under complex conjugation (see Fig. \ref{fig:All-possible-of} (a) and (b)).
However, this alternate possibility is forbidden by our contour choice because the real part of the effective potential \eqref{eq:89} must be minimized on $D_{\r}\cap \mC_{\pm}$, where $D_{\r}$ is the support of the spectrum \cite{Eynard:2015aea}.
However, a general fact of matrix models \cite{Eynard:2015aea} is that $\Re V_{\text{eff}}$
decreases along the direction normal to $D_{\r}$ (that is, $\Re V_{\text{eff}}$
decreases for some amount along the blue and green dashed
lines on the real axis in Fig. \ref{fig:All-possible-of} (b)).

Let the support of the one-cut spectrum be $[a,b]$. For a one-cut
solution of a polynomial potential $V$ with degree $d+1$, we will
closely follow the techniques in Section 3.3.3 of \cite{Eynard:2015aea}. As $R(x)$
is a two-fold covering of the complex plane with a cut along $[a,b]$, the topology
of $R(x)$ is a Riemann sphere. A biholomorphism between the complex plane (with coordinate $z$) and the Riemann sphere (with coordinate $x$) is given as follows:
\begin{equation}
z\mapsto x=\f{a+b}2+\d(z+1/z),\quad\d=\f{a-b}4\label{eq:app1}
\end{equation}
This biholomorphism is called the Joukowsky map. The upper (lower) sheet
is mapped to the exterior (interior) of the unit disc, and the cut is
mapped to the unit circle $|z|=1$. Sending $z\ra1/z$ amounts to switching the sheet for fixed
$x$. The inverse map is
\begin{equation}
z=\f 1{2\d}\left(x-\f{a+b}2\pm\sqrt{\left(x-\f{a+b}2\right)^{2}-4\d^{2}}\right)
\end{equation}
where the plus (minus) sign is for the upper (lower) sheet. Using (\ref{eq:app1}),
we can write
\begin{equation}
\sqrt{\s(x)}\equiv\sqrt{(x-a)(x-b)}=\d(z-1/z)
\end{equation}
where we choose the sign for the square root such that near $x\ra\infty$,
$\sqrt{\s(x)}\ra x\sim\d z$ on the upper sheet, and $\sqrt{\s(x)}\ra-x\sim-\d/z$
in the lower sheet. It follows that 
\begin{equation}
R(x)=\f 12(V'(x)-M(x)\sqrt{\s(x)})
\end{equation}
is a rational function of $z$ with poles at $z=0$ and $z=\infty$
because both $V'(x)$ and $M(x)$ are polynomials in $x$. Let us
denote $\bar{R}(z)=R(x(z))$. From the definition \eqref{eq:dfnR}, $R(x)\ra1/x+O(1/x^{2})$
for $x\ra\infty$ on upper sheet, which implies that $\bar{R}(z)\in\C[1/z]$,
or
\begin{equation}
\bar{R}(z)=\sum_{k=0}^{d}v_{k}z^{-k},\qquad\text{with }v_{0}=0,\;v_{1}=1/\d\label{eq:app5}
\end{equation}
where we used the fact that both $V'(x)^{2}$ and $M(x)^{2}\s(x)$
have degree $2d$. As $V'(x)=R(x+i\e)+R(x-i\e)$ for $x\in[a,b]$,
we have
\begin{equation}
V'(\f{a+b}2+\d(z+1/z))=\bar{R}(z)+\bar{R}(1/z)=\sum_{k=0}^{d}v_{k}(z^{k}+z^{-k})\label{eq:app6}
\end{equation}
Using (\ref{eq:app5}) and (\ref{eq:app6}), one can determine $a$
and $b$ and thus the spectrum.

Taking our potential into (\ref{eq:app6}), we have
\begin{align}
v_{0} & =\f 18(3a^{2}+2ab+3b^{2}-4(a+b)t)\\
v_{1} & =\f 14(a-b)(a+b-t)\\
v_{2} & =\f 1{16}(a-b)^{2}
\end{align}
Solving with the condition (\ref{eq:app5}) leads to $a=a(t)$ and $b=b(t)$,
though an explicit analytic expression is not available. It follows that
\begin{equation}
M(x)=\f{V'(x)-2R(x)}{\sqrt{\s(x)}}=\f{\sum_{k=0}^{d}v_{k}(z^{k}-z^{-k})}{\d(z-1/z)}=x+\f{a+b-2t}2.
\end{equation}
For $x\in[a,b]$, the spectral density is given by
\begin{equation}
\r(x)=\f 1{2\pi}M(x)\sqrt{-\s(x)}=\f 1{2\pi}\left(x+\f{a+b-2t}2\right)\sqrt{(x-a)(b-x)}.
\end{equation}
For $x>b$ or $x<a$, the derivative of the effective potential is given by an analytic continuation of $\rho(x)$,
\be
V'_{\text{eff}}(x)=M(x)\sqrt{\s(x)},
\ee
from which we can compute the effective potential. The critical point occurs when $x+\f{a+b-2t}2=0$ for $x=a$, which
means that near $a$, the spectrum scales like $(x-a)^{3/2}$. Solving
this condition with $a(t)$ and $b(t)$, we find two cases:
\begin{equation}
\begin{cases}
a=(3\sqrt{3}-5)^{1/3}\app0.581,\quad b=(3(9+5\sqrt{3}))^{1/3}\app3.756 & t=t_{cr}\\
a=-(3\sqrt{3}+5)^{1/3}\app-2.168,\quad b=(3(9-5\sqrt{3}))^{1/3}\app1.006 & t=-t_{cr}
\end{cases}\label{eq:app12}
\end{equation}
where $t_{cr}=2^{2/3}\sqrt{3}\app2.749$. For $t\in(-t_{cr},t_{cr})$,
one can show that no real solution exists for $a(t)$ and $b(t)$. We plot the one-cut spectrum for all cases with $|t|\geq t_{cr}$ in Fig. \ref{fig:Potential--(blue)}, where we see that for $|t|>t_{cr}$, the effective potential $V_{\text{eff}}(x)$ (dashed red lines) in the region $x < a$ increases and then decreases. For $|t| = t_{cr}$,  $V_{\text{eff}}(x)$ monotonically decreases for $x < a$ and has zero derivative at $x=a$. This is similar to our matrix model of JT gravity with EOW branes in \eqref{eq:115-1}.

Because a one-cut solution does not exist for $t\in(-t_{cr},t_{cr})$, we
need to consider two-cut solutions. There are three possibilities that respect the symmetry of $\mC_{\pm}$: two cuts symmetric under complex
conjugation (Fig. \ref{fig:All-possible-of} (c)), one cut along the real
axis plus one cut along a curve symmetric under complex conjugation
(Fig. \ref{fig:All-possible-of} (d)), and a special ``Y'' shaped case of the previous
type obtained by joining the two cuts at a junction (Fig.
\ref{fig:All-possible-of} (e)).\footnote{The case of two cuts along the real axis is forbidden because there is
no double-well in the potential along the real axis.} The first case cannot be stable because the Coulomb repulsive
force between the two cuts cannot be balanced by $V(x)$. The second
case is not allowed due to the requirement that $\Re V_{\text{eff}}$ is minimized
on $D_{\r}\cap\mC_{\pm}$. The third case is the only choice. Physically,
for $|t|$ small, the potential well along the real axis is not deep enough to hold all of the eigenvalues; this naturally leads to
complex eigenvalues along $\mC_{\pm}$.

\begin{figure}
\begin{centering}
\subfloat[\label{fig:app3a}]{\begin{centering}
\includegraphics[height=3.7cm]{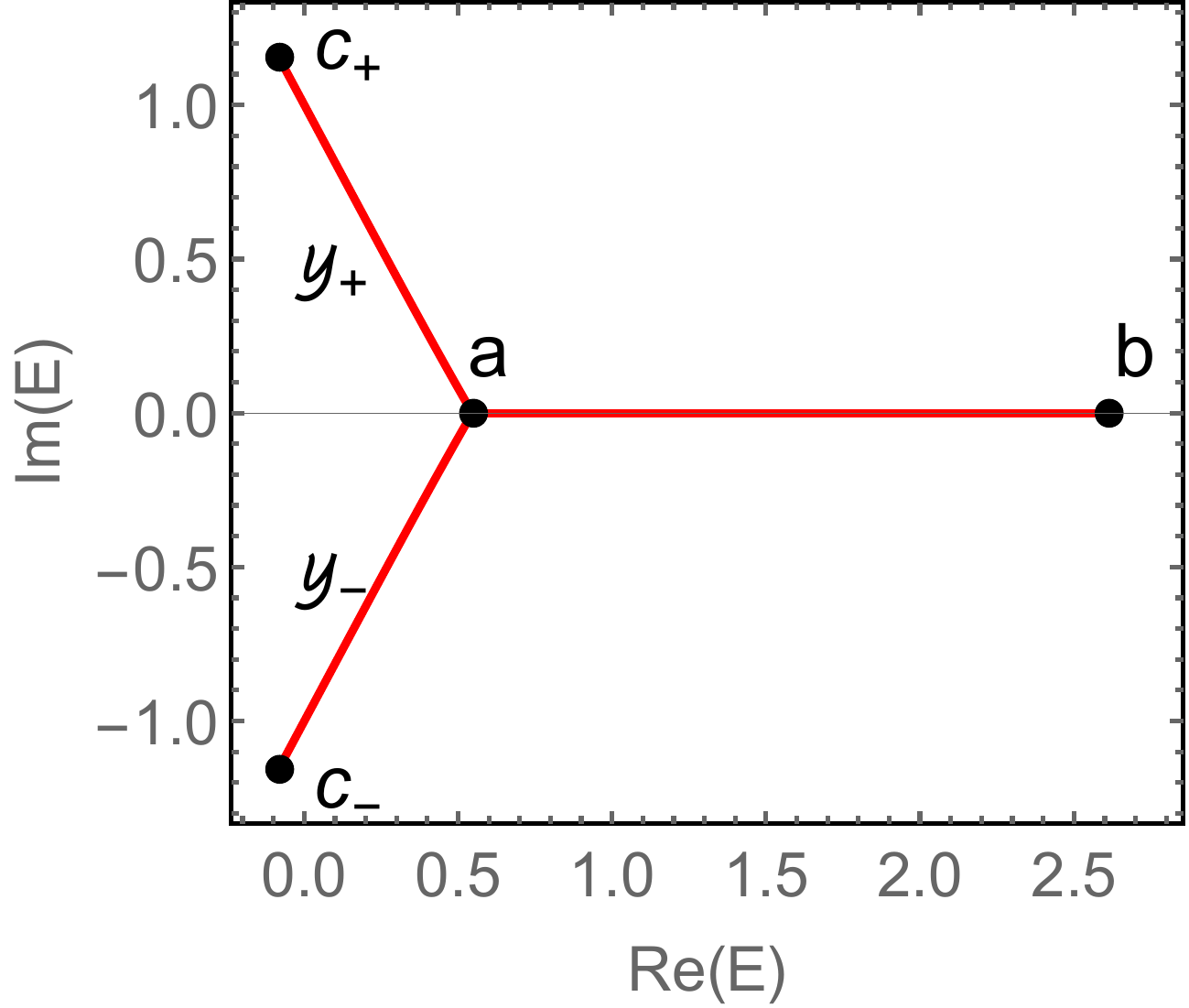}
\par\end{centering}
}\subfloat[\label{fig:app3b}]{\begin{centering}
\includegraphics[height=3.7cm]{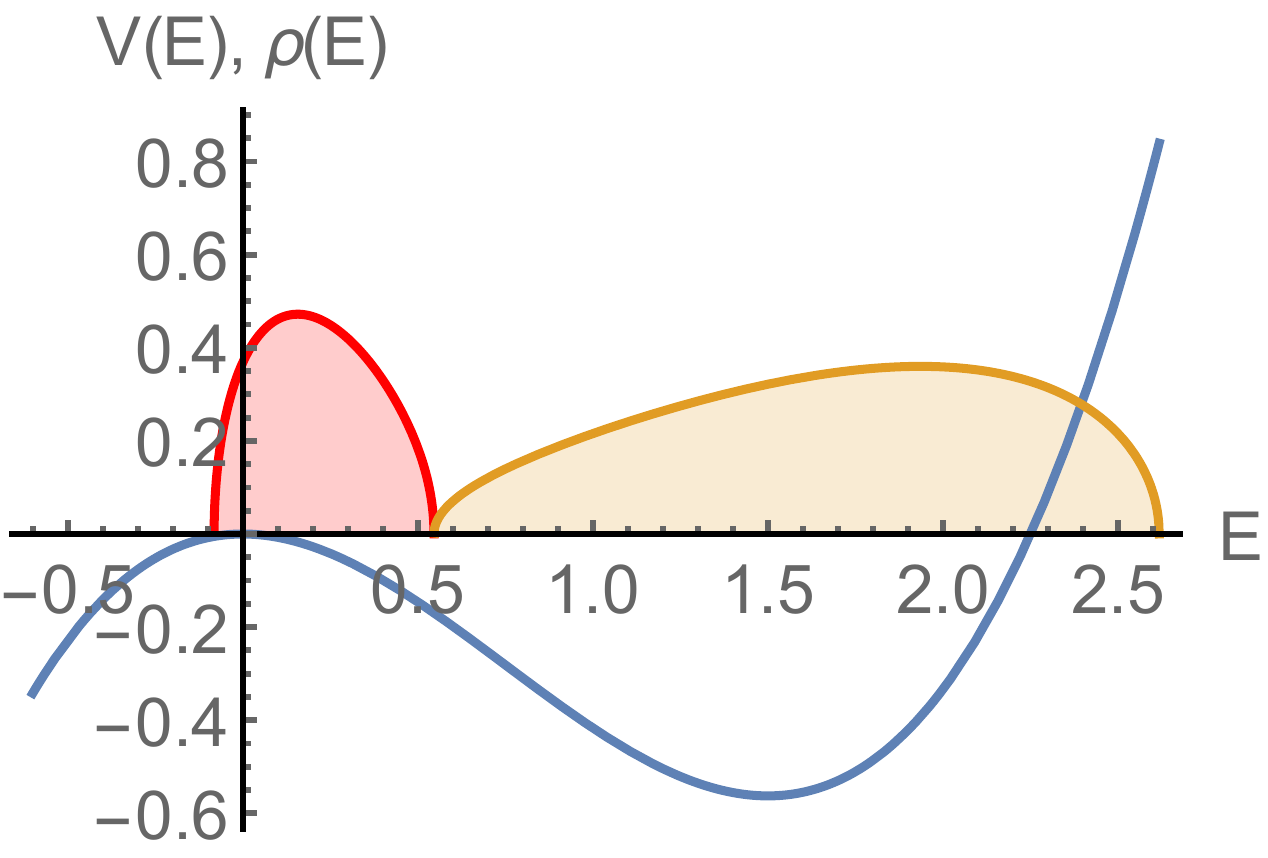}
\par\end{centering}
}\subfloat[\label{fig:app3c}]{\begin{centering}
\includegraphics[height=3.7cm]{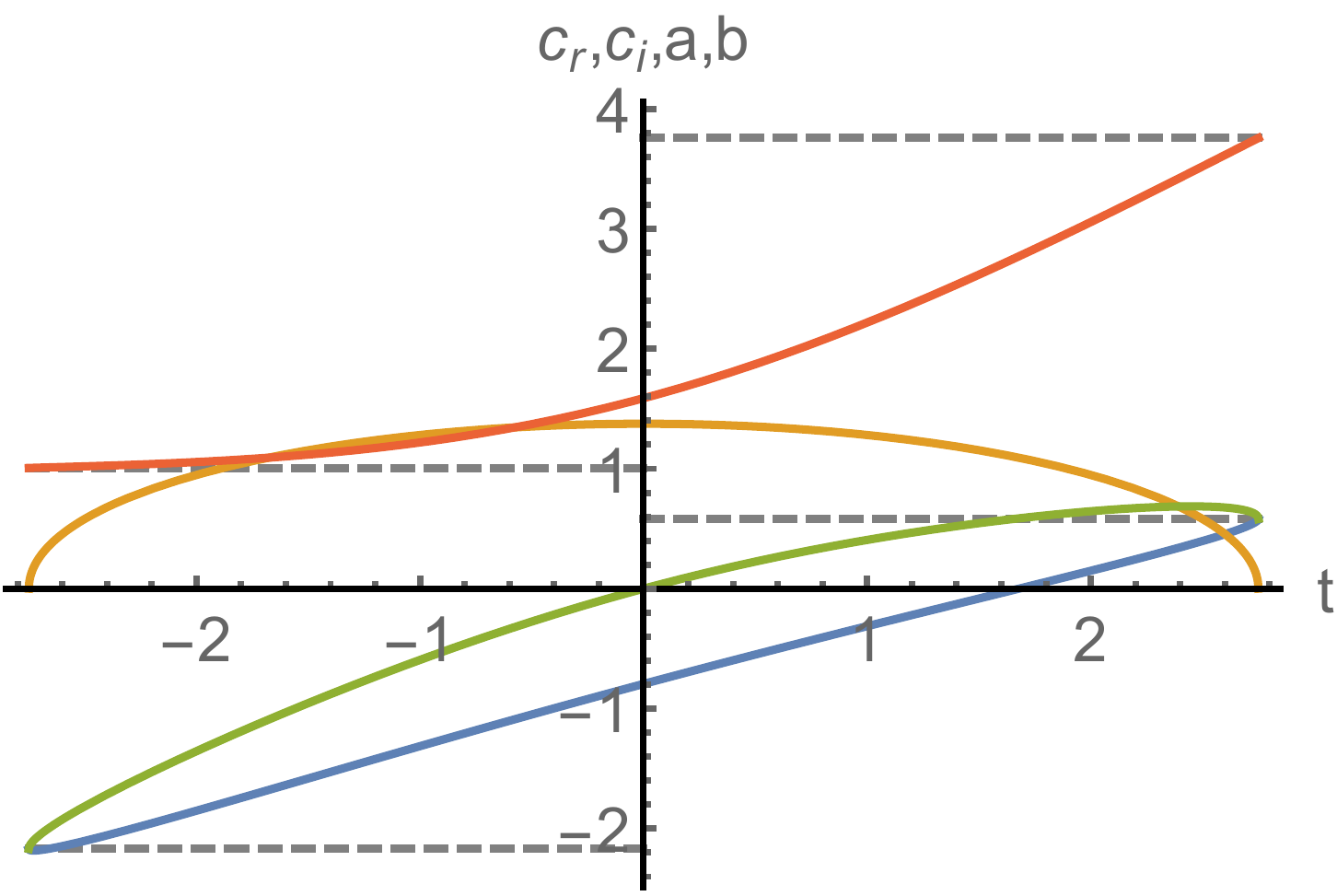}
\par\end{centering}
}
\par\end{centering}
\caption{(a) The ``Y'' shaped solution $D_{\protect\r}$ on the complex $E$ plane
for $t=3/2$. Note that $\protect\mY_{\pm}$ are not exactly straight lines.
(b) The potential $V(E)$ (blue) and ``Y'' shaped spectral density
(shaded yellow and red) for $t=3/2$. The potential well on the real axis is not deep enough to hold all of the eigenvalues in a real cut. The shaded red region is the plot
of the spectral density on $\protect\mY_{+}$ projected to the real axis as a function of $E$ (see \eqref{eq:c20}). (c) The plot of $c_{r}$ (blue), $c_{i}$
(yellow), $a$ (green) and $b$ (red) as functions of $t\in(-t_{cr},t_{cr})$.
The gray dashed lines mark the values that match with (\ref{eq:app12}).
\label{fig:The-plot-of ccc}}
\end{figure}
Let $c_{\pm}$ label the two complex ends of the ``Y'' shaped solution, let $b$ label the real end, and let $a$ label the junction (see Fig. \ref{fig:app3a}). Let $\mY_{\pm}$ denote the two complex branches.
As the degree of $M^{2}\s$ is $d=2$, $M$ must be a constant, $\t$.
It follows that 
\begin{equation}
R(x)=\f 12(x^{2}-tx-\t\sqrt{(x-c_{+})(x-c_{-})(x-a)(x-b)})\label{eq:app13}.
\end{equation}
Let $c_{\pm}=c_{r}\pm ic_{i}$. The asymptotic condition for $R(x)$
when $x\ra\infty$ requires that $R(x)\ra1/x$, which implies that
\begin{align}
\t & =1,\\
c_{i} & =\sqrt{c_{r}(c_{r}-t)+\f 2{t-2c_{r}}},\\
a,b & =t-c_{r}\mp\sqrt{c_{r}(t-c_{r})+\f 2{t-2c_{r}}}.
\end{align}
To solve for $c_{r}$, we still need one more equation. This is given
by requiring that $\r(x)dx$ is a real and nonnegative measure along $\mY_{\pm}$.
This condition fixes the locus of $\mY_{\pm}$ and also implies that 
\begin{equation}
\Im\int_{a}^{c_{+}}\r(x)dx=\f 1{2\pi}\Im\int_{a}^{c_{+}}\sqrt{(x-c_{+})(x-c_{-})(x-a)(b-x)}dx=0.
\end{equation}
Note that the above integral is invariant under deformations of the integration
path. We choose a straight line for computational simplicity.
We plot $c_{r},c_{i},a,b$ as functions of $t\in(-t_{cr},t_{cr})$
in Fig. \ref{fig:The-plot-of ccc}. From the plot, we see that at the
critical point $t=\pm t_{cr}$, $c_{i}=0$ and $c_{r}=a$, which means
that the ``Y'' shaped phase continuously connects to the one-cut phase.
Given a $t\in(-t_{cr},t_{cr})$, the locations of the branches $\mY_{\pm}$ are determined by the differential equation
\begin{equation}
\Im(\r(x)dx)=0\iff dx_{r}/dx_{i}=-\r_{r}(x)/\r_{i}(x)\label{eq:app18},
\end{equation}
where $x=x_{r}+ix_{i}$ and $\r(x)=\r_{r}(x)+i\r_{i}(x)$, and
the boundary conditions $x_{r}(c_{i})=c_{r}$ (or $ x_{r}(-c_{i})=c_{r}$)
and $x_{r}(0)=a$. As an example, we plot $D_{\r}$ of the ``Y'' shape
for $t=3/2$ in Fig. \ref{fig:app3a} and the spectral density in Fig. \ref{fig:app3b}.
Note that on $\mY_{\pm}$ we can project the density measure to the real
axis as follows:
\begin{equation}
\r(x)dx=|\r(x)|^{2}/\r_{r}(x)dx_{r}
\label{eq:c20}
\end{equation}
where we have used (\ref{eq:app18}).

\section{``Y'' shaped spectrum\label{sec:Y-shape-spectrum}}

\subsection*{Universal piece}

For the universal piece, we have 
\begin{equation}
\d R_{U}(E)=\f{e^{S_{0}}\phi_{b}}{4\pi^{2}i}\oint_{\mC}\f{d\lambda\sin(2\pi\sqrt{2\phi_{b}(-\lambda)})}{\lambda-E}\f{\sqrt{(E-E_{+})(E-E_{-})(E_{0}-E)}}{\sqrt{(\lambda-E_{+})(\lambda-E_{-})(E_{0}-\lambda)}},
\end{equation}
where $\mC$ is the contour circling the branch cut $\g_{+}\cup\g_{-}\cup[E_{0},0]$
anti-clockwise. We expand the sine function in a Taylor series and evaluate each term by moving the contour to infinity. It turns
out that the integral becomes the sum over the residues at $E$ and $\infty$,
\begin{equation}
\d R_{U}(E)=-\f{e^{S_{0}}\phi_{b}}{2\pi}\left[\sin(2\pi\sqrt{2\phi_{b}(-E)})-\f 1{2\pi i}\sum_{n=0}^{\infty}\f{(2\pi\sqrt{2\phi_{b}})^{2n+1}}{(2n+1)!}\oint_{\infty}d\lambda\f{\lambda^{n+1/2}}{\lambda-E}\f{\sqrt{(E-E_{+})(E-E_{-})(E_{0}-E)}}{\sqrt{(\lambda-E_{+})(\lambda-E_{-})(E_{0}-\lambda)}}\right].
\end{equation}
The first term contributes to $\d\r_{U}(E)=-\r_{JT}(E)$ for $E>0$
and cancels out $\r_{JT}(E)$ as in the one-cut case. Therefore
the spectral density only comes from the second integral around
infinity. Under the coordinate transformation $\lambda\ra1/z$, it becomes
\begin{align}
R_{U}(E)\simeq & \f{e^{S_{0}}\phi_{b}}{4\pi^{2}i}\sum_{n=0}^{\infty}\f{(2\pi\sqrt{2\phi_{b}})^{2n+1}}{(2n+1)!}\oint_{0}\f{dzz^{-n}}{1-Ez}\f{\sqrt{(E-E_{+})(E-E_{-})(E_{0}-E)}}{\sqrt{(1-E_{+}z)(1-E_{-}z)(1-E_{0}z)}}\nonumber \\
= & \f{e^{S_{0}}\phi_{b}\sqrt{(E-E_{+})(E-E_{-})(E_{0}-E)}}{2\pi}\sum_{n=1}^{\infty}\f{(2\pi\sqrt{2\phi_{b}})^{2n+1}E^{n-1}}{(2n+1)!}\nonumber \\
 & \times\sum_{\substack{n_{0}+n_{+}+n_{-}\leq n-1\\
n_{0},n_{+},n_{-}=0
}
}\f{(1/2)_{n_{0}}(1/2)_{n_{+}}(1/2)_{n_{-}}}{n_{0}!n_{+}!n_{-}!}\left(\f{E_{0}}E\right)^{n_{0}}\left(\f{E_{+}}E\right)^{n_{+}}\left(\f{E_{-}}E\right)^{n_{-}}\label{eq:168}
\end{align}
where in the second step we used a Taylor expansion and picked out the
coefficient of $z^{n-1}$. Let us denote the sum over all $n_{i}$
as $S_{n}(E_{i}/E)$ ($i=0,\pm$). We will rewrite it in a form similar
to the one-cut case.

First, we expand each $(E_{i}/E)^{n_{i}}=((E_{i}/E-1)+1)^{n_{i}}$
in powers of $(E_{i}/E-1)$
\begin{align}
S_{n}(E_{i}/E) & =\sum_{\substack{\sum n_{i}\leq n-1\\
n_{i}=0
}
}\sum_{k_{i}=0}^{n_{i}}\prod_{i=0,\pm}\f{(1/2)_{n_{i}}}{(n_{i}-k_{i})!k_{i}!}(E_{i}/E-1)^{k_{i}}\nonumber \\
 & =\sum_{\substack{\sum k_{i}\leq n-1\\
k_{i}=0
}
}\prod_{i=0,\pm}\f{(1/2)_{k_{i}}}{k_{i}!}(E_{i}/E-1)^{k_{i}}\sum_{\substack{\sum s_{i}\leq n-1-\sum_{i}k_{i}\\
k_{i}=0
}
}\prod_{i=0}^{2}\f{(1/2+k_{i})_{s_{i}}}{s_{i}!}\nonumber \\
 & =\sum_{\substack{\sum k_{i}\leq n-1\\
k_{i}=0
}
}\f{\G(n+3/2)}{\G(5/2+\sum_{i}k_{i})\G(n-\sum_{i}k_{i})}\prod_{i=0,\pm}\f{(1/2)_{k_{i}}}{k_{i}!}(E_{i}/E-1)^{k_{i}}\nonumber \\
 & =\sum_{\substack{\sum k_{i}\leq n-1\\
k_{i}=0
}
}\f{\G(n+3/2)}{\G(1/2)^{3}\G(n-\sum_{i}k_{i})}\int_{V_{3}}t_{0}^{k_{0}-1/2}t_{+}^{k_{+}-1/2}t_{-}^{k_{-}-1/2}dt_{0}dt_{+}dt_{-}\prod_{i=0,\pm}\f{(E_{i}/E-1)^{k_{i}}}{k_{i}!}\nonumber \\
 & =\f{\G(n+3/2)}{\G(1/2)^{3}(n-1)!}\int_{V_{3}}\f{\left(1+\sum_{i=0,\pm}t_{i}(E_{i}/E-1)\right)^{n-1}}{t_{0}^{1/2}t_{+}^{1/2}t_{-}^{1/2}}dt_{0}dt_{+}dt_{-}\label{eq:169}
\end{align}
where in the second line we defined $s_{i}=n_{i}-k_{i}$ and changed
the order of sum. In the third line we used the identity
\begin{equation}
\sum_{\sum k_{i}\leq m}\f{(a_{1})_{k_{1}}\cdots(a_{n})_{k_{n}}}{k_{1}!\cdots k_{n}!}=\f{\G(1+m+\sum_{i=1}^{n}a_{i})}{\G(1+\sum_{i=1}^{n}a_{i})\G(1+m)},
\end{equation}
and in the fourth line we used the identity
\begin{equation}
\f{\G(k_{1})\cdots\G(k_{n})}{\G(1+k_{1}+\cdots+k_{n})}=\int_{V_{n}}t_{1}^{k_{1}-1}\cdots t_{n}^{k_{n}-1}\prod_{i=1}^{n}dt_{i},\qquad V_{n}\equiv\{(t_{0},\cdots,t_{n-1})|t_{i}\geq0,\sum_{i=0}^{n-1}t_{i}\leq1\}
\end{equation}
and in the last line we computed the sum over $k_{i}$ which gives a simple
power function of order $n-1$. The sum over $n$ is straightforward,
\begin{equation}
\sum_{n=1}^{\infty}\f{(2\pi\sqrt{2\phi_{b}})^{2n+1}E^{n-1}}{(2n+1)!}S_{n}(E_{i}/E)=2\pi\phi_{b}\int_{V_{3}}\f{I_{1}\left(2\pi\sqrt{2\phi_{b}[E+\sum_{i=0,\pm}t_{i}(E_{i}-E)]}\right)}{\sqrt{E+\sum_{i=0,\pm}t_{i}(E_{i}-E)}t_{0}^{1/2}t_{+}^{1/2}t_{-}^{1/2}}dt_{0}dt_{+}dt_{-}.
\end{equation}
As $I_{1}(x)$ is an entire function, the integrand is analytic for
all $E$. Therefore, the discontinuity of $\d R_{U}(E)$ is purely
determined by the factor $\sqrt{(E-E_{+})(E-E_{-})(E_{0}-E)}$. We
choose the branch cut to be along $\g_{+}\cup\g_{-}\cup[E_{0},+\infty)$.
For $E\in[E_{0},+\infty)$, we have 
\begin{equation}
\r_{U}(E)=\f{e^{S_{0}}\phi_{b}^{2}\sqrt{(E-E_{+})(E-E_{-})(E-E_{0})}}{\pi}\int_{V_{3}}\f{I_{1}\left(2\pi\sqrt{2\phi_{b}[E+\sum_{i=0,\pm}t_{i}(E_{i}-E)]}\right)}{\sqrt{E+\sum_{i=0,\pm}t_{i}(E_{i}-E)}t_{0}^{1/2}t_{+}^{1/2}t_{-}^{1/2}}dt_{0}dt_{+}dt_{-}\label{eq:173}
\end{equation}
For $E\in\g_{\pm}$, $\r(E)$ differs from the above expression by a phase
that depends on the locus of $\g_{\pm}$, which is determined by requiring that
$\r(E)dE$ is real and nonnegative along $\g_{\pm}$. 

\subsection*{Potential related piece}

Using a similar trick, we have
\begin{align}
\d R_{K}(E) & =\f i{4\pi}\oint_{D_{\r}}d\lambda\f{K\d V'(\lambda)}{\lambda-E}\f{\sqrt{(E-E_{+})(E-E_{-})(E_{0}-E)}}{\sqrt{(\lambda-E_{+})(\lambda-E_{-})(E_{0}-\lambda)}}\nonumber \\
 & =\f{iK\phi_{b}}{2\pi}\int_{\mD}d\a m(\a)\oint_{\mC'}\f{d\lambda}{(\lambda-E)(\a^{2}+2\phi_{b}\lambda)}\f{\sqrt{(E-E_{+})(E-E_{-})(E_{0}-E)}}{\sqrt{(\lambda-E_{+})(\lambda-E_{-})(E_{0}-\lambda)}}\label{eq:D3}
\end{align}
where $D_{\r}\equiv\g_{+}\cup\g_{-}\cup[E_{0},+\infty)$ and $\mC '$ circles around $D_\r$ anti-clockwise.\footnote{For the unregulated measure \eqref{eq:malpha}, similar to the one-cut case discussed below \eqref{eq:32}, we assume that $-\a^2/(2\phi_b)$ is outside of $\mC '$ for all $\a\in\mD$.} Similarly,
we can move the contour $\mC '$ and this leads to the residues at $E$ and
$-\a^{2}/2\phi_{b}$. The residue at $E$ obviously does not contribute
to $\d\r_{K}(E)$. The residue at $-\a^{2}/2\phi_{b}$ gives
\begin{equation}
\d R_{K}(E)=K\phi_{b}\int_{\mD}d\a m(\a)\f{\sqrt{(E-E_{+})(E-E_{-})(E_{0}-E)}}{(\a^{2}+2\phi_{b}E)\sqrt{(\a^{2}/2\phi_{b}+E_{+})(\a^{2}/2\phi_{b}+E_{-})(\a^{2}/2\phi_{b}+E_{0})}}
\end{equation}
where $\sqrt{(\lambda-E_{+})(\lambda-E_{-})}$ in the denominator contributes
a minus sign by our branch cut prescription. For $E\in[E_{0},+\infty)$,
this leads to
\begin{equation}
\r_{K}(E)=\f{K\phi_{b}}{\pi}\int_{\mD}d\a m(\a)\f{\sqrt{(E-E_{+})(E-E_{-})(E-E_{0})}}{(\a^{2}+2\phi_{b}E)\sqrt{(\a^{2}/2\phi_{b}+E_{0})(\a^{2}/2\phi_{b}+E_{+})(\a^{2}/2\phi_{b}+E_{-})}}\label{eq:176}
\end{equation}
and for $E\in\g_{\pm}$ the expression depends on the locus of $\g_{\pm}$.

\subsection*{Large $E$ limit}

Taking the large $E$ limit of $\r(E)-\r_{JT}(E)$ only requires (\ref{eq:173})
and (\ref{eq:176}) and not the specific solution of $\g_{\pm}$.
One way to see the large $E$ behavior is to go back to (\ref{eq:168}).
Using the identity
\begin{align}
\sinh(2\pi\sqrt{2\phi_{b}E})= & \sqrt{(E-E_{+})(E-E_{-})(E-E_{0})}\sum_{n=0}^{\infty}\f{(2\pi\sqrt{2\phi_{b}})^{2n+1}E^{n-1}}{(2n+1)!}\nonumber \\
 & \times\sum_{n_{i}=0}^{\infty}\f{(1/2)_{n_{0}}(1/2)_{n_{+}}(1/2)_{n_{-}}}{n_{0}!n_{+}!n_{-}!}\left(\f{E_{0}}E\right)^{n_{0}}\left(\f{E_{+}}E\right)^{n_{+}}\left(\f{E_{-}}E\right)^{n_{-}},
\end{align}
we have
\begin{align}
\d\r_{U}(E) & =-\f{e^{S_{0}}\phi_{b}\sqrt{(E-E_{+})(E-E_{-})(E-E_{0})}}{2\pi^{2}}\sum_{n=0}^{\infty}\f{(2\pi\sqrt{2\phi_{b}})^{2n+1}E^{n-1}}{(2n+1)!}\nonumber \\
 & \times\sum_{n_{0}+n_{+}+n_{-}\geq n}\f{(1/2)_{n_{0}}(1/2)_{n_{+}}(1/2)_{n_{-}}}{n_{0}!n_{+}!n_{-}!}\left(\f{E_{0}}E\right)^{n_{0}}\left(\f{E_{+}}E\right)^{n_{+}}\left(\f{E_{-}}E\right)^{n_{-}}.
\end{align}
The large $E$ limit is easy to see from this formula. The leading
order is $E^{1/2}$ which comes from terms where $n_{0}+n_{+}+n_{-}=n$, and the next order
$E^{-1/2}$ involves terms where $n_{0}+n_{+}+n_{-}=n+1$. The higher order terms are
not relevant for checking the normalization of the spectrum. Define 
\begin{equation}
P_{n}(E_{i})\equiv\sum_{n_{0}+n_{+}+n_{-}=n}\f{(1/2)_{n_{0}}(1/2)_{n_{+}}(1/2)_{n_{-}}}{n_{0}!n_{+}!n_{-}!}E_{0}^{n_{0}}E_{+}^{n_{+}}E_{-}^{n_{-}},
\end{equation}
which, similarly to (\ref{eq:169}), can be written as
\begin{align}
P_{n}(E_{i}) & \equiv E_{0}^{n}\sum_{n_{0}=0}^{n}\f{(1/2)_{n_{0}}}{n_{0}!}\sum_{\sum_{\pm}k_{i}\leq n-n_{0}}\prod_{i=\pm}\f{(1/2)_{k_{i}}(E_{i0}/E_{0})^{k_{i}}}{k_{i}!}\sum_{\sum_{\pm}s_{i}=n-n_{0}-\sum_{\pm}k_{i}}\prod_{i=\pm}\f{(1/2+k_{i})_{s_{i}}}{s_{i}!}\nonumber \\
 & =E_{0}^{n}\sum_{n_{0}=0}^{n}\f{(1/2)_{n_{0}}}{n_{0}!}\sum_{\sum_{\pm}k_{i}\leq n-n_{0}}\f{(n-n_{0})!}{(k_{+}+k_{-})!(n-n_{0}-k_{+}-k_{-})!}\prod_{i=\pm}\f{(1/2)_{k_{i}}(E_{i0}/E_{0})^{k_{i}}}{k_{i}!}\nonumber \\
 & =E_{0}^{n}\sum_{n_{0}=0}^{n}\f{(1/2)_{n_{0}}}{n_{0}!}\sum_{\sum_{\pm}k_{i}\leq n-n_{0}}\f{(n-n_{0})!}{\pi(n-n_{0}-k_{+}-k_{-})!}\prod_{i=\pm}\f{(E_{i0}/E_{0})^{k_{i}}}{k_{i}!}\int_{0}^{1}u^{k_{+}-1/2}(1-u)^{k_{-}-1/2}du\nonumber \\
 & =\f 1{\pi}E_{0}^{n}\sum_{n_{0}=0}^{n}\f{(1/2)_{n_{0}}}{n_{0}!}\int_{0}^{1}\f{(E_{-}/E_{0}+E_{+-}u/E_{0})^{n-n_{0}}}{u^{1/2}(1-u)^{1/2}}du\nonumber \\
 & =\f{\G(n+3/2)}{\pi^{3/2}n!}\int_{0}^{1}\f{dw}{(1-w)^{1/2}}\int_{0}^{1}\f{du}{u^{1/2}(1-u)^{1/2}}(E_{0}+w(E_{+-}u-E_{0-}))^{n}.
\end{align}
In the second line we used
\begin{equation}
\sum_{\sum k_{i}=m}\f{(a_{1})_{k_{1}}\cdots(a_{n})_{k_{n}}}{k_{1}!\cdots k_{n}!}=\f{\G(m+\sum_{i=1}^{n}a_{i})}{\G(\sum_{i=1}^{n}a_{i})\G(1+m)}.
\end{equation}
In the third line we used the integral representation of the beta function.
Summing over $n$ leads to some Bessel functions and we have
\begin{align}
\d\r_{U}(E) \simeq & -E^{1/2}\cdot\f{e^{S_{0}}\phi_{b}^{3/2}}{\sqrt{2}\pi^{2}}\int_{0}^{1}\f{dw}{(1-w)^{1/2}}\int_{0}^{1}du\f{I_{0}(2\pi\sqrt{2\phi_{b}(E_{0}+w(E_{+-}u-E_{0-}))})}{u^{1/2}(1-u)^{1/2}}\nonumber \\
 & +E^{-1/2}\cdot\f{e^{S_{0}}\phi_{b}^{1/2}}{2\sqrt{2}\pi^{2}}\left[\int_{0}^{1}\f{dw}{(1-w)^{1/2}}\int_{0}^{1}du\f{I_{0}(2\pi\sqrt{2\phi_{b}(E_{0}+w(E_{+-}u-E_{0-}))})}{(\phi_{b}E_{0}+\phi_{b}E_{+-}(1-2u))^{-1}u^{1/2}(1-u)^{1/2}} \right. \nonumber \\
 &\left. -\sqrt{2\phi_{b}E_{0}}I_{1}(2\pi\sqrt{2\phi_{b}E_{0}})\right]
\end{align}
The large $E$ limit of $\r_{K}(E)$ is straightforward
\begin{align}
\r_{K}(E)\app & E^{1/2}\cdot\f K{2\pi}\int_{\mD}d\a\f{m(\a)}{\sqrt{(\a^{2}/2\phi_{b}+E_{0})(\a^{2}/2\phi_{b}+E_{+})(\a^{2}/2\phi_{b}+E_{-})}}\nonumber \\
 & -E^{-1/2}\cdot\f K{4\pi}\int_{\mD}d\a\f{ m(\a)[2\a^{2}E+(\a^2+2\phi_b E)(E_{0}+E_{+}+E_{-})]}{(\a^2+2\phi_b E)\sqrt{(\a^{2}/2\phi_{b}+E_{0})(\a^{2}/2\phi_{b}+E_{+})(\a^{2}/2\phi_{b}+E_{-})}}.
\end{align}
We can take the large $E$ limit inside the $d\alpha$ integral on the second line only when using a regulated measure.

\subsection*{Integration of $\protect\r(E)$ from $E_{0}$ to $E_{+}$ along the
straight line}

Define $E_{\pm}=E_{r}\pm iE_{i}$, $E_{r,i}\in\R$ and $E=uE_{+}+(1-u)E_{0}$.
The variable in the Bessel function in (\ref{eq:173}) is
\begin{equation}
E+\sum_{i=0,\pm}t_{i}(E_{i}-E)=E_{0}-E_{0r}y+iE_{i}x,
\end{equation}
where 
\begin{equation}
x\equiv t_{+-}+(1-t_{0}-t_{+}-t_{-})u,\qquad y\equiv t_{+}+t_{-}+(1-t_{0}-t_{+}-t_{-})u.
\end{equation}
Define $\mF(x)\equiv I_{1}(2\pi\sqrt{2\phi_{b}x})/\sqrt{x}$. Solving
$t_{\pm}$ in terms of $x$ and $y$, the integral of the universal piece
$\r_{U}(E)$ from $E_{0}$ to $E_{+}$ along the straight line is
\begin{align}
 & \int_{E_{0}}^{E_{+}}dE\r_{U}(E)\nonumber \\
= & -\f{e^{S_{0}}\phi_{b}^{2}}{\pi}E_{0+}^{5/2}\int_{0}^{1}dy\int_{-y}^{y}dx\int_{0}^{1-y}dt_{0}\int_{0}^{\f{x+y}{2-2t_{0}+x-y}}du\f{u^{1/2}(u-E_{0-}/E_{0+})^{1/2}\mF(E_{0}-E_{0r}y+iE_{i}x)}{t_{0}^{1/2}(2-2t_{0}+x-y)^{1/2}(y-x)^{1/2}\left(\f{x+y}{2-2t_{0}+x-y}-u\right)^{1/2}}\nonumber \\
= & -\f{e^{S_{0}}\phi_{b}^{2}}2E_{0+}^{2}E_{-0}^{1/2}\int_{0}^{1}dy\int_{-y}^{y}dx\int_{0}^{1-y}dt_{0}\f{\mF(E_{0}-E_{0r}y+iE_{i}x)(x+y){}_{2}F_{1}(-\f 12,\f 32,2,\f{x+y}{2-2t_{0}+x-y}\f{E_{0+}}{E_{0-}})}{t_{0}^{1/2}(2-2t_{0}+x-y)^{3/2}(y-x)^{1/2}}\nonumber \\
= & -\f{e^{S_{0}}\phi_{b}^{2}}{2\sqrt{2}}E_{0+}^{2}E_{-0}^{1/2}\int_{0}^{1}dy\int_{-y}^{y}dx\f{(x+y)^{1/2}\mF(E_{0}-E_{0r}y+iE_{i}x)}{(y-x)^{1/2}(2+x-y)^{1/2}}\int_{\f{x+y}{2+x-y}}^{1}dq\f{_{2}F_{1}(-\f 12,\f 32,2,q\f{E_{0+}}{E_{0-}})}{(q-\f{x+y}{2+x-y})^{1/2}}\label{eq:184}
\end{align}
where we defined $q=\f{x+y}{2-2t_{0}+x-y}$ in the last line. The
integral of the potential related piece is 
\begin{equation}
\int_{E_{0}}^{E_{+}}dE\r_{K}(E)=-\f K{16}E_{0+}^{2}E_{-0}^{1/2}\int_{\mD}d\a\f{m(\a)F_{1}(\f 32;-\f 12,-\f 12;3;\f{E_{0+}}{E_{0-}},\f{E_{0+}}{\a^{2}/2\phi_{b}+E_{0}})}{(\a^{2}/2\phi_{b}+E_{0})^{3/2}\sqrt{(\a^{2}/2\phi_{b}+E_{+})(\a^{2}/2\phi_{b}+E_{-})}}
\end{equation}
where $F_{1}$ is Appell function.

\bibliographystyle{JHEP.bst}
\bibliography{EOWbranes.bib}

\providecommand{\href}[2]{#2}\begingroup\raggedright\begin{thebibliography}{10}

\bibitem{Almheiri:2019qdq}
A.~Almheiri, T.~Hartman, J.~Maldacena, E.~Shaghoulian and A.~Tajdini,
  \emph{{Replica Wormholes and the Entropy of Hawking Radiation}},
  \href{https://doi.org/10.1007/JHEP05(2020)013}{\emph{JHEP} {\bfseries 05}
  (2020) 013} [\href{https://arxiv.org/abs/1911.12333}{{\ttfamily
  1911.12333}}].

\bibitem{Penington:2019kki}
G.~Penington, S.H.~Shenker, D.~Stanford and Z.~Yang, \emph{{Replica wormholes
  and the black hole interior}},
  \href{https://arxiv.org/abs/1911.11977}{{\ttfamily 1911.11977}}.

\bibitem{Goto:2020wnk}
K.~Goto, T.~Hartman and A.~Tajdini, \emph{{Replica wormholes for an evaporating
  2D black hole}},  \href{https://arxiv.org/abs/2011.09043}{{\ttfamily
  2011.09043}}.

\bibitem{Saad:2018bqo}
P.~Saad, S.H.~Shenker and D.~Stanford, \emph{{A semiclassical ramp in SYK and
  in gravity}},  \href{https://arxiv.org/abs/1806.06840}{{\ttfamily
  1806.06840}}.

\bibitem{Marolf:2020rpm}
D.~Marolf and H.~Maxfield, \emph{{Observations of Hawking radiation: the Page
  curve and baby universes}},
  \href{https://arxiv.org/abs/2010.06602}{{\ttfamily 2010.06602}}.

\bibitem{Engelhardt:2020qpv}
N.~Engelhardt, S.~Fischetti and A.~Maloney, \emph{{Free energy from replica
  wormholes}}, \href{https://doi.org/10.1103/PhysRevD.103.046021}{\emph{Phys.
  Rev. D} {\bfseries 103} (2021) 046021}
  [\href{https://arxiv.org/abs/2007.07444}{{\ttfamily 2007.07444}}].

\bibitem{Giddings:2020yes}
S.B.~Giddings and G.J.~Turiaci, \emph{{Wormhole calculus, replicas, and
  entropies}}, \href{https://doi.org/10.1007/JHEP09(2020)194}{\emph{JHEP}
  {\bfseries 09} (2020) 194}
  [\href{https://arxiv.org/abs/2004.02900}{{\ttfamily 2004.02900}}].

\bibitem{Jackiw:1984je}
R.~Jackiw, \emph{{Lower Dimensional Gravity}},
  \href{https://doi.org/10.1016/0550-3213(85)90448-1}{\emph{Nucl. Phys. B}
  {\bfseries 252} (1985) 343}.

\bibitem{Teitelboim:1983ux}
C.~Teitelboim, \emph{{Gravitation and Hamiltonian Structure in Two Space-Time
  Dimensions}}, \href{https://doi.org/10.1016/0370-2693(83)90012-6}{\emph{Phys.
  Lett. B} {\bfseries 126} (1983) 41}.

\bibitem{Saad:2019lba}
P.~Saad, S.H.~Shenker and D.~Stanford, \emph{{JT gravity as a matrix
  integral}},  \href{https://arxiv.org/abs/1903.11115}{{\ttfamily 1903.11115}}.

\bibitem{Harlow:2018tqv}
D.~Harlow and D.~Jafferis, \emph{{The Factorization Problem in
  Jackiw-Teitelboim Gravity}},
  \href{https://doi.org/10.1007/JHEP02(2020)177}{\emph{JHEP} {\bfseries 02}
  (2020) 177} [\href{https://arxiv.org/abs/1804.01081}{{\ttfamily
  1804.01081}}].

\bibitem{Goel:2020yxl}
A.~Goel, L.V.~Iliesiu, J.~Kruthoff and Z.~Yang, \emph{{Classifying boundary
  conditions in JT gravity: from energy-branes to $\alpha$-branes}},
  \href{https://arxiv.org/abs/2010.12592}{{\ttfamily 2010.12592}}.

\bibitem{Witten:2020ert}
E.~Witten, \emph{{Deformations of JT Gravity and Phase Transitions}},
  \href{https://arxiv.org/abs/2006.03494}{{\ttfamily 2006.03494}}.

\bibitem{Witten:2020wvy}
E.~Witten, \emph{{Matrix Models and Deformations of JT Gravity}},
  \href{https://doi.org/10.1098/rspa.2020.0582}{\emph{Proc. Roy. Soc. Lond. A}
  {\bfseries 476} (2020) 20200582}
  [\href{https://arxiv.org/abs/2006.13414}{{\ttfamily 2006.13414}}].

\bibitem{Maxfield:2020ale}
H.~Maxfield and G.J.~Turiaci, \emph{{The path integral of 3D gravity near
  extremality; or, JT gravity with defects as a matrix integral}},
  \href{https://arxiv.org/abs/2006.11317}{{\ttfamily 2006.11317}}.

\bibitem{Maldacena:2016upp}
J.~Maldacena, D.~Stanford and Z.~Yang, \emph{{Conformal symmetry and its
  breaking in two dimensional Nearly Anti-de-Sitter space}},
  \href{https://doi.org/10.1093/ptep/ptw124}{\emph{PTEP} {\bfseries 2016}
  (2016) 12C104} [\href{https://arxiv.org/abs/1606.01857}{{\ttfamily
  1606.01857}}].

\bibitem{gradshteyn2014table}
I.S.~Gradshteyn and I.M.~Ryzhik, \emph{Table of integrals, series, and
  products}, Academic press (2014).

\bibitem{Mertens:2020hbs}
T.G.~Mertens and G.J.~Turiaci, \emph{{Liouville quantum gravity -- holography,
  JT and matrices}},  \href{https://arxiv.org/abs/2006.07072}{{\ttfamily
  2006.07072}}.

\bibitem{Johnson:2020lns}
C.V.~Johnson and F.~Rosso, \emph{{Solving Puzzles in Deformed JT Gravity: Phase
  Transitions and Non-Perturbative Effects}},
  \href{https://arxiv.org/abs/2011.06026}{{\ttfamily 2011.06026}}.

\bibitem{Eynard:2015aea}
B.~Eynard, T.~Kimura and S.~Ribault, \emph{{Random matrices}},
  \href{https://arxiv.org/abs/1510.04430}{{\ttfamily 1510.04430}}.

\bibitem{Johnson:2019eik}
C.V.~Johnson, \emph{{Nonperturbative Jackiw-Teitelboim gravity}},
  \href{https://doi.org/10.1103/PhysRevD.101.106023}{\emph{Phys. Rev. D}
  {\bfseries 101} (2020) 106023}
  [\href{https://arxiv.org/abs/1912.03637}{{\ttfamily 1912.03637}}].

\bibitem{Marino:2009dp}
M.~Marino, S.~Pasquetti and P.~Putrov, \emph{{Large N duality beyond the genus
  expansion}}, \href{https://doi.org/10.1007/JHEP07(2010)074}{\emph{JHEP}
  {\bfseries 07} (2010) 074} [\href{https://arxiv.org/abs/0911.4692}{{\ttfamily
  0911.4692}}].

\bibitem{bleher2016mother}
P.~Bleher and G.~Silva, \emph{The mother body phase transition in the normal
  matrix model},  \href{https://arxiv.org/abs/1601.05124}{{\ttfamily
  1601.05124}}.

\bibitem{PhysRevD.49.5227}
D.~Louis-Martinez and G.~Kunstatter, \emph{Birkhoff's theorem in
  two-dimensional dilaton gravity},
  \href{https://doi.org/10.1103/PhysRevD.49.5227}{\emph{Phys. Rev. D}
  {\bfseries 49} (1994) 5227}.

\bibitem{Hawking:1982dh}
S.~Hawking and D.N.~Page, \emph{{Thermodynamics of Black Holes in anti-De
  Sitter Space}}, \href{https://doi.org/10.1007/BF01208266}{\emph{Commun. Math.
  Phys.} {\bfseries 87} (1983) 577}.

\bibitem{Blommaert:2019wfy}
A.~Blommaert, T.G.~Mertens and H.~Verschelde, \emph{{Eigenbranes in
  Jackiw-Teitelboim gravity}},
  \href{https://arxiv.org/abs/1911.11603}{{\ttfamily 1911.11603}}.

\bibitem{Johnson:2020heh}
C.V.~Johnson, \emph{{JT Supergravity, Minimal Strings, and Matrix Models}},
  \href{https://arxiv.org/abs/2005.01893}{{\ttfamily 2005.01893}}.

\bibitem{Johnson:2020exp}
C.V.~Johnson, \emph{{Explorations of Non-Perturbative JT Gravity and
  Supergravity}},  \href{https://arxiv.org/abs/2006.10959}{{\ttfamily
  2006.10959}}.

\bibitem{Johnson:2020mwi}
C.V.~Johnson, \emph{{Low Energy Thermodynamics of JT Gravity and
  Supergravity}},  \href{https://arxiv.org/abs/2008.13120}{{\ttfamily
  2008.13120}}.

\bibitem{Banks:1989df}
T.~Banks, M.R.~Douglas, N.~Seiberg and S.H.~Shenker, \emph{{Microscopic and
  Macroscopic Loops in Nonperturbative Two-dimensional Gravity}},
  \href{https://doi.org/10.1016/0370-2693(90)91736-U}{\emph{Phys. Lett. B}
  {\bfseries 238} (1990) 279}.

\bibitem{Jafferis:2019wkd}
D.L.~Jafferis and D.K.~Kolchmeyer, \emph{{Entanglement Entropy in
  Jackiw-Teitelboim Gravity}},
  \href{https://arxiv.org/abs/1911.10663}{{\ttfamily 1911.10663}}.

\bibitem{Marolf:2020xie}
D.~Marolf and H.~Maxfield, \emph{{Transcending the ensemble: baby universes,
  spacetime wormholes, and the order and disorder of black hole information}},
  \href{https://doi.org/10.1007/JHEP08(2020)044}{\emph{JHEP} {\bfseries 08}
  (2020) 044} [\href{https://arxiv.org/abs/2002.08950}{{\ttfamily
  2002.08950}}].

\end{thebibliography}\endgroup

\end{document}